\documentclass{report}

\usepackage[utf8]{inputenc} % For text encoding
\usepackage{amsmath}        % For advanced math environments
\usepackage{amssymb}        % For math symbols like \mathcal
\usepackage{graphicx}       % To include graphics
\usepackage{booktabs}       % For professional-looking tables
\usepackage{algorithm}      % For algorithm environment
\usepackage{algpseudocode}  % For algorithmic pseudocode
\usepackage{bm}             % For bold math symbols
\usepackage{pdfpages}
\usepackage{siunitx}

\usepackage[
    backend=biber,
    style=ieee,
    sorting=none
]{biblatex}
\newcommand{\matr}[1]{\mathbf{#1}}
\newcommand{\vect}[1]{\mathbf{#1}}
\newcommand{\reals}{\mathbb{R}}

\begin{document}

\includepdf[pages=1, fitpaper=true]{Frontispiece_Data_Science_thesisV2_2024-2025_A.Y.pdf}

\chapter*{Abstract}
The optimal assignment of Large Language Models (LLMs) to specialized roles in multi-agent systems is a significant challenge, defined by a vast combinatorial search space, expensive black-box evaluations, and an inherent trade-off between performance and cost. Current optimization methods focus on single-agent settings and lack a principled framework for this multi-agent, multi-objective problem.

This thesis introduces MALBO (Multi-Agent LLM Bayesian Optimization), a systematic framework designed to automate the efficient composition of LLM-based agent teams. We formalize the assignment challenge as a multi-objective optimization problem, aiming to identify the Pareto front of configurations between task accuracy and inference cost. The methodology employs multi-objective Bayesian Optimization (MOBO) with independent Gaussian Process surrogate models. By searching over a continuous feature-space representation of the LLMs, this approach performs a sample-efficient exploration guided by the expected hypervolume improvement.

The primary contribution is a principled and automated methodology that yields a Pareto front of optimal team configurations. Our results demonstrate that the Bayesian optimization phase, compared to an initial random search, maintained a comparable average performance while reducing the average configuration cost by over 45\%. Furthermore, MALBO identified specialized, heterogeneous teams that achieve cost reductions of up to 65.8\% compared to homogeneous baselines, all while maintaining maximum performance. The framework thus provides a data-driven tool for deploying cost-effective and highly specialized multi-agent AI systems.

\clearpage

\clearpage

\tableofcontents
\clearpage
\chapter{Introduction}
\label{ch:introduction}

\section{From Monolithic Models to Multi-Agent Systems}
\label{sec:intro_context}

The field of artificial intelligence has undergone a paradigm shift, catalyzed by the advent of Large Language Models (LLMs). Architectures like the Transformer \cite{vaswani2017attention} have enabled the creation of foundational models with remarkable capabilities in language understanding, reasoning, and generation. Initially, the focus of this revolution was on scaling these models to be larger and more powerful, treating them as monolithic entities to be prompted for a wide array of tasks.

More recently, the focus has evolved. The first step in this evolution was to enhance the \textit{agentic capabilities} of individual models, equipping them with advanced tool-calling functionalities to interact with external systems. This focus on agentic capabilities has become particularly prominent since the introduction of models like Anthropic's Claude 3.5, which established new benchmarks for sophisticated, multi-step tool use \cite{anthropic2024claude35haiku}. The current frontier, however, pushes beyond single-agent execution. Major research labs and the open-source community are now releasing dedicated multi-agent frameworks such as Google's Agent Development Kit, OpenAI's Agents SDK, and frameworks like LangGraph designed to orchestrate teams of specialized agents. This paradigm, where complex problems are decomposed and solved collaboratively, marks a significant step towards more autonomous and capable AI systems but also introduces novel challenges in their design and deployment.

\section{The Agent Composition Challenge}
\label{sec:intro_problem}

The efficacy of these multi-agent systems is critically dependent on their composition. The central question is no longer just how to prompt a single model, but which LLM to assign to each distinct agent role. An orchestrating 'manager' agent may require a model with strong planning skills, a 'tool-using' agent may need proficiency in code generation, and a 'verifier' agent may prioritize factual accuracy.

This assignment task gives rise to a vast combinatorial design space. With a pool of $M$ available LLMs and a team of $N$ agents, the number of possible configurations is $M^N$. To the best of our knowledge, no systematic or mathematical method for solving this specific problem has been documented in the literature. The selection is typically performed manually, guided by heuristics. This often results in one of two suboptimal strategies: assigning the most capable (and expensive) model from public benchmarks to all roles, or assigning a single, cost-effective model across the board. Both approaches ignore the potential for optimization through specialized, heterogeneous team composition. Navigating this space systematically is hindered by two fundamental obstacles:

\begin{enumerate}
    \item \textbf{The Black-Box Evaluation Problem:} The performance of any given team configuration can only be determined by executing it on a representative task, a process that is computationally expensive and time-consuming. The relationship between a configuration and its outcome is an opaque, black-box function with no accessible analytical form or gradient.
    \item \textbf{The Multi-Objective Imperative:} A viable solution must balance two inherently conflicting objectives. We aim to \textbf{maximize the collective performance} of the agent team on its designated task while simultaneously \textbf{minimizing the aggregate operational cost}, which is typically dominated by API inference fees.
\end{enumerate}

This intersection of a combinatorial search space, expensive black-box evaluations, and conflicting objectives defines a challenging new optimization frontier.

\section{The Gap in Existing Research}
\label{sec:intro_gap}

The current state of the art in LLM optimization has primarily focused on single-agent or monolithic pipeline settings. Research has yielded powerful techniques for prompt engineering, instruction tuning, and hyperparameter optimization for Retrieval-Augmented Generation (RAG) systems \cite{Archetti2023PromptBO, Sabbatella2024BOInG, Barker2025RAG}. While these methods are valuable, they do not address the unique challenges posed by multi-agent systems. To the best of our knowledge, the literature still lacks a principled and automated framework for the multi-objective, multi-agent LLM assignment problem. The composition of agent teams is thus often left to manual heuristics, trial-and-error, or intuition, a process that is neither scalable nor guaranteed to find optimal solutions.

\section{Proposed Solution: Introducing MALBO}
\label{sec:intro_thesis_statement}
\begin{figure}
    \centering
    \includegraphics[width=1\linewidth]{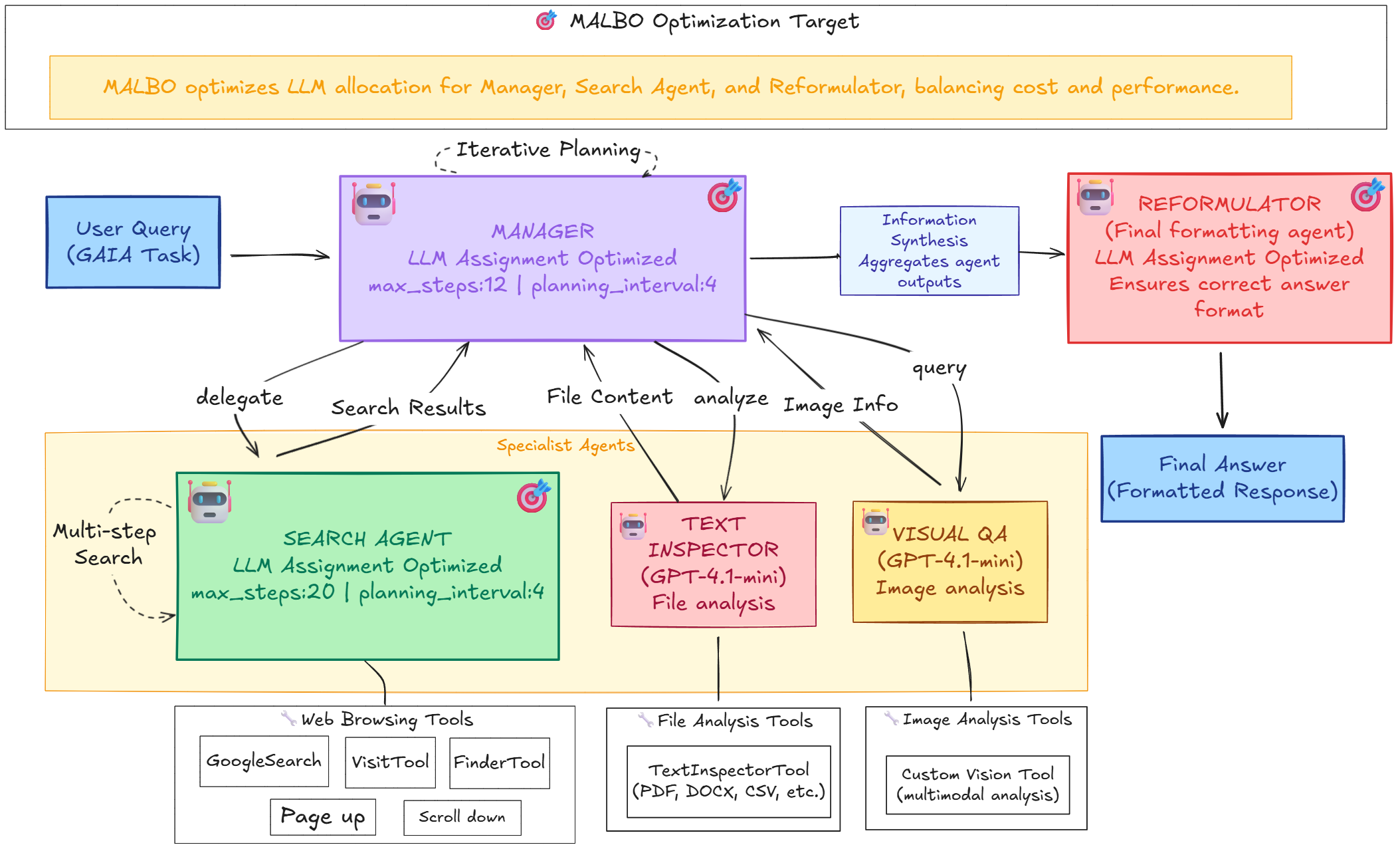}
    \caption{Overview of the multi-agent LLM system architecture. The \textbf{Manager} orchestrates interactions among specialized agents, the \textbf{Search Agent}, \textbf{Visual QA}, and \textbf{Text Inspector}, each equipped with specific toolsets for web browsing, image analysis, and file analysis, respectively. Both the \textbf{Manager} and the \textbf{Search Agent} operate in iterative loops: the Manager performs re-planning every 2 steps, while the Search Agent refines its search every 4 steps. The \textbf{Manager} aggregates all agent outputs through an information synthesis phase before passing them to the \textbf{Reformulator}, which produces the final formatted response. The \textbf{MALBO} optimization process dynamically assigns LLM configurations to balance cost and performance across key agents. In the figure, agents are represented by a robot icon at the top left, and those optimized by MALBO are marked with a red target icon at the top right.}

    \label{fig:malboAgentsv3}
\end{figure}
This thesis introduces and evaluates \textbf{MALBO (Multi-Agent LLM Bayesian Optimization)}, a novel and systematic framework designed to automate the efficient composition of LLM-based agent teams (see Fig. \ref{fig:malboAgentsv3} for a graphical illustration of the DeepResearch Agent Team used in the MALBO validation tests). We formalize the agent assignment challenge as a multi-objective, black-box optimization problem, with the explicit goal of identifying the set of configurations that form the optimal Pareto front between task performance and inference cost.

The MALBO methodology is built on a continuous relaxation of the discrete assignment problem. We first represent each available LLM as a vector in a continuous feature space, capturing its capabilities and costs. A complete team configuration is thus a point in a high-dimensional continuous space. We employ Bayesian Optimization with independent Gaussian Process surrogate models to probabilistically model the two conflicting objectives. The search is guided by the q-Expected Hypervolume Improvement (qEHVI) acquisition function, which intelligently selects new candidate teams to evaluate, ensuring a sample-efficient exploration of the search space. A projection function then maps these "ideal" continuous solutions back to discrete, deployable LLM assignments for evaluation.

\section{Research Questions and Objectives}
\label{sec:intro_rq}

To guide our investigation, we formulate the following primary research questions, which address the core challenges and intended outcomes of this work:

\begin{enumerate}
    \item \textbf{RQ1: Formalization.} How can the problem of assigning LLMs to different roles in a multi-agent system be formally cast as a multi-objective, black-box optimization problem suitable for Bayesian Optimization?
    
    \item \textbf{RQ2: Efficiency and Optimization.} Can a Bayesian Optimization framework efficiently explore the design space to not only identify a Pareto front, but specifically to improve the \textbf{cost-efficiency} of high-performing configurations within a constrained evaluation budget?
    
    \item \textbf{RQ3: Structural Insights.} What structural insights can be extracted from the optimization process? Specifically, can this data-driven approach identify which agent roles and which model features are the most \textbf{influential drivers} of system performance and cost?
\end{enumerate}

\section{Key Contributions}
\label{sec:intro_contributions}

The primary contributions of this thesis are delivered across three integrated stages. We begin by introducing a novel formalization of the LLM-to-agent assignment challenge, providing a structured foundation where previously only heuristics existed. Building upon this, we develop and implement the MALBO framework, a practical and sample-efficient methodology for automatically discovering the Pareto front of optimal team configurations. We then demonstrate the framework's validity through an empirical study that yields quantifiable insights into the architectural drivers of performance and cost.

\section{Thesis Outline}
\label{sec:intro_outline}

The remainder of this thesis is structured as follows:

\begin{itemize}
    \item \textbf{Chapter 2: Theoretical Background} provides a comprehensive overview of the foundational concepts underpinning this research, including the Transformer architecture, the evolution of Large Language Models, the paradigms of Multi-Agent Systems, and the principles of Bayesian Optimization.
    \item \textbf{Chapter 3: Related Works} reviews the existing literature on the application of Bayesian Optimization to LLM-related problems, contextualizing our work and highlighting the research gap that MALBO aims to fill.
    \item \textbf{Chapter 4: MALBO: Methodology and Mathematical Formulation} presents a detailed, formal description of our proposed framework, detailing the vector representation of LLMs, the problem formulation, and the components of the Bayesian optimization loop.
    \item \textbf{Chapter 5: Experimental Setup} describes the complete design of our empirical evaluation, including the software platform, the evaluation benchmark, the performance metrics, the pool of LLMs, and the configuration of the optimizer.
    \item \textbf{Chapter 6: Results and Analysis} presents and analyzes the empirical findings from our experiments, examining both the convergence of the optimization process and the practical insights derived from the optimal configurations discovered.
    \item \textbf{Chapter 7: Conclusion, Limitations, and Future Prospective} summarizes the key findings of this thesis, discusses the limitations of our current work, and proposes promising directions for future research.
\end{itemize}

\chapter{Theoretical background}
\section{The Transformer Architecture and Foundations of Large Language Models (LLMs)}

The advent of the Transformer architecture, introduced by Vaswani et al. in their seminal 2017 paper "Attention Is All You Need," marked a paradigm shift in natural language processing (NLP) \cite{vaswani2017attention}. Before the Transformer, sequence modeling tasks were dominated by recurrent neural network (RNN) architectures, such as Long Short-Term Memory (LSTM) networks. While effective, RNNs process sequential data step-by-step, an inherently sequential computation that limits parallelization and poses challenges in capturing long-range dependencies due to the "information bottleneck" of a single hidden state vector. While the Transformer architecture, discussed next, abandoned recurrence entirely, it is noteworthy that recent research has begun to. These limitations have recently spurred new research into hybrid architectures, such as Mamba, which combine principles from both recurrent and parallelizable models to offer alternative scaling paradigms \cite{gu2023mamba}.

The Transformer architecture proposed a novel solution: to dispense with recurrence entirely and rely solely on an attention mechanism to draw global dependencies between input and output. This design not only yielded superior performance on tasks like machine translation but also enabled significantly more parallelization, allowing for training on much larger datasets than was previously feasible. This scalability is the cornerstone upon which modern Large Language Models (LLMs) are built.

In this section, we provide a detailed overview of the foundational components of the Transformer architecture. We begin with the initial processing of input data and then delve into the core mechanisms that define its operation: the self-attention mechanism, multi-head attention, position-wise feed-forward networks, and the use of residual connections with layer normalization.
\begin{figure}
    \centering
    \includegraphics[width=0.4\linewidth]{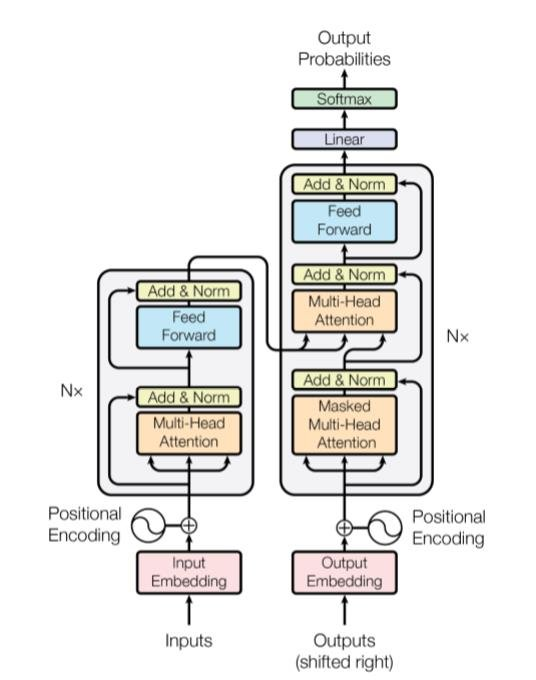}
    \caption{The Transformer architecture by \cite{vaswani2017attention}, illustrating the transformer architecture, including embedding, positional encoding, self-attention mechanism, and Feed-Forward Networks.}
    \label{fig:transformer_architecture}
\end{figure}

\subsection{Input Processing: Embeddings and Positional Encoding}

Unlike RNNs that ingest tokens sequentially, the Transformer processes an entire sequence of tokens at once. To do so, the input text must first be converted into a numerical representation. This involves two primary steps.

\paragraph{Tokenization and Token Embedding}
First, the raw text is segmented into a sequence of tokens using a subword tokenization algorithm like Byte-Pair Encoding (BPE) \cite{sennrich2016neural}. Each token in the vocabulary is then mapped to a unique integer ID. These IDs are used to retrieve a corresponding dense vector representation from an embedding matrix, $W_e \in \mathbb{R}^{d_{\text{model}} \times N_v}$, where $N_v$ is the vocabulary size and $d_{\text{model}}$ is the dimensionality of the embedding vectors.

\paragraph{Positional Encoding}
Because the model contains no recurrence, the self-attention mechanism is inherently permutation-invariant; it has no sense of the order of tokens in the sequence. To provide the model with this crucial information, a positional encoding vector is added to each token embedding. The original Transformer paper employed sinusoidal functions of different frequencies for this purpose:
\begin{align}
    \text{PE}_{(pos, 2i)} &= \sin(pos / 10000^{2i/d_{\text{model}}}) \\
    \text{PE}_{(pos, 2i+1)} &= \cos(pos / 10000^{2i/d_{\text{model}}})
\end{align}
where $pos$ is the position of the token in the sequence and $i$ is the dimension index of the embedding. This method allows the model to learn to attend to relative positions, as the positional encoding for any position can be represented as a linear function of any other. Other models, such as BERT, use learned positional embeddings instead \cite{devlin2018bert}. The final input representation for each token is the sum of its token embedding and its positional encoding.

\subsection{The Self-Attention Mechanism}

The core innovation of the Transformer is the self-attention mechanism. It allows the model to weigh the importance of different tokens in the input sequence when producing a representation for each token. The mechanism is based on the concepts of Query ($Q$), Key ($K$), and Value ($V$). For each input token, we create three vectors: a Query vector, a Key vector, and a Value vector by multiplying its embedding by three distinct, learnable weight matrices ($W_Q$, $W_K$, $W_V$).

\begin{figure}[h!]
    \centering
    \includegraphics[width=0.7\textwidth]{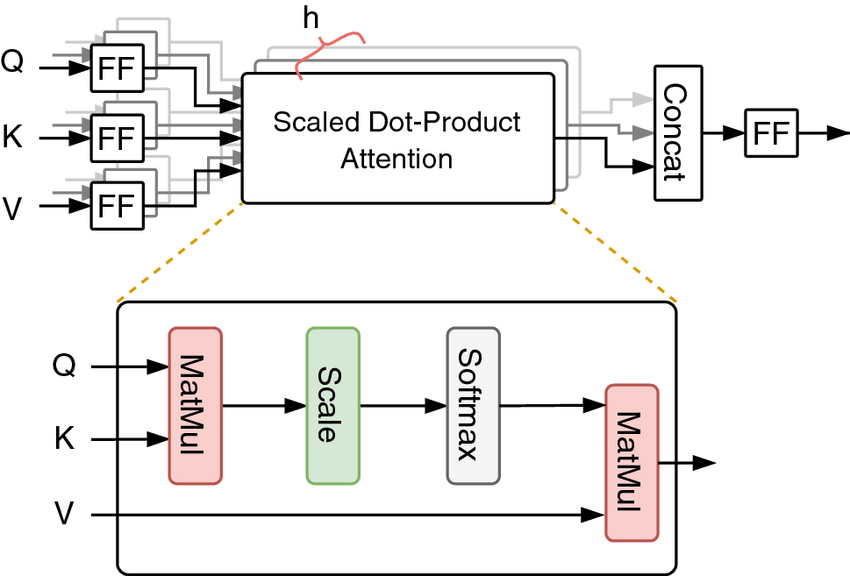}
    \caption{The Scaled Dot-Product Attention mechanism. The dot product of queries and keys is scaled and passed through a softmax function to obtain weights for the value vectors.}
    \label{fig:scaled-dot-product}
\end{figure}

The attention score is computed as the dot product of the Query vector of the current token with the Key vectors of all other tokens in the sequence. This score determines how much attention the current token should pay to every other token. These scores are then scaled, passed through a softmax function to create a probability distribution, and used to compute a weighted sum of the Value vectors. The complete operation, known as Scaled Dot-Product Attention, is concisely expressed as:
\begin{equation}
    \text{Attention}(Q, K, V) = \text{softmax}\left(\frac{QK^T}{\sqrt{d_k}}\right)V
    \label{eq:attention}
\end{equation}
Here, $Q, K, V$ are matrices containing the stacked query, key, and value vectors for all tokens in the sequence. The scaling factor $\sqrt{d_k}$, where $d_k$ is the dimension of the key vectors, is crucial for stabilizing gradients during training. A visualization of this mechanism is shown in Figure \ref{fig:scaled-dot-product}. From a theoretical standpoint, recent work has begun to interpret the self-attention dynamics as a form of Wasserstein gradient flow, providing a deeper mathematical grounding for the architecture's behavior \cite{chewi2024statistical}.

\subsection{Multi-Head Attention}

Rather than performing a single attention function, the authors found it beneficial to linearly project the queries, keys, and values $h$ times with different, learned linear projections. This allows the model to jointly attend to information from different representation subspaces at different positions. This mechanism is called Multi-Head Attention (MHA).

Each of these projected versions of queries, keys, and values is fed to an attention function in parallel, producing $h$ output vectors. These are then concatenated and once again projected with a final weight matrix $W_O$ to produce the final output of the MHA layer.
\begin{align}
    \text{MultiHead}(Q, K, V) &= \text{Concat}(\text{head}_1, \dots, \text{head}_h)W^O \\
    \text{where head}_i &= \text{Attention}(QW_i^Q, KW_i^K, VW_i^V)
\end{align}
The projection matrices $W_i^Q$, $W_i^K$, and $W_i^V$ are unique for each attention head $i$. MHA enhances the model's ability to focus on different positions and different types of relationships (e.g., syntactic vs. semantic).

\paragraph{Position-wise Feed-Forward Networks}
In addition to attention sub-layers, each layer of the encoder and decoder contains a fully connected feed-forward network (FFN), which is applied to each position separately and identically. This consists of two linear transformations with a ReLU activation in between:
\begin{equation}
    \text{FFN}(x) = \max(0, xW_1 + b_1)W_2 + b_2
\end{equation}
This component adds non-linearity to the model, increasing its expressive power.

The Transformer architecture is composed of a stack of these core components, organized into an encoder and a decoder.

\paragraph{Residual Connections and Layer Normalization}
Each sub-layer in the model (both the MHA and the feed-forward network) is followed by a residual connection and a layer normalization step. The output of each sub-layer is thus $\text{LayerNorm}(x + \text{Sublayer}(x))$, where $\text{Sublayer}(x)$ is the function implemented by the sub-layer itself. These residual connections are vital for training very deep networks by allowing gradients to flow more directly through the network.

\subsection{Architectural Variants and Foundational Models}

The core components of the Transformer (Multi-Head Attention, Position-wise Feed-Forward Networks, residual connections, and layer normalization) are stacked to form deeper, more powerful models. The specific arrangement and utilization of the encoder and decoder blocks give rise to three primary architectural families, each suited for different classes of tasks \cite{ji2025overview}.

\paragraph{Encoder-Decoder Architectures}
The full Transformer architecture, which incorporates both the encoder and decoder stacks, is primarily employed for sequence-to-sequence (seq2seq) tasks. In this configuration, the encoder processes the entire input sequence to generate a set of contextual representations. The decoder then attends to these representations (via cross-attention) while autoregressively generating the output sequence. This architecture is the standard for tasks like machine translation and text summarization. Foundational models built on this design include \textbf{T5 (Text-to-Text Transfer Transformer)} \cite{raffel2020exploring} and \textbf{BART (Bidirectional and Auto-Regressive Transformers)} \cite{lewis2020bart}.

\paragraph{Encoder-Only Architectures}
This variant utilizes only the encoder stack. The model processes an input sequence and outputs a rich, contextualized embedding for each token. These embeddings serve as powerful features for downstream tasks that require a deep understanding of the input text, such as text classification, named entity recognition, or sentiment analysis. The canonical example of this family is \textbf{BERT (Bidirectional Encoder Representations from Transformers)} \cite{devlin2018bert}, which is pre-trained using a Masked Language Modeling (MLM) objective. By predicting randomly masked tokens, BERT learns deep bidirectional context, a significant departure from the unidirectional context of traditional language models. Its successor, \textbf{RoBERTa}, further optimized the pre-training process to achieve improved performance \cite{liu2019roberta}.

\begin{figure}[h!]
    \centering
    \includegraphics[width=0.9\textwidth]{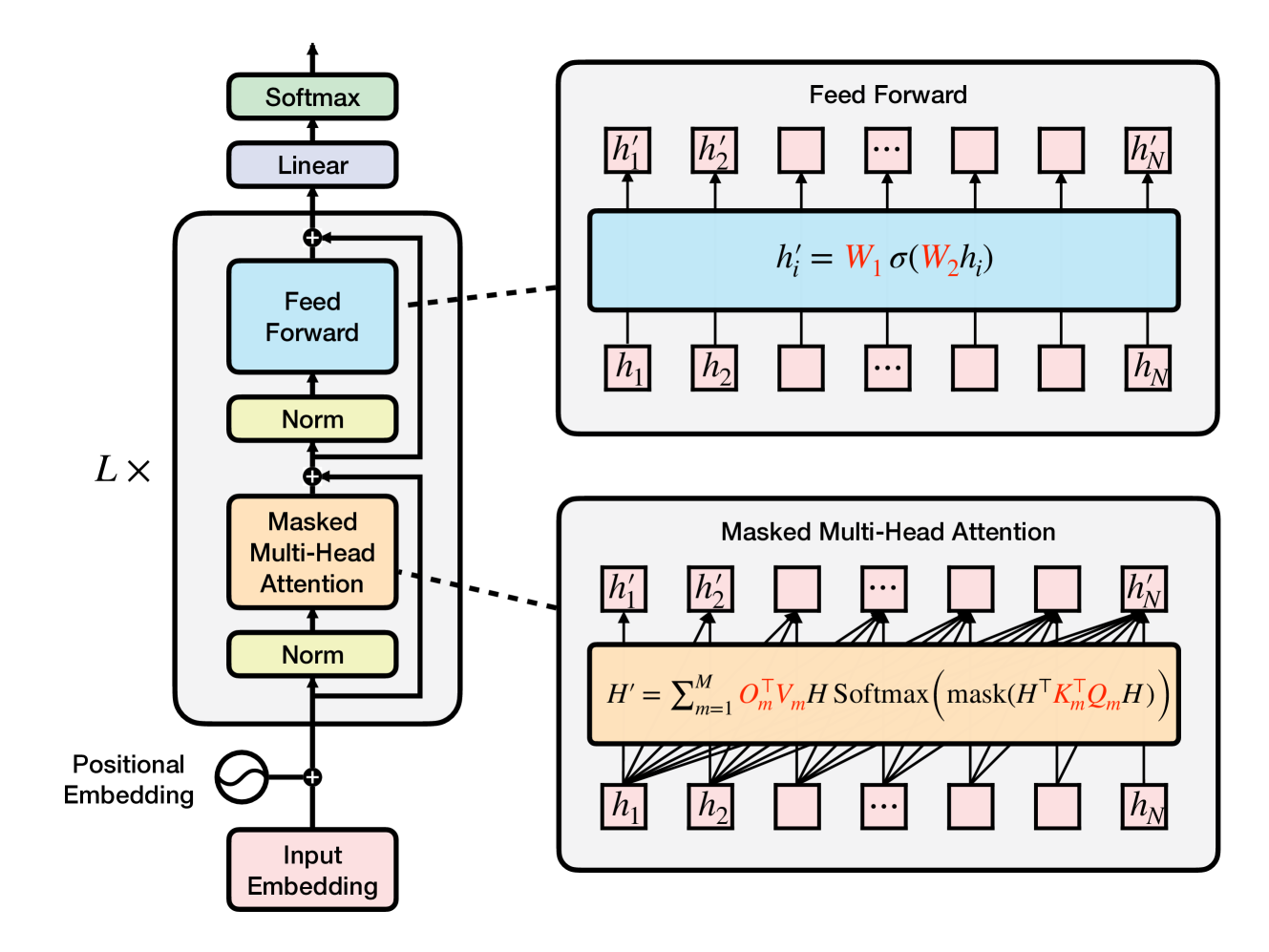}
    \caption{The architecture of a decoder-only Transformer, which forms the basis of most modern LLMs. The model consists of a stack of decoder layers, each containing a masked multi-head self-attention module and a feed-forward network. The cross-attention module is omitted. In the diagram, $L$ represents the number of stacked decoder layers, $W_1$ and $W_2$ are the weight matrices of the position-wise feed-forward network, while $Q_m$, $K_m$, $V_m$, and $O_m$ denote the query, key, value, and output projection matrices of the multi-head attention mechanism, respectively. Figure adapted from \cite{ji2025overview}.}
    \label{fig:decoder_only_transformer}
\end{figure}

\paragraph{Decoder-Only Architectures}
This architecture, illustrated in Figure \ref{fig:decoder_only_transformer}, uses only the decoder stack, with the cross-attention mechanism removed. These models are inherently generative and are pre-trained on an autoregressive language modeling objective: predicting the next token given all previous tokens. The unification of diverse NLP tasks into this straightforward format of the next-token prediction, along with the scalability of this architecture, has led to its widespread architecture being used across various language tasks.

The seminal model in this family is the \textbf{GPT (Generative Pre-trained Transformer)} series \cite{radford2018improving, radford2019language, brown2020language}. The success of this paradigm has inspired a vast ecosystem of decoder-only models, including Meta's \textbf{Llama} series \cite{touvron2023llama, touvron2023llama2} and Mistral AI's models \cite{jiang2023mistral}. These models were instrumental in pioneering the field of high-performance open-source LLMs, which has since expanded to include significant contributions from other major research labs such as Google, Microsoft, OpenAI, Alibaba, and DeepSeek AI. The ability of these models to perform new tasks in a zero-shot or few-shot manner, significantly enhanced by techniques like instruction tuning \cite{wei2022finetuned}, has cemented the decoder-only architecture as the prevailing design for modern LLMs. The theoretical underpinnings of this success are an active area of research, with recent studies modeling transformers as universal in-context learners that operate over probability distributions, further explaining their expressive power \cite{furuya2025transformers}.

These pre-trained models, regardless of their architectural family, serve as the foundation of transfer learning in NLP. They can be adapted to specific downstream tasks via fine-tuning on relatively small, task-specific datasets, achieving state-of-the-art performance across a wide range of benchmarks.

\subsection{From Autocomplete to Assistants}

The fundamental Transformer architectures generate what are called "base models," which are potent engines for pattern completion and possess vast knowledge about the world. The transition to the practical LLM "assistants" that characterize the present landscape was driven by the development of methods to align the model's behavior with human intentions. This alignment is essential for integrating the model's intrinsic skills, acquired in pre-training, with its function as an effective and safe conversational agent.

\paragraph{Pre-training: Train a Base Model}
The first step in creating an LLM is pre-training. In this phase, a Transformer architecture is trained on a colossal corpus of text, often comprising trillions of tokens. The training objective is typically self-supervised. For decoder-only models, this objective is autoregressive language modeling: predicting the next token in a sequence. The result of this process is a \textbf{base model}, a powerful pattern-completion engine that possesses immense world knowledge but lacks an inherent understanding of user instructions.

From a statistical perspective, this self-supervised objective of predicting the next token is equivalent to minimizing the negative log-likelihood of the training corpus. Formally, the loss function for a sequence of tokens $\vect{x}$ is given by:
\begin{equation}
    \mathcal{L}_{\text{LM}} = - \sum_{t=1}^{T} \log P(x_t | x_{<t}; \theta)
\end{equation}
where $\theta$ represents the model's parameters. This formulation grounds the pre-training process in the well-established principle of Maximum Likelihood Estimation (MLE), where the model parameters are optimized to maximize the probability of observing the training data \cite{ji2025overview}.

\paragraph{Pre-training Data}
\begin{figure}
    \centering
    \includegraphics[width=0.7\linewidth]{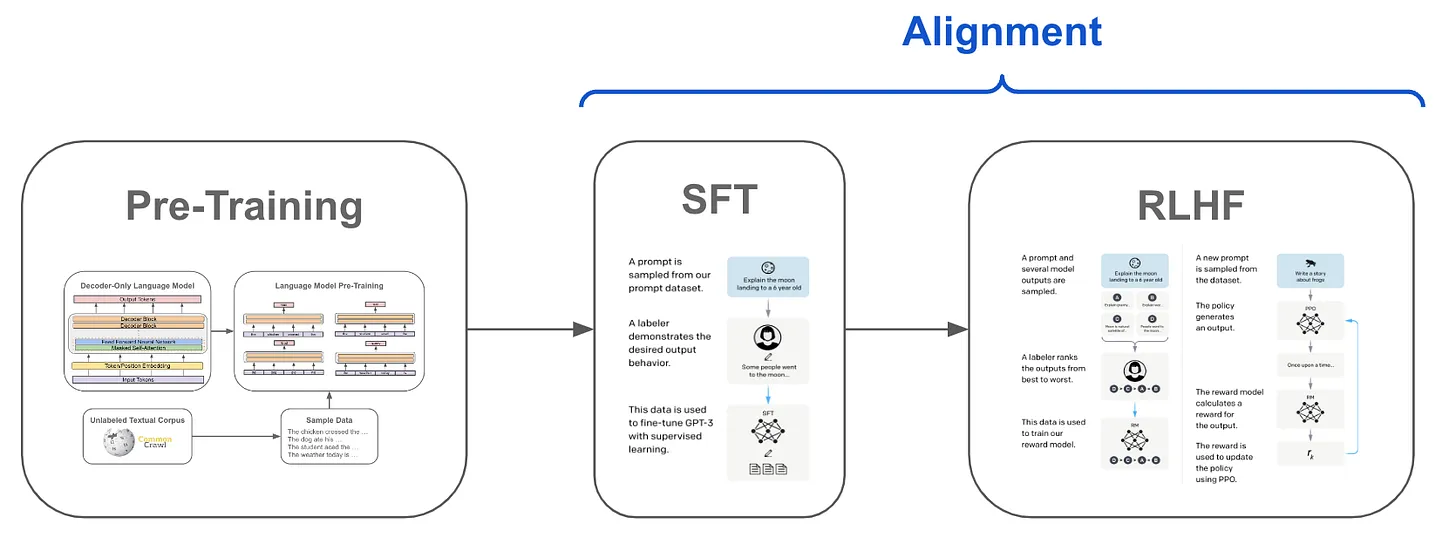}
    \caption{Overview of the training pipeline for instruction-following language models, adapted from Ouyang et al. (2022) \cite{ouyang2022training}. The process consists of three main stages: (1) \textbf{Pre-Training}, where a decoder-only Transformer is trained on large unlabeled text corpora; (2) \textbf{Supervised Fine-Tuning (SFT)}, where human labelers provide examples of desired model behavior; and (3) \textbf{Reinforcement Learning from Human Feedback (RLHF)}, where human preferences are used to train a reward model and iteratively improve the policy via reinforcement learning.}
    \label{fig:pretraining_allinment}
\end{figure}

The capabilities of a base model are fundamentally shaped by the data on which it is trained. Pre-training corpora are vast, heterogeneous mixtures of data sourced from the public web (e.g., Common Crawl), books, academic articles (e.g., arXiv), and source code (e.g., GitHub) \cite{liu2024datasetslargelanguagemodels}. The composition of this data mixture is a decisive factor for model performance.. For instance, including a significant portion of high-quality code data has been shown to improve not only programming abilities but also general reasoning skills \cite{aryabumi2024codecodeexploringimpact}.

The process of constructing these datasets involves sophisticated data selection and mixture strategies. Data is typically filtered and weighted at multiple levels of granularity (token, sample, and source group) to optimize the final data mixture. While early methods relied on manual mixing ratios \cite{gao2020pile800gbdatasetdiverse}, more recent approaches employ learned mixture strategies, where proxy models or the LLM itself are used to determine the optimal composition of data sources to maximize performance under a fixed computational budget \cite{xie2023doremi, wettig2024quratingselectinghighqualitydata}. This curation of data is as important to the final model's capabilities as the architectural design itself.

% equations from  3.5.1 arxiv 2502.17814
\paragraph{Alignment: From Base Models to Instruction-Tuned Models}
Transforming a base model into a useful and safe assistant requires a process known as alignment. This process fine-tunes the model to follow instructions and adhere to desired human behaviors. Alignment is typically a multi-stage process, with preference tuning being the most critical phase.

\begin{itemize}
    \item \textbf{Supervised Fine-Tuning (SFT):} The first stage involves fine-tuning the base model on a curated dataset of high-quality instruction-response pairs. SFT teaches the model the general format of following instructions and provides a strong foundation for helpfulness, yielding an initial policy denoted as $\pi^{\text{SFT}}$.

    \item \textbf{Preference Tuning:} After SFT, we further refine the model's behavior using human preference data. The two dominant paradigms for this stage are Reinforcement Learning from Human Feedback (RLHF) and Direct Preference Optimization (DPO).

    \subparagraph{Reinforcement Learning from Human Feedback (RLHF)}
    RLHF is a reward-based method that involves three phases \cite{ouyang2022training}. First, a reward model (RM), $r_\phi(\vect{x}, y)$, is trained to predict human preferences. To collect data for the RM, the SFT model is prompted with an input $\vect{x}$ to generate a pair of responses $(y_1, y_2)$. A human labeler then indicates which response is preferred, denoted $y_w \succ y_l$. The preferences are assumed to be generated by a latent reward function $r^*$, and are often modeled using the Bradley-Terry model:
    \begin{equation}
        P^*(y_w \succ y_l | \vect{x}) = \frac{\exp(r^*(\vect{x}, y_w))}{\exp(r^*(\vect{x}, y_w)) + \exp(r^*(\vect{x}, y_l))}
    \end{equation}
    The reward model $r_\phi$ is trained via maximum likelihood on a dataset of such preferences $\mathcal{D} = \{(\vect{x}^{(i)}, y_w^{(i)}, y_l^{(i)})\}_{i=1}^N$. This is equivalent to minimizing the binary cross-entropy loss:
    \begin{equation}
        \mathcal{L}_R(r_\phi, \mathcal{D}) = -\mathbb{E}_{(\vect{x}, y_w, y_l) \sim \mathcal{D}} \left[ \log \sigma \left( r_\phi(\vect{x}, y_w) - r_\phi(\vect{x}, y_l) \right) \right]
    \end{equation}
    where $\sigma$ is the logistic function. In the final phase, the language model policy, $\pi_\theta$, is optimized using reinforcement learning (e.g., Proximal Policy Optimization, PPO) to maximize the expected reward from the learned RM, while a KL-divergence penalty term prevents the policy from deviating too far from the initial SFT policy $\pi^{\text{SFT}}$:
    \begin{equation}
        \max_{\pi_\theta} \mathbb{E}_{\vect{x} \sim \mathcal{D}, y \sim \pi_\theta(y|\vect{x})} [r_\phi(\vect{x}, y)] - \beta D_{KL}[\pi_\theta(y|\vect{x}) || \pi^{\text{SFT}}(y|\vect{x})]
    \end{equation}

    \subparagraph{Direct Preference Optimization (DPO)}
    DPO is a more recent, reward-free method that simplifies the alignment process by eliminating the need to explicitly train a separate reward model \cite{rafailov2023direct}. It leverages a specific parameterization of the reward model that allows the optimal policy to be extracted in closed form. This insight enables direct optimization of the language model policy on the preference dataset $\mathcal{D}$. The DPO loss function is formulated as:
    \begin{equation}
        \mathcal{L}_{\text{DPO}}(\pi_\theta; \pi_{\text{ref}}) = -\mathbb{E}_{(\vect{x}, y_w, y_l) \sim \mathcal{D}} \left[ \log \sigma \left( \beta \log \frac{\pi_\theta(y_w|\vect{x})}{\pi_{\text{ref}}(y_w|\vect{x})} - \beta \log \frac{\pi_\theta(y_l|\vect{x})}{\pi_{\text{ref}}(y_l|\vect{x})} \right) \right]
    \end{equation}
    Here, $\pi_{\text{ref}}$ is a reference policy, typically the SFT model $\pi^{\text{SFT}}$. By minimizing this loss, the model $\pi_\theta$ learns to increase the likelihood of preferred responses ($y_w$) and decrease the likelihood of dispreferred responses ($y_l$) relative to the reference policy, thus achieving alignment more directly and with greater training stability than RLHF.
\end{itemize}
Models that have undergone this alignment pipeline are known as \textbf{instruction-tuned} or \textbf{preference-tuned} models and form the basis of most practical applications, including the agent-based systems we analyze in this work.

% AGGIUNGI QUESTO SOTTOPARAGRAFO ALLA FINE DELLA SEZIONE SULL'ALLINEAMENTO
\subparagraph{Self-Alignment and Its Challenges}
As models approach or surpass human-level capabilities in specific domains, the reliance on human feedback for alignment becomes a significant bottleneck. This has motivated a new research direction in \textbf{self-alignment}, where an LLM is leveraged to generate its own training signals, reducing the dependence on human annotators.

One prominent approach is the use of \textbf{self-rewarding language models} \cite{yuan2024selfrewarding}. In this paradigm, a model iteratively improves itself by generating responses to synthetic instructions and then scoring these responses using an "LLM-as-a-Judge" mechanism \cite{zheng2023judging}. The highest and lowest-scoring responses are used as preference pairs to further fine-tune the model with DPO, creating an autonomous improvement loop. Extensions like meta-rewarding models refine this process by also fine-tuning the model's ability to judge, improving both its generation and evaluation skills concurrently \cite{wu2024metarewarding}.

However, this reliance on synthetic data and self-generated rewards introduces its own set of challenges. One critical issue is \textbf{reward misspecification}, where over-optimization against an imperfect, LLM-generated reward signal can amplify biases. These include verbosity bias (favoring longer answers) and self-enhancement bias (favoring its own style of response) \cite{gao2023scaling, zheng2023judging}. Another significant challenge is \textbf{distributional shift}, where training exclusively on model-generated data can lead to a loss of diversity and a degradation of performance, a phenomenon known as "model collapse" \cite{shumailov2023curse}. Current research suggests that maintaining a mixture of human-authored and synthetic data is crucial for mitigating these risks and preserving model quality \cite{gerstgrasser2024is}.

\subsection{Architectural Innovations for Scalability and Efficiency}

The immense scale of modern LLMs necessitated further evolution of the Transformer architecture itself. These innovations focus on improving training stability and, particularly for our work, enhancing inference efficiency to manage computational and memory costs.

\paragraph{The Dense and Efficient Lineage}
Many state-of-the-art models follow a dense architecture, where all parameters are used for every token. Meta's LLaMA 3 models exemplify this trend by incorporating \textbf{Grouped-Query Attention (GQA)} \cite{ainslie2023gqa}. GQA reduces the computational and memory burden of the Key-Value (KV) cache, a primary bottleneck during inference, thus enabling longer sequences and larger batch sizes.

\paragraph{The Sparse Path: Mixture of Experts (MoE)}
The Mixture of Experts (MoE) architecture decouples a model's knowledge capacity from its inference cost \cite{shazeer2017outrageously}. An MoE layer replaces the dense FFN with a collection of parallel "expert" networks and a learned "router" that directs each token to a small subset of them. This creates two critical metrics: \textbf{total parameters} (reflecting knowledge capacity) and \textbf{active parameters} (reflecting inference cost).

This architecture is not only central to leading open-source models but is also widely understood to be the design behind proprietary state-of-the-art models like GPT-4. In the open space, Mixtral 8x7B demonstrated the efficacy of this approach, achieving the performance of a 70B dense model with only 13B active parameters \cite{jiang2024mixtral}. The more recent \textbf{DeepSeek-V3} pushes this paradigm further, employing a 671B total parameter MoE architecture that activates 37B parameters per token. It introduces advanced routing strategies to balance expert utilization, achieving high performance while managing computational load \cite{deepseekv3_report}.

\subsection{The Economics of LLM Inference: Beyond Price-per-Token}

The architectural choices described above have direct and significant consequences on the cost of using these models. The total cost of an LLM query is a function of its pricing model and its efficiency for a given task. The fundamental cost equation is:
\begin{equation}
    \text{Total Cost} = (\text{Input Tokens} \times \text{Price}_{\text{in}}) + (\text{Output Tokens} \times \text{Price}_{\text{out}})
\end{equation}

However, a model's price-per-token is an incomplete measure of its true operational cost. A more critical factor is its \textbf{token efficiency}: the number of input and output tokens a model requires to successfully complete a specific task. As demonstrated by industry benchmarks like the Artificial Analysis Intelligence Index, there is often a divergence between a model's token pricing and its total cost to perform a complex evaluation suite.

This divergence is particularly pronounced in tasks that require sophisticated \textbf{reasoning}. Such tasks often necessitate longer, more structured prompts (e.g., using chain-of-thought or reasoning models), which inflates the input token count. The model, in turn, may need to generate verbose, step-by-step reasoning to arrive at a correct answer, increasing the output token count. Some providers even use tiered pricing, charging a premium for models or endpoints optimized for reasoning. An example is DeepSeek, whose models such as DeepSeek-V1 and DeepSeek-V3 share the same architecture but differ significantly in inference cost per token. Therefore, even with a lower per-token cost, a model that lacks token efficiency may incur higher overall expenses for complex tasks, as it may require longer reasoning traces compared to a more advanced, though higher-priced, alternative.

This interplay between architecture (which determines active parameters and capabilities), pricing, and token efficiency for a given task creates the complex optimization space that this thesis addresses. The goal is to find configurations that minimize total cost while satisfying performance constraints, a problem that cannot be solved by evaluating token prices alone.

\begin{figure}[H]
    \centering
    \includegraphics[width=0.8\linewidth]{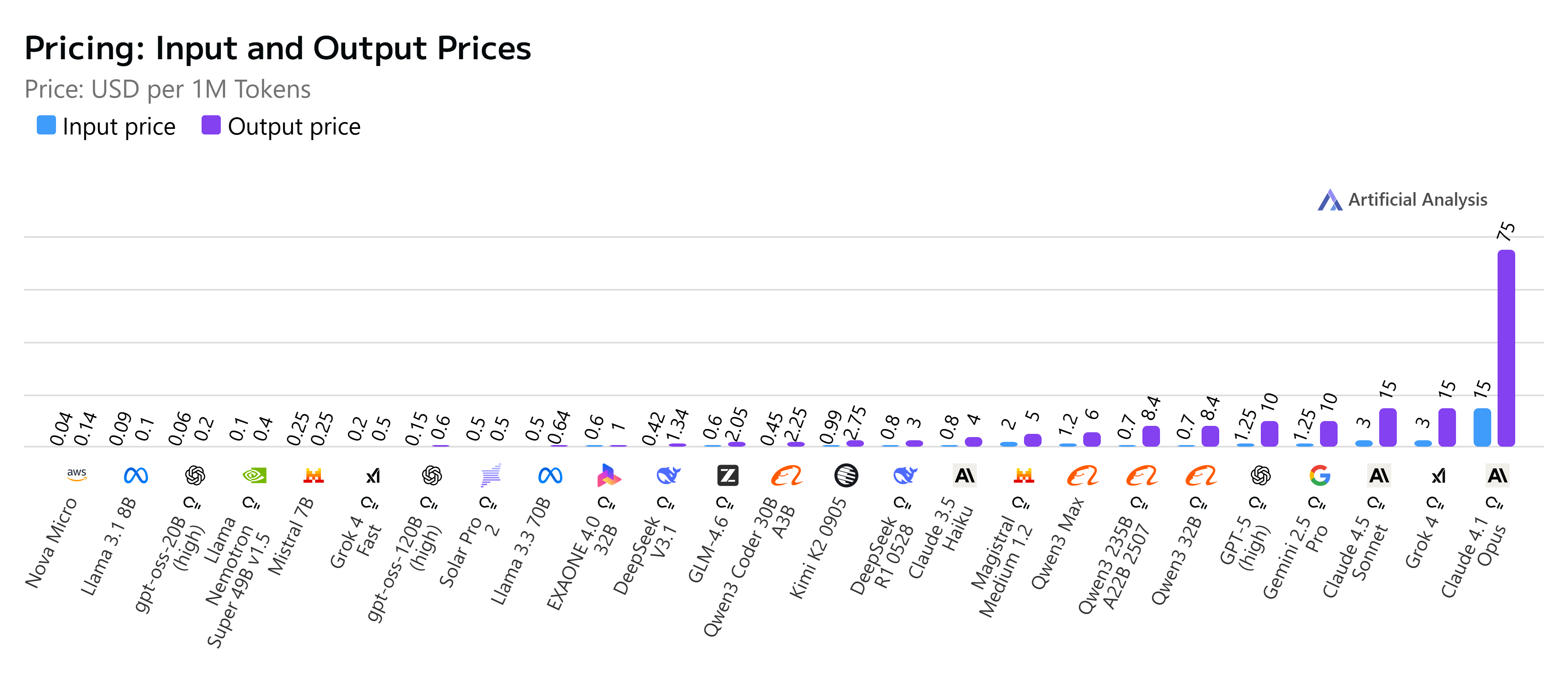}
    \caption{Input and output prices per token (in USD) across leading LLMs as reported by the Artificial Analysis Intelligence Index. 
    Despite large differences in per-token pricing --- for example, Grok-4 charges \$0.003/\$0.015 per input/output token, compared to Opus-4.1’s \$0.015/\$0.075 --- these prices alone do not reflect true inference cost. 
    Models such as Qwen-3 and Sonnet-4.5 show that lower token prices do not necessarily lead to cheaper execution when token usage efficiency is considered.}
    \label{fig:price_per_token}
\end{figure}

\begin{figure}[H]
    \centering
    \includegraphics[width=0.8\linewidth]{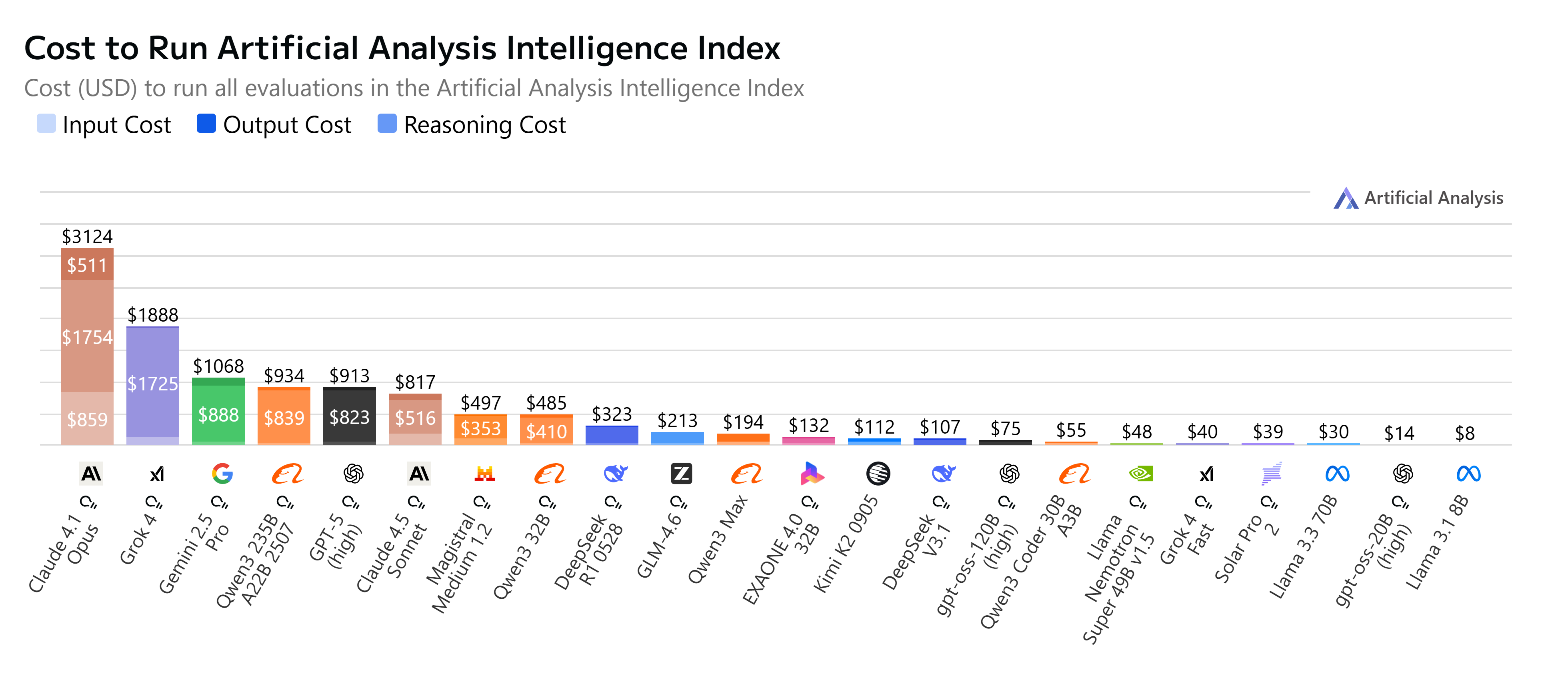}
    \caption{Total benchmark execution costs on the Artificial Analysis Intelligence Index for the same models. 
    While Grok-4 and Sonnet-4.5 share similar token pricing, Sonnet completes the benchmark with a total cost of roughly \$817, 
    over \$1{,}000 less than Grok-4’s \$1{,}888. 
    Similarly, Qwen-3 (\$934) costs more than Sonnet-4.5 despite cheaper per-token rates, underscoring the importance of \textbf{token efficiency} and reasoning quality over nominal pricing.}
    \label{fig:benchmark_total_cost}
\end{figure}

\subsection{The Role of Scaling Laws}
The evolution of LLMs is inextricably linked to the empirical discovery of \textbf{scaling laws}, which describe how a model's performance improves as a function of its size, the dataset size, and the computational budget for training \cite{kaplan2020scaling}. These laws revealed that the test loss of an LLM decreases predictably as a power-law of these three factors, catalyzing a trend towards building ever-larger models.

A pivotal refinement to this understanding came with the "Chinchilla" scaling laws, which demonstrated that for optimal performance under a fixed computational budget, both model size and the number of training tokens must be scaled in tandem \cite{hoffmann2022training}. This work showed that many large models at the time were "undertrained," and that smaller models trained on more data could outperform them.

More recently, a new paradigm of \textbf{inference-time scaling laws} has emerged, pioneered by models like OpenAI's o1 and DeepSeek-R1 \cite{openai2024o1, deepseek2025r1}. This principle posits that performance can be significantly improved not by increasing model parameters, but by allocating more computational resources at inference time. This is achieved through techniques such as generating multiple candidate responses and selecting the best one via a reward model or verifier. These findings introduce a new dimension to the optimization landscape: the trade-off is no longer solely between training cost and performance, but also between inference-time cost/latency and performance. Our work operates within this complex context, where the "cost" objective implicitly captures the consequences of both architectural choices (e.g., active parameters in MoE models) and the verbosity required for reasoning, which is a form of inference-time computation, also referred to as test-time scaling.

\begin{figure}
    \centering
    \includegraphics[width=0.9\linewidth]{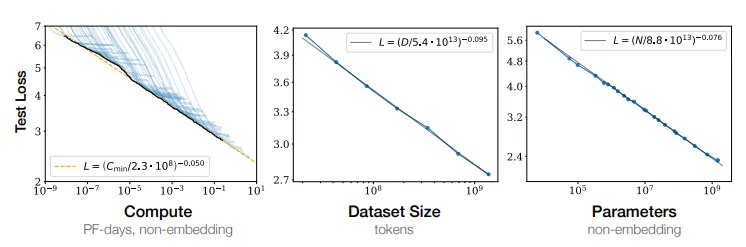}
    \caption{Compute–optimal scaling relationships for large language models, from Hoffmann et al. (2022) \cite{hoffmann2022training}. 
The plots illustrate (left) the training loss as a function of total compute (FLOPs) across models of different sizes, 
(center) the relation between compute and model parameters, and (right) the relation between compute and the number of training tokens. 
These scaling laws show that, for a fixed compute budget, optimal performance is achieved when model size and the number of training tokens are scaled proportionally.}
    \label{fig:scalingLawsPanel2022}
\end{figure}

\subsection{A Statistical Perspective on Modern LLMs}
\label{statisticalPerspective}
While advances in deep learning architectures and computational scale have been the primary drivers of LLM success, a deeper engagement with the field of statistics is required to address emerging challenges related to their trustworthiness and reliability \cite{ji2025overview}. This work identifies two primary directions for this synergy: applying statistical principles to improve LLMs and, conversely, leveraging LLMs to augment statistical workflows.

\paragraph{Improving LLMs with Statistical Rigor}
While the probabilistic nature of LLM outputs provides a foundation for reasoning about their behavior, it does not inherently guarantee statistical validity. A critical challenge is that models can generate factually incorrect statements, or \textit{hallucinations}, with high confidence as measured by token probability \cite{ji2023hallucination}. This discrepancy between probabilistic confidence and factual correctness necessitates the application of rigorous statistical methods to build trustworthy systems. We identify three key areas where such methods are essential:

\begin{itemize}
    \item \textbf{Uncertainty Quantification (UQ):} The primary goal of UQ is to develop principled methods for quantifying the reliability of LLM outputs. While classical metrics like entropy can be computed over the next-token distribution \cite{malinin2021uncertainty}, they often fail to capture semantic or factual uncertainty. More advanced approaches aim to address this by incorporating semantic features or internal model states \cite{kadavath2022language}. A particularly robust framework for UQ is \textbf{Conformal Prediction (CP)} \cite{vovk2005algorithmic}. CP is a distribution-free method that can construct prediction sets with formal, finite-sample statistical guarantees on coverage. Its application to LLMs, while challenging due to large output spaces, has shown promise in controlling hallucination risks and providing calibrated confidence estimates for tasks ranging from question answering to machine translation \cite{yadkori2024mitigating, zerva2024conformalizing}.

    \item \textbf{Interpretability and Fairness:} LLMs often function as opaque, black-box models, making it difficult to understand their decision-making processes. Statistical tools for model interpretation, such as those that identify influential training data or internal model "circuits," can be adapted to probe their mechanisms. This is a prerequisite for diagnosing and mitigating the societal biases (e.g., related to gender or race) that are invariably inherited from large-scale, unstructured training data.

    \item \textbf{Principled Alignment:} The process of aligning LLM behavior with human values, exemplified by Reinforcement Learning from Human Feedback (RLHF) and Direct Preference Optimization (DPO), is fundamentally a problem of statistical inference from preference data. A rigorous statistical framework, often grounded in preference models like the Bradley-Terry model, is necessary to develop more robust, sample-efficient, and theoretically understood alignment techniques.
\end{itemize}

\paragraph{Augmenting Statistical Workflows with LLMs}
Conversely, LLMs can serve as powerful components within traditional statistical analysis pipelines. Their capabilities in natural language understanding enable new applications in:
\begin{itemize}
    \item \textbf{Automated Data Processing:} LLMs can be employed for tasks such as automated data cleaning, feature engineering, and the extraction of structured variables from unstructured text sources like clinical records or financial reports.
    \item \textbf{Synthetic Data Generation:} When real data is scarce, private, or imbalanced, LLMs can be used to generate high-fidelity synthetic datasets. These datasets can then be used to train downstream statistical models while preserving privacy or correcting for class imbalances \cite{ji2025overview}.
\end{itemize}
The integration of these two fields is essential for advancing both the theoretical foundations and the practical, trustworthy application of these transformative models.

The application of rigorous statistical methods extends beyond improving LLMs to drawing parallels with other complex learning systems, such as the human brain. Recent work has shown strong correlations between the internal representations of LLMs and human fMRI brain responses during language tasks, suggesting a degree of representational similarity that is specific to models trained on human language \cite{parra2025neural}. Furthermore, analytical frameworks rooted in Bayesian optimization and Pareto rationality have been successfully applied to model human decision-making under uncertainty. These studies analyze how human learners manage the exploration-exploitation trade-off, using tools like Wasserstein distance to represent their behavioral patterns as probability distributions \cite{candelieri2023wasserstein, candelieri2023uncertainty}. 

\section{Multi-Agent Systems: Paradigms and Frameworks}

The concept of distributing intelligence among multiple autonomous entities is a foundational paradigm in computer science, originating from the field of Distributed Artificial Intelligence (DAI) \cite{weiss2013multiagent}. A Multi-Agent System (MAS) is a computerized system composed of multiple interacting, intelligent agents designed to solve problems that are difficult or impossible for a single agent or a monolithic system to solve \cite{wikipediaMAS}. This section provides a formal definition of MAS, explores key taxonomies for classifying them, and presents an overview of the open-source frameworks that enable their practical implementation, particularly in the context of LLMs.

\subsection{Definitions and Taxonomies}

An ``agent'' is generally defined as an autonomous entity that can perceive its environment, act upon it to achieve goals, and communicate with other agents \cite{rocha2017introductory}. A system becomes a MAS when it features multiple such agents interacting within a shared environment. Key characteristics of a MAS include:
\begin{itemize}
    \item \textbf{Autonomy:} Each agent has control over its own actions and internal state \cite{wooldridge2009introduction}.
    \item \textbf{Local Views:} No single agent possesses a global view of the system; its knowledge is limited to its own state and communications.
    \item \textbf{Decentralization:} There is typically no central controller; collective behavior emerges from local interactions \cite{guttmann2025taxonomy}.
\end{itemize}

It is critical to distinguish Multi-Agent Systems from Agent-Based Models (ABMs). While both involve simulating agents, their objectives differ. ABMs are primarily used as a scientific tool to \textit{understand} the emergent collective behavior of agents, often in natural or social systems \cite{wikipediaABM}. In contrast, MAS are an engineering paradigm used to \textit{build} complex, distributed systems to solve specific practical problems \cite{wikipediaMAS}.

The advent of Large Language Models has catalyzed a new era for MAS. LLMs serve as the reasoning ``brain'' for each agent, enabling them to perform complex planning, tool use, and, most importantly, coordinate through natural language \cite{ibmWhatIsMAS}. This has shifted the primary challenge from designing rigid communication protocols to orchestrating sophisticated, language-based collaboration \cite{guo2024surveyLLMMA}.

To navigate the design space of these systems, several taxonomies have been proposed. At a high level, MAS architectures can be classified by their control structure \cite{katalinic2014classification}:
\begin{itemize}
    \item \textbf{Centralized:} A single coordinating agent or ``orchestrator'' assigns tasks and manages the information flow between other agents.
    \item \textbf{Decentralized:} Decision-making authority is dispersed, and agents coordinate through peer-to-peer interactions \cite{bronsdon2025centralized}.
\end{itemize}

For LLM-powered systems specifically, a more granular taxonomy is required to capture the nuances of their architecture. Recent work proposes a multi-dimensional framework for analyzing these systems across four key axes: \textbf{task management}, \textbf{agent composition}, \textbf{collaboration}, and \textbf{context interaction} \cite{pinzger2023taxonomy}. This framework helps to classify how a system decomposes problems, defines agent roles, manages communication, and maintains a shared understanding of the task.

Complementing these formal taxonomies are practical design philosophies from industry leaders. Anthropic, for instance, advocates for building agentic systems using simple, composable patterns, differentiating between fixed ``workflows'' and dynamic ``agents'' that direct their own processes \cite{anthropicEffectiveAgents}. Their multi-agent research system utilizes a hierarchical structure with a lead agent for planning and sub-agents for parallel information gathering, demonstrating a practical application of a centralized control paradigm \cite{anthropicMultiAgentSystem}.

\subsection{Overview of Open-Source Frameworks}

The theoretical paradigms of multi-agent systems are put into practice through software frameworks that abstract the complexities of agent creation, communication, and orchestration. The open-source community has produced a diverse ecosystem of such frameworks, each with a distinct architectural philosophy. We introduce two representative examples, OpenManus and SmolAgents, and then provide a comparative overview of other prominent frameworks.

\textbf{OpenManus} is an open-source framework designed to replicate and democratize the capabilities of advanced autonomous AI agents \cite{pankaj2025openmanus}. Its architecture is explicitly modular, comprising distinct layers for Agents, Tools, Prompts, and LLM Interaction \cite{llmmultiagentsOpenManus}. A key feature of OpenManus is its support for two execution modes: a flexible \textit{Direct Agent Execution} mode and a more structured \textit{Flow Orchestration Execution} mode, which separates task planning from execution \cite{li2025openmanus}. This dual-mode design allows it to handle both simple, reactive tasks and complex, pre-defined workflows.

\begin{figure}
    \centering
    \includegraphics[width=0.8\linewidth]{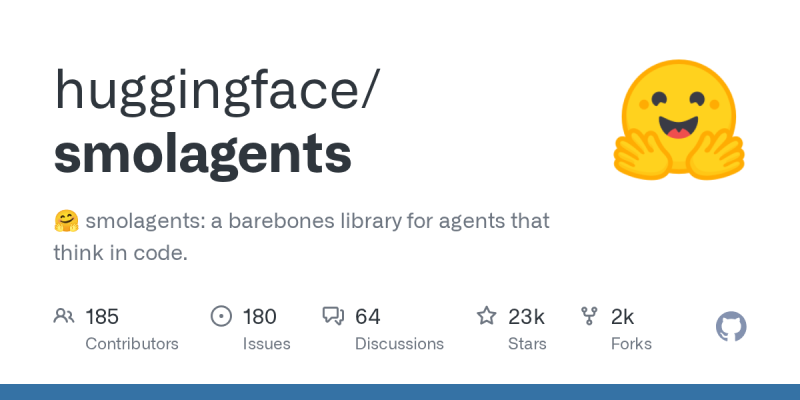}
    \caption{The open-source repository for the Smol-Agents framework. The multi-agent architecture employed in this thesis for validating MALBO is derived from the `open-deep-research` code provided within this repository. The codebase was subsequently forked and adapted to integrate our optimization loop and configuration-based model swapping.}
    \label{fig:smolagents_repo}
\end{figure}

\textbf{SmolAgents}, developed by Hugging Face (fig. \ref{fig:smolagents_repo}), represents a minimalist and code-centric design philosophy \cite{huggingfaceSmolagentsIntro}. The core idea is that agents ``think in code''; instead of relying on complex tool-calling APIs, agents generate and execute Python code to perform actions \cite{gupta2025smolagents}. This approach aims to improve accuracy and efficiency by leveraging the structured and expressive nature of a programming language. The framework is intentionally lightweight, LLM-agnostic, and designed for simplicity, making it a powerful tool for developers who prefer direct, code-based control over agent behavior \cite{smolagentsOrg}.

Beyond these examples, the landscape of open-source frameworks is rich with different architectural approaches. Table~\ref{tab:mas_frameworks} provides a comparative summary of several influential frameworks, highlighting their core paradigms and key features. This diversity underscores the active research and development in the field, as different architectures are suited for different types of problems, from structured workflows to dynamic, conversational problem-solving \cite{silfverskiold2025agenticAI}. The design of MALBO, as detailed in Chapter 3, is informed by this existing landscape, drawing on principles of modularity and orchestration to address our optimization challenge.

\begin{table}[h!]
\centering
\caption{Comparative Overview of Prominent Open-Source Multi-Agent Frameworks.}
\label{tab:mas_frameworks}
\begin{tabular}{@{}p{2.5cm} p{3cm} p{6cm}@{}}
\toprule
\textbf{Framework} & \textbf{Paradigm} & \textbf{Features \& Use Cases} \\ \midrule
\textbf{AutoGen} \cite{microsoftAutoGen} & Conversation-Driven & Agents solve tasks via automated conversations. Highly flexible and extensible. Supports human-in-the-loop. \\
\textbf{LangGraph} \cite{langchainLangGraph} & Graph-Based, State Machine & Represents workflows as a graph. Enables cyclical control flows, explicit state management, and long-running agents. \\
\textbf{CrewAI} \cite{crewaiOpenSource} & Role-Playing \& Hierarchical & Agents are assigned specific roles and goals, collaborating like a human team. Uses a manager to orchestrate tasks. \\
\textbf{MetaGPT} \cite{metagptGithub} & SOPs \& Software Company Simulation & Implements Standardized Operating Procedures (SOPs) to simulate a software development team for code generation. \\
\textbf{ChatDev} \cite{chatdevGithub} & Waterfall Model Simulation & Simulates a virtual software company with distinct roles (CEO, programmer, tester) following a structured waterfall process. \\
\bottomrule
\end{tabular}
\end{table}

% This would be part of Chapter 1 in your thesis.
\section{Bayesian Optimization (BO) for Black-Box Functions}

Many foundational problems in science and engineering involve optimizing a function that is computationally or financially expensive to evaluate. Examples range from tuning hyperparameters of deep learning models to discovering novel materials or optimizing the design of complex systems \cite{shahriari2016taking, frazier2018tutorial}. In these scenarios, the objective function $f(\mathbf{x})$ lacks a known analytical form and its derivatives are typically unavailable. We can only evaluate it at a point $\mathbf{x}$ to receive a (potentially noisy) observation $y$. This defines the problem of \textbf{black-box optimization}.

When function evaluations are expensive, standard optimization methods like grid search are infeasible due to the prohibitive number of evaluations required. Bayesian Optimization (BO) is a sequential, model-based strategy designed specifically for the global optimization of such expensive black-box functions \cite{snoek2012practical}.

The core strategy of BO is to build a probabilistic surrogate model of the objective function, which captures our beliefs about $f(\mathbf{x})$. This surrogate is cheap to evaluate and is updated with each new observation from the true function. To decide where to sample next, BO uses an \textbf{acquisition function} that leverages the surrogate's predictions and, crucially, its uncertainty estimates. This function guides the search by balancing \textbf{exploitation} (sampling in regions predicted to have high-performing outcomes) and \textbf{exploration} (sampling in regions where uncertainty is high). The point that maximizes the acquisition function is chosen for the next expensive evaluation of $f(\mathbf{x})$. This intelligent, sequential search process allows BO to find a global optimum with a significantly smaller number of function evaluations compared to other methods. The framework is composed of two primary components: the surrogate model and the acquisition function.

\subsection{Gaussian Processes as Surrogate Models}

The most common and effective surrogate model used in Bayesian Optimization is the \textbf{Gaussian Process (GP)} \cite{rasmussen2006gaussian}. A GP is a non-parametric model that defines a distribution over functions. It is a generalization of the multivariate Gaussian distribution to an infinite-dimensional space of functions.

Formally, a Gaussian Process is a collection of random variables, any finite number of which have a joint Gaussian distribution \cite{rasmussen2006gaussian}. A GP is fully specified by a mean function $m(\mathbf{x})$ and a covariance function, or \textbf{kernel}, $k(\mathbf{x}, \mathbf{x'})$:
\begin{equation}
    f(\mathbf{x}) \sim \mathcal{GP}(m(\mathbf{x}), k(\mathbf{x}, \mathbf{x'}))
\end{equation}
The mean function $m(\mathbf{x})$ represents the expected value of the function at input $\mathbf{x}$, and is often assumed to be zero for simplicity. The kernel $k(\mathbf{x}, \mathbf{x'})$ models the covariance between the function values at two points, $\mathbf{x}$ and $\mathbf{x'}$. The choice of kernel is critical as it encodes our prior beliefs about the properties of the function, such as its smoothness or periodicity. A common choice is the Matérn family of kernels, which allows for controlling the smoothness of the modeled function.

Given a set of $n$ observations $\mathcal{D}_n = \{(\mathbf{x}_i, y_i)\}_{i=1}^n$, the GP framework allows us to compute a posterior distribution over the function $f$. A key property of GPs is that the posterior predictive distribution for the function value $f_*$ at a new test point $\mathbf{x}_*$ is also a Gaussian distribution:
\begin{equation}
    P(f_* | \mathbf{x}_*, \mathcal{D}_n) = \mathcal{N}(\mu_n(\mathbf{x}_*), \sigma_n^2(\mathbf{x}_*))
\end{equation}
Here, $\mu_n(\mathbf{x}_*)$ is the posterior mean and $\sigma_n^2(\mathbf{x}_*)$ is the posterior variance. The posterior mean serves as the current best estimate of the function at $\mathbf{x}_*$, while the posterior variance provides a measure of uncertainty about that estimate. This principled quantification of uncertainty is what enables the intelligent exploration-exploitation trade-off managed by the acquisition function.

\subsection{Acquisition Functions}

The acquisition function, $\alpha(\mathbf{x})$, uses the posterior distribution provided by the GP to quantify the utility of evaluating the black-box function at a candidate point $\mathbf{x}$. The next point to be evaluated is selected by maximizing this function:
\begin{equation}
    \mathbf{x}_{n+1} = \arg\max_{\mathbf{x} \in \mathcal{X}} \alpha(\mathbf{x} | \mathcal{D}_n)
\end{equation}
While numerous acquisition functions exist for single-objective optimization, such as Expected Improvement (EI) and Upper Confidence Bound (UCB), the focus of this thesis is on the more complex multi-objective setting.

In Multi-Objective Optimization (MOO), the goal is not to find a single optimal point but to identify the set of optimal trade-offs known as the \textbf{Pareto front}. A solution is Pareto-optimal if no objective can be improved without degrading at least one other objective. A standard metric for evaluating the quality of a Pareto front approximation is the \textbf{Hypervolume (HV)} indicator, which measures the volume of the objective space that is dominated by the front and bounded by a reference point \cite{zitzler2003performance}.

A principled acquisition function for MOO is, therefore, the \textbf{Expected Hypervolume Improvement (EHVI)}. EHVI measures the expected increase in the hypervolume of the current approximate Pareto front that would result from evaluating a new candidate point (or a batch of points) \cite{emmerich2006single, daulton2020differentiable}. By maximizing EHVI, the BO algorithm is directly guided to select points that are most likely to expand the dominated hypervolume, thus efficiently mapping the true Pareto front. While the computation of EHVI has historically been a significant bottleneck, particularly for parallel evaluations, recent advances in differentiable programming have made its optimization tractable and highly effective, forming a core component of the methodology proposed in this thesis \cite{daulton2020differentiable}.

\chapter{Related works - LLM and Bayesian Optimization}
\label{ch:soa}

\section{Optimization in the Age of Large Language Models}
\label{sec:soa_optimization_llms}

The proliferation of pre-trained Large Language Models (LLMs) has marked a significant shift in natural language processing. These foundation models exhibit remarkable general-purpose capabilities, yet their performance on specific downstream tasks is highly sensitive to how they are prompted and configured. The initial approach for adapting LLMs to a specific task was model fine-tuning. While effective, this process presents considerable computational overhead and data privacy challenges, hindering its practical application in many scenarios \cite{Sabbatella2024BOInG}.

This has led to the rise of \textit{prompt engineering}, a more lightweight and flexible paradigm for model adaptation. A prompt is a sequence of symbols or tokens, selected from a vocabulary, which is prepended or concatenated to a user's query to guide the model's output. The challenge of discovering an optimal prompt sequence can be framed as a complex combinatorial optimization problem. The search space, defined by the vocabulary size $|V|$ raised to the power of the prompt length $L$ (i.e., $|V|^L$), is often intractably large, necessitating efficient search strategies \cite{Archetti2023PromptBO}.

Prompt optimization methods are broadly categorized into two families:
\begin{enumerate}
    \item \textbf{Soft Prompt Tuning (SPT):} These "white-box" methods require access to the model's internal states and gradients. They operate in the continuous embedding space of the model, directly optimizing prompt embeddings while keeping the core model parameters frozen. While parameter-efficient, SPT is incompatible with the growing trend of accessing LLMs via restricted, black-box APIs.
    \item \textbf{Hard Prompt Tuning (HPT):} This "black-box" approach directly searches for an optimal sequence of discrete tokens. HPT is critically important in the \textit{Model-as-a-Service (MaaS)} ecosystem, where users only have query-level access to powerful proprietary models. This black-box constraint aligns with both user needs for simplicity and provider needs for security and intellectual property protection \cite{Sabbatella2024BOInG}.
\end{enumerate}

The optimization challenge, however, extends beyond prompts. The modern LLM landscape is rich with complex decision-making problems, including the dynamic routing of queries to the most suitable model \cite{Shirkavand2025CSCR}, the efficient merging of model checkpoints to combine skills, and the enforcement of complex logical constraints on text generation. Across these varied problems, a central theme emerges: the need for sample-efficient optimization algorithms that can navigate vast, high-dimensional, and often discrete search spaces under a black-box constraint. Among the various techniques available, Bayesian Optimization has emerged as a dominant and principled approach to address these challenges.

\section{The Evolution of Bayesian Optimization for LLMs}

\begin{figure}[ht]
    \centering
    \includegraphics[width=0.8\linewidth]{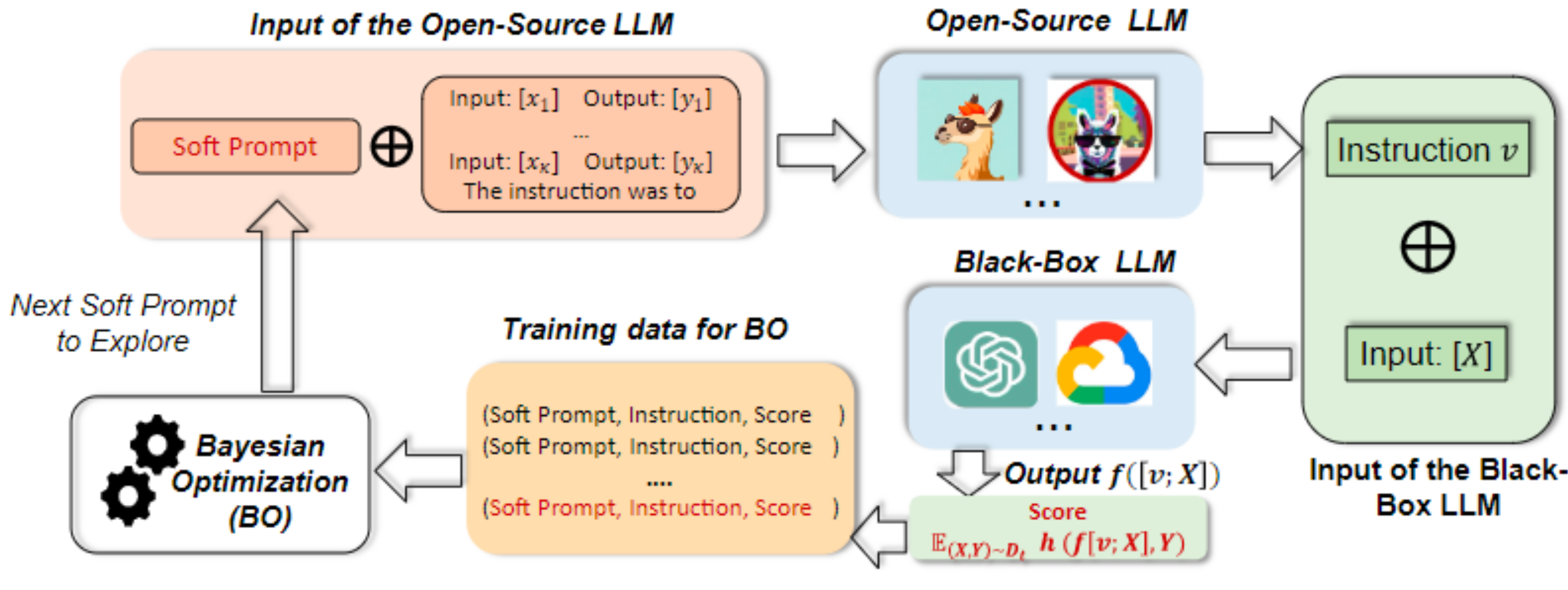}
    \caption{The iterative workflow of the InstructZero \cite{Chen2023InstructZero} and BOInG \cite{Sabbatella2024BOInG} frameworks, an example of "white-box" instruction optimization. Bayesian Optimization is used to tune a continuous soft prompt for an open-source LLM, which acts as an instruction generator. The quality of the generated instruction is then evaluated by a separate, black-box LLM (the task solver). The resulting performance score provides the feedback signal to guide the optimization loop. Figure adapted from Chen et al. \cite{Chen2023InstructZero}.}
    \label{fig:instructzero_workflow}
\end{figure}

Bayesian Optimization (BO) has become the de-facto standard for sample-efficient black-box optimization, particularly when function evaluations are expensive—a condition that is quintessentially true for queries to large-scale LLMs. Its primary advantage lies in its strategy for managing the exploration-exploitation trade-off. By building a probabilistic surrogate model of the objective function (typically a Gaussian Process, GP), BO can quantify uncertainty and intelligently select the next point to evaluate, minimizing the number of costly queries required to find an optimal solution \cite{Archetti2019Bayesian}.

\subsection{Foundational Applications: Prompt and Instruction Optimization}
\label{subsec:bo_foundations}

\begin{figure}[ht]
    \centering
    \includegraphics[width=0.9\textwidth]{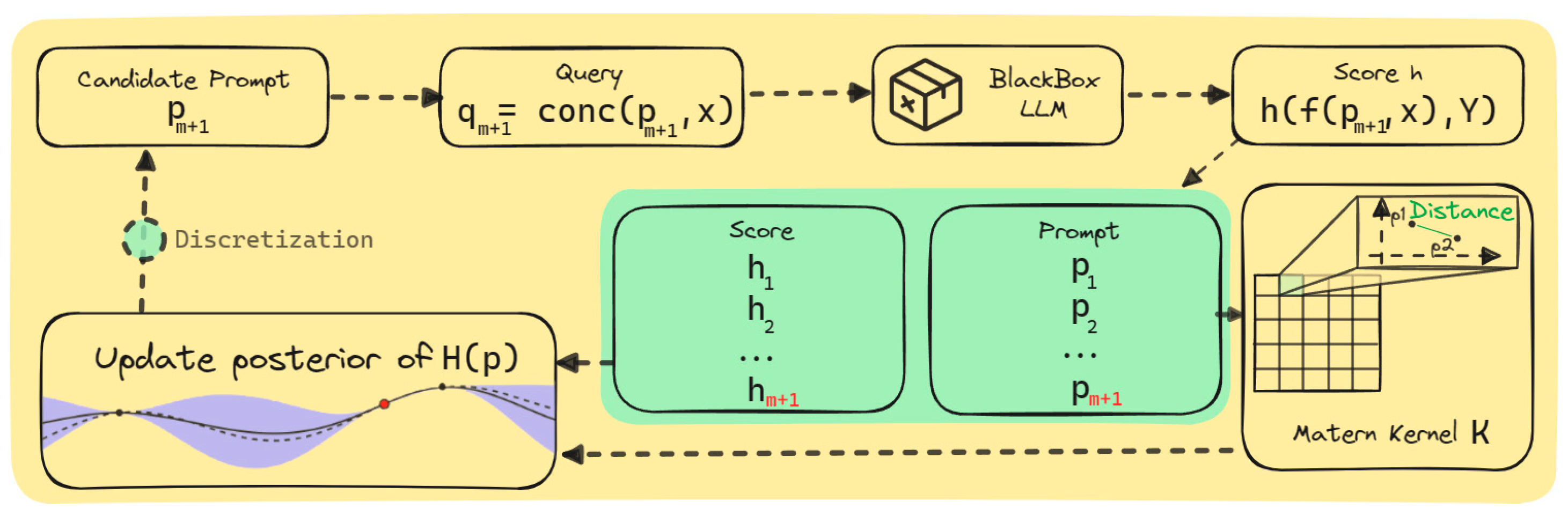} 
    \caption{The general workflow of Hard Prompt Tuning (HPT) via Bayesian Optimization. A surrogate model (GP) is iteratively updated with new prompt evaluations to guide the search for an optimal discrete prompt. Figure adapted from \cite{Archetti2023PromptBO}.}
    \label{fig:prompt_bo_loop}
\end{figure}

Initial applications of BO in the LLM space focused squarely on the problem of Hard Prompt Tuning. In our prior work \cite{Sabbatella2024BOInG, Archetti2023PromptBO, archetti2025bayesian, sabbatella2023bayesian}, we demonstrated the feasibility of using a "vanilla" BO algorithm for HPT. The core methodological innovation was a \textbf{continuous relaxation} of the discrete token search space. Instead of searching directly over tokens, the algorithm searches over a continuous space of token indices, optimizes a continuous acquisition function, and then rounds the resulting solution back to the nearest integer indices to retrieve the discrete tokens. This technique, illustrated in Figure \ref{fig:prompt_bo_loop}, unlocked the power of standard GP-based BO for a fundamentally combinatorial problem.

\begin{figure}[ht]
    \centering
    \includegraphics[width=1.0\textwidth]{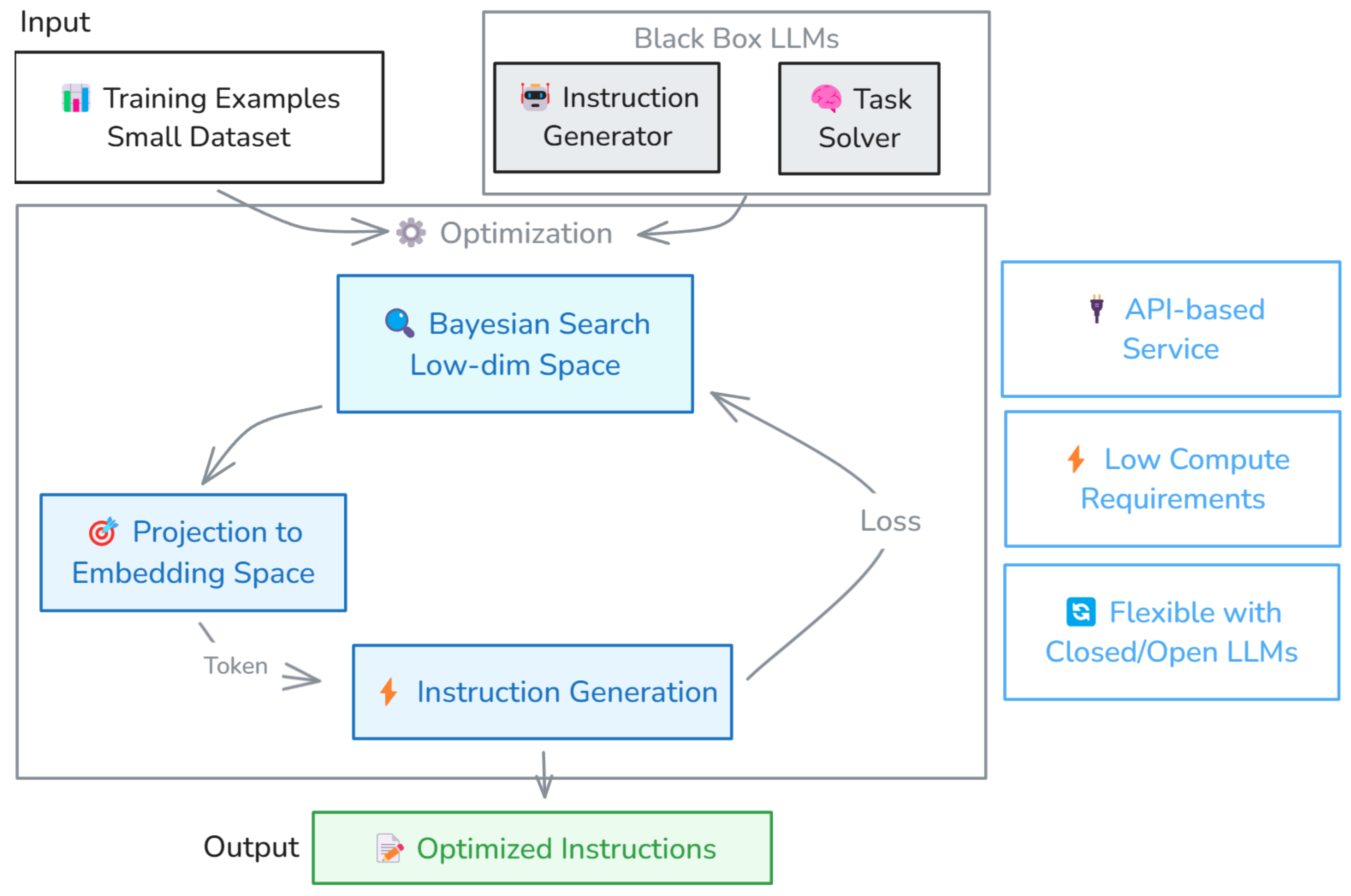} % Replace with your actual image file
    \caption{High-level workflow of the BOInG framework. Bayesian Optimization is used to find an optimal hard prompt for an Instruction Generator LLM, which in turn provides instructions to a Task Solver LLM. Both models can be treated as black-boxes. Figure adapted from \cite{Sabbatella2024BOInG}.}
    \label{fig:boing_workflow}
\end{figure}

Building on this foundation, more sophisticated methods have emerged. Our work on \textbf{BOInG (Bayesian Optimization for Instruction Generation)} \cite{Sabbatella2024BOInG} leverages two black-box LLMs: an \textit{instruction generator} and a \textit{task solver}. BO is used to find an optimal hard prompt for the generator, which in turn produces a natural language instruction for the solver. Its key innovation is a penalty term incorporated into the BO acquisition function, which pushes the search towards continuous representations that are close to the embeddings of known, valid tokens, thereby improving the coherence of the generated instructions. This black-box-centric approach (Figure \ref{fig:boing_workflow}) stands in contrast to methods like InstructZero \cite{Chen2023InstructZero}, which also employ a two-LLM setup but require at least one model to be "white-box" to access its internal states.

\subsection{Expanding the Scope: BO for Complex Workflows and Trustworthiness}
\label{subsec:bo_expansion}

As the field has matured, the focus of BO has expanded beyond optimizing single prompts to configuring entire LLM-powered systems and addressing the inherent challenges of reliability and cost. This progression reflects a move towards more complex, real-world deployment scenarios.

\begin{figure}[ht]
    \centering
    \includegraphics[width=1.0\textwidth]{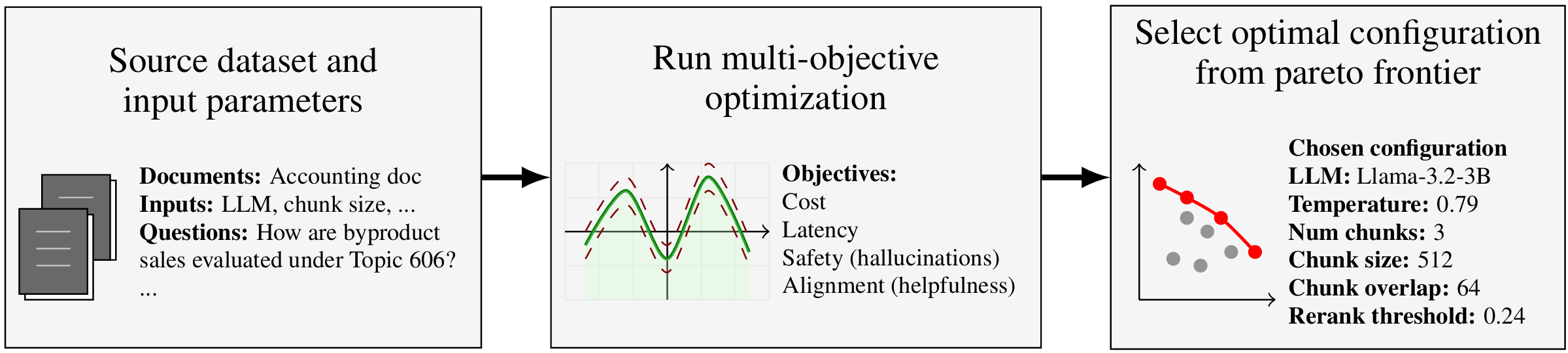} % Replace with an actual image file
    \caption{A high-level overview of multi-objective Bayesian Optimization for a RAG pipeline. The optimization process considers system-level parameters, including the choice of LLM, to find a Pareto front of optimal configurations balancing cost, latency, and performance. Figure adapted from \cite{Barker2025RAG}.}
    \label{fig:rag_bo_overview}
\end{figure}

A significant step in this direction involves optimizing multi-component pipelines, such as those used in Retrieval-Augmented Generation (RAG). Barker et al. \cite{Barker2025RAG} introduce a framework for the multi-objective optimization of a complete RAG system. The search space is no longer limited to prompt tokens but includes a wide array of hyperparameters, such as the choice of the LLM and embedding models, chunk size, and re-ranker thresholds, as illustrated in Figure \ref{fig:rag_bo_overview}. They frame this as a multi-objective problem to simultaneously optimize for cost, latency, and performance metrics like safety and alignment. We note that they employ a BO approach with the qLogNEHVI acquisition function, which is specifically designed to handle the noisy objective evaluations inherent in stochastic LLM outputs. This work demonstrates that BO can effectively navigate the high-dimensional, mixed-variable search space of a full pipeline to identify a Pareto-optimal set of configurations.

\begin{figure}[ht]
    \centering
    \includegraphics[width=0.8\textwidth]{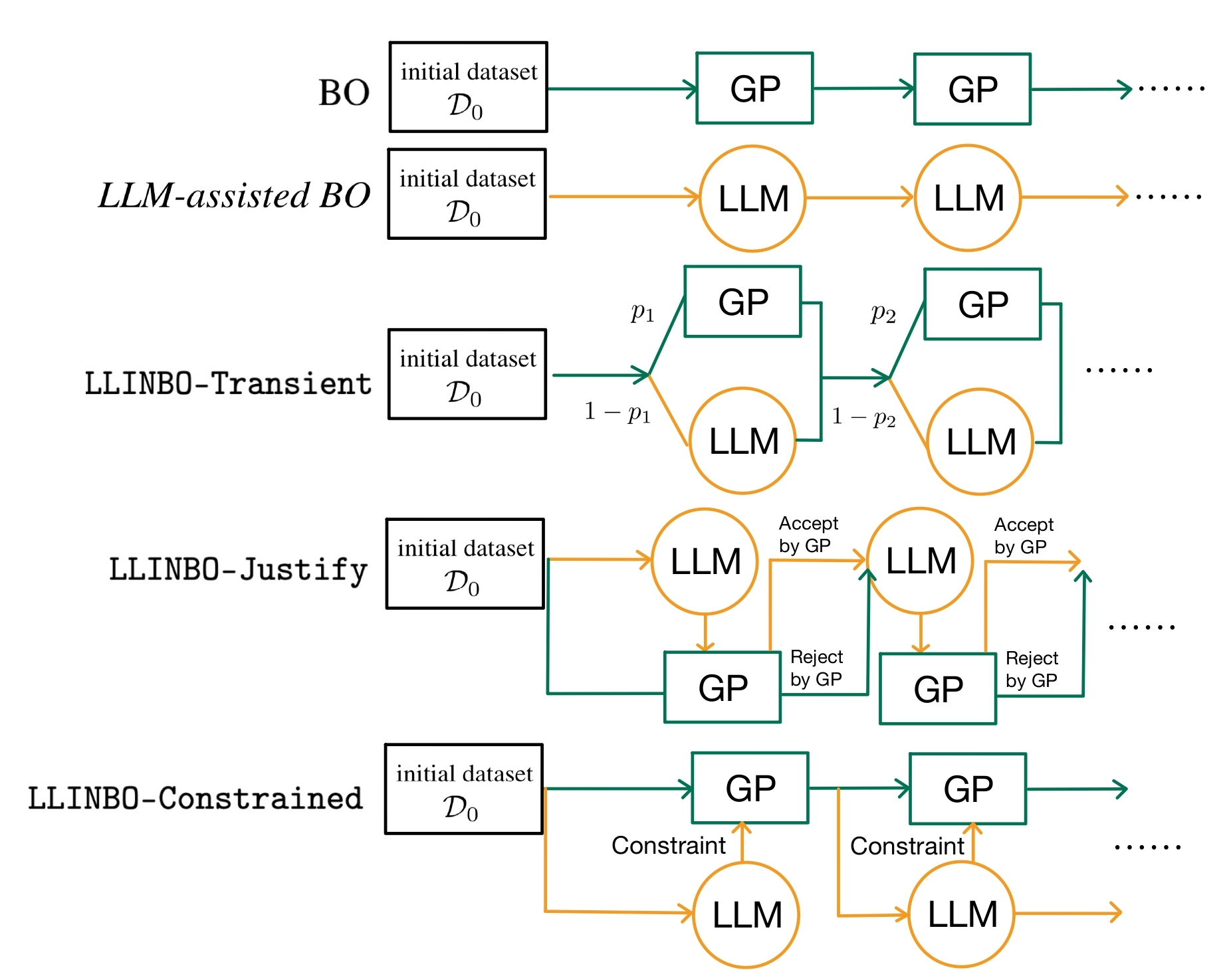}
    \caption{Diagram illustrating the three hybrid mechanisms proposed in the LLINBO framework, which combine suggestions from a traditional BO loop (with a GP) and an LLM-assisted loop. Figure adapted from \cite{chang2025llinbo}.}
    \label{fig:llinbo_mechanisms}
\end{figure}

Parallel to optimizing system complexity, a second research thrust has focused on improving the trustworthiness of LLM-driven optimization. Relying solely on an LLM as an optimizer is risky due to its lack of calibrated uncertainty and its opaque internal reasoning, which undermines theoretical tractability \cite{Chang2025LLINBO}. To address this, Chang et al. \cite{Chang2025LLINBO} propose \textbf{LLINBO}, a hybrid framework that combines the strengths of LLMs with the principled uncertainty quantification of statistical surrogates like GPs. Their core philosophy is to "leverage contextual reasoning strengths of LLMs for early exploration, while relying on principled statistical models to guide efficient exploitation". They introduce three distinct mechanisms for this collaboration (Figure \ref{fig:llinbo_mechanisms}):
\begin{itemize}
    \item \textbf{LLINBO-Transient:} A strategy that initially favors LLM suggestions and gradually transitions to GP-guided suggestions as more data becomes available.
    \item \textbf{LLINBO-Justify:} A rejection-sampling mechanism where the GP-based acquisition function acts as a verifier, discarding LLM suggestions that are deemed substantially suboptimal.
    \item \textbf{LLINBO-Constrained:} A method that treats the LLM's suggestion as a soft constraint, refining the GP posterior to favor regions around the suggested point.
\end{itemize}
This hybrid approach demonstrates that integrating LLMs as collaborators within a principled BO loop, rather than as standalone optimizers, leads to more robust and reliable performance.

A third direction has focused on optimizing the dynamic selection of models from a pool of available experts, a task known as LLM routing. Shirkavand et al. \cite{Shirkavand2025CSCR} frame this as a similarity search problem and introduce \textbf{CSCR (Cost-Spectrum Contrastive Routing)}. Their method learns a shared embedding space where both prompts and models are represented. Models are described by compact, efficiently computed "fingerprints" (either from logits for open models or perplexity scores for black-box APIs). A contrastive learning objective, named Cost-Spectrum InfoNCE, is trained to pull prompt embeddings towards the cheapest expert that can solve the task correctly. At inference time, routing becomes a highly efficient k-Nearest Neighbors (k-NN) lookup. This work explicitly incorporates cost into the representation learning process, creating a metric space that inherently balances the accuracy-cost trade-off.

Pushing the paradigm even further, recent research has explored using LLMs for meta-optimization—that is, to automatically generate the optimization algorithm itself. Li et al. \cite{Li2025LLaMEABO} introduce \textbf{LLaMEA-BO}, a framework that uses an evolution strategy to guide an LLM to write, combine, and mutate full Python implementations of BO algorithms. The LLM generates candidate algorithms, which are then evaluated on a benchmark test suite. The performance scores are used to select the best candidates, which are then "bred" (via crossover prompts) or "mutated" (via refinement prompts) to create the next generation of algorithms. This work demonstrates that LLM-generated algorithms can outperform state-of-the-art, human-designed BO baselines, positioning LLMs not just as tools within an optimization loop, but as co-designers of the optimization process itself. These advanced applications collectively show the field's rapid progression towards tackling more complex, system-level, and even meta-level optimization problems.

\section{Limitations of the State of the Art and Motivation for MALBO}
\label{sec:soa_limitations}

While the works we have reviewed demonstrate considerable progress in applying Bayesian Optimization to LLM-related tasks, they predominantly operate under a single-agent or monolithic pipeline paradigm. The existing literature largely overlooks the distinct and significantly more complex challenge of configuring and deploying an entire \textbf{team of collaborative LLM agents}.

The limitations of the current state of the art can be summarized in three key areas:

\begin{enumerate}
    \item \textbf{The Multi-Agent Dimension:} Prior methods focus on optimizing a single artifact, be it a prompt, a set of hyperparameters, or a routing policy for a single query. They do not address the combinatorial complexity inherent in assigning specific models to distinct roles within a multi-agent system. The problem is no longer finding the "best prompt" or "best model," but finding the \textit{optimal configuration of models} across a team of agents, where inter-agent dependencies and collaborative task structures are paramount. The search space explodes from optimizing a single vector to optimizing a matrix of assignments ($X \in \mathbb{R}^{N \times D}$, where $N$ is the number of agents and $D$ is the feature dimension of the models), a challenge existing methods are not designed to handle.

    \item \textbf{The Multi-Objective Imperative:} While some recent work has begun to address multiple objectives like cost and latency \cite{Barker2025RAG, Shirkavand2025CSCR}, this is not yet a standard approach in the field. Many frameworks still optimize for a single performance metric. A principled approach to multi-agent systems must have multi-objective optimization at its core, as the central challenge is precisely the trade-off between the collective performance of the team and its aggregate operational cost. This necessitates a framework capable of discovering a Pareto front of solutions.

    \item \textbf{A Fundamentally Different Search Space:} The search space in prior work is typically tied to the token vocabulary (for HPT) or a flat list of system hyperparameters. In the multi-agent assignment problem we address, the decision variables are not tokens or simple scalar values but abstract, continuous representations of entire LLMs. The search space is a structured, high-dimensional space of \textit{model capabilities}, where each point represents a potential team configuration. A new approach is required to effectively model this abstract configuration space.
\end{enumerate}

In summary, the state of the art lacks a framework that can perform sample-efficient, black-box, multi-objective optimization for the multi-agent LLM assignment problem. This thesis introduces \textbf{Multi-Agent LLM Bayesian Optimization (MALBO)} to fill this gap. We design MALBO to tackle the combinatorial complexity of assigning different LLMs to various agent roles, while simultaneously optimizing for the critical trade-off between task performance and API costs. The following chapters will detail the mathematical formulation and methodology of the approach.

\chapter{MALBO: Methodology and Mathematical Formulation}

Leveraging teams of LLM-based agents to solve complex problems is a promising frontier in artificial intelligence \cite{guo2024surveyLLMMA}. However, the effective composition of these teams presents a significant optimization challenge. The effectiveness of a multi-agent system hinges on the specific LLM assigned to each agent role, resulting in a vast and combinatorially complex design space. Furthermore, evaluating the performance of any given team configuration is an expensive black-box operation, requiring extensive simulation or real-world deployment.

This chapter introduces our proposed methodology, Multi-Agent LLM Bayesian Optimization (MALBO), designed to navigate this complex trade-off space. We formalize the problem of optimal LLM assignment as a multi-objective black-box optimization task. Our approach seeks to identify a set of agent configurations that represent the optimal frontier between task performance and the associated computational and financial costs. We detail each component of the MALBO framework, from the vector representation of LLMs to the multi-objective Bayesian optimization loop that drives the search for optimal solutions.

\subsection{Problem Description: Optimal LLM Assignment in an Agent Team}

We consider a multi-agent system composed of $N$ agents, where each agent is assigned a distinct role (e.g., planner, tool-user, verifier). We have access to a pool of $M$ unique Large Language Models, each with different capabilities, performance profiles, and inference costs. The core problem is to assign one LLM from this pool to each of the $N$ agent roles to optimize the overall system's performance on a given task.

This assignment problem is fundamentally combinatorial. With $M$ available models and $N$ agent roles, the total number of possible team configurations is $M^N$. Evaluating a single configuration requires executing the entire multi-agent workflow on a benchmark task and measuring its performance and cost, which is an expensive, black-box process. The goal is not to find a single best team but rather to uncover the set of assignments that offer the best trade-offs between performance and cost. This framing naturally leads to a multi-objective optimization problem, where we aim to co-optimize two conflicting objectives: maximizing task accuracy and minimizing operational cost \cite{blondin2020multi, paria2022strategies}.

\subsection{Vector Representation of LLMs and the Configuration Space}

To apply Bayesian optimization, we must first map the discrete choice of LLMs into a continuous space. We achieve this by representing each LLM as a point in a multi-dimensional feature space, a technique validated in hyperparameter optimization and universal black-box optimization \cite{feurer2019hyperparameter, tan2025towards}.

\subsubsection{The Feature Space $\mathcal{F} \subset \reals^D$}

We define a feature space $\mathcal{F} \subset \reals^D$, where each of the $D$ dimensions corresponds to a key, quantifiable characteristic of a Large Language Model. These features are chosen to capture the aspects of a model that are most likely to influence its performance and cost within an agentic system. The features include:
\begin{itemize}
    \item \textbf{Performance Metrics:} Standardized scores on academic benchmarks (e.g., MMLU-Pro, LiveCodeBench, GPQA Diamond) that proxy for reasoning, coding, and language understanding capabilities.
    \item \textbf{Architectural Properties:} Parameters such as context window length, total and active parameter counts (for MoE models), and model type (e.g., dense, sparse).
    \item \textbf{Economic Factors:} API inference costs, measured in price per input and output token.
\end{itemize}
Each LLM is thus represented by a feature vector $\mathbf{f} \in \mathcal{F}$ that quantifies its profile.

\subsubsection{The Matrix of Available Models $\mathbf{L} \in \reals^{M \times D}$}

The set of $M$ concrete, deployable LLMs available for assignment is represented as a matrix $\mathbf{L} \in \reals^{M \times D}$. Each row $\vect{l}_j$ of $\mathbf{L}$ is the $D$-dimensional feature vector for the $j$-th LLM in our pool. This matrix constitutes the ground truth of our discrete search space.

\subsection{Multi-Objective Optimization Problem Formulation}

With the continuous feature space defined, we can now formally state the optimization problem. A complete team configuration is a selection of one LLM for each of the $N$ agent roles. In our continuous relaxation, a configuration is a point $\vect{x} \in \mathcal{F}^N$, which is an $N \times D$ matrix where each row represents the feature vector of the LLM assigned to the corresponding agent role. The search space for our optimization is therefore $\mathcal{X} = \mathcal{F}^N \subset \reals^{N \times D}$.

\subsubsection{The Black-Box Objective Function $f: \mathcal{X} \rightarrow \reals^2$}

We define a vector-valued black-box objective function $f$ that maps a team configuration $\vect{x}$ to a two-dimensional output vector representing its performance and cost:
\begin{equation}
    f(\vect{x}) = [y_{\text{accuracy}}(\vect{x}), y_{\text{cost}}(\vect{x})]
\end{equation}
The components of this function are:
\begin{itemize}
    \item $y_{\text{accuracy}}(\vect{x})$: A performance metric (e.g., task success rate on the GAIA benchmark) which we aim to \textbf{maximize}.
    \item $y_{\text{cost}}(\vect{x})$: The total operational cost (e.g., cumulative API costs in dollars) incurred to complete the task, which we aim to \textbf{minimize}.
\end{itemize}
Since we seek to maximize accuracy and minimize cost, we formulate the optimization problem as maximizing the vector $[y_{\text{accuracy}}(\vect{x}), -y_{\text{cost}}(\vect{x})]$.

\subsubsection{Defining the Pareto Front}

Because the objectives of accuracy and cost are conflicting, there is typically no single configuration that is optimal for both. Instead, we seek the set of Pareto-optimal solutions \cite{paria2022strategies}. An objective vector $\vect{v}_1$ \textit{Pareto-dominates} another vector $\vect{v}_2$, denoted $\vect{v}_1 \succ \vect{v}_2$, if it is better or equal on all objectives and strictly better on at least one. A configuration $\vect{x}$ is Pareto-optimal if no other configuration $\vect{x}'$ exists such that $f(\vect{x}') \succ f(\vect{x})$.

The set of all such non-dominated objective vectors constitutes the \textbf{Pareto front}, denoted $\mathcal{P}^*$. The goal of MALBO is to find a diverse set of configurations whose objective values provide a high-quality approximation of this true Pareto front.

\subsection{The Bayesian Optimization Loop for MALBO}

\begin{figure}[H]
    \centering
    \includegraphics[width=\textwidth]{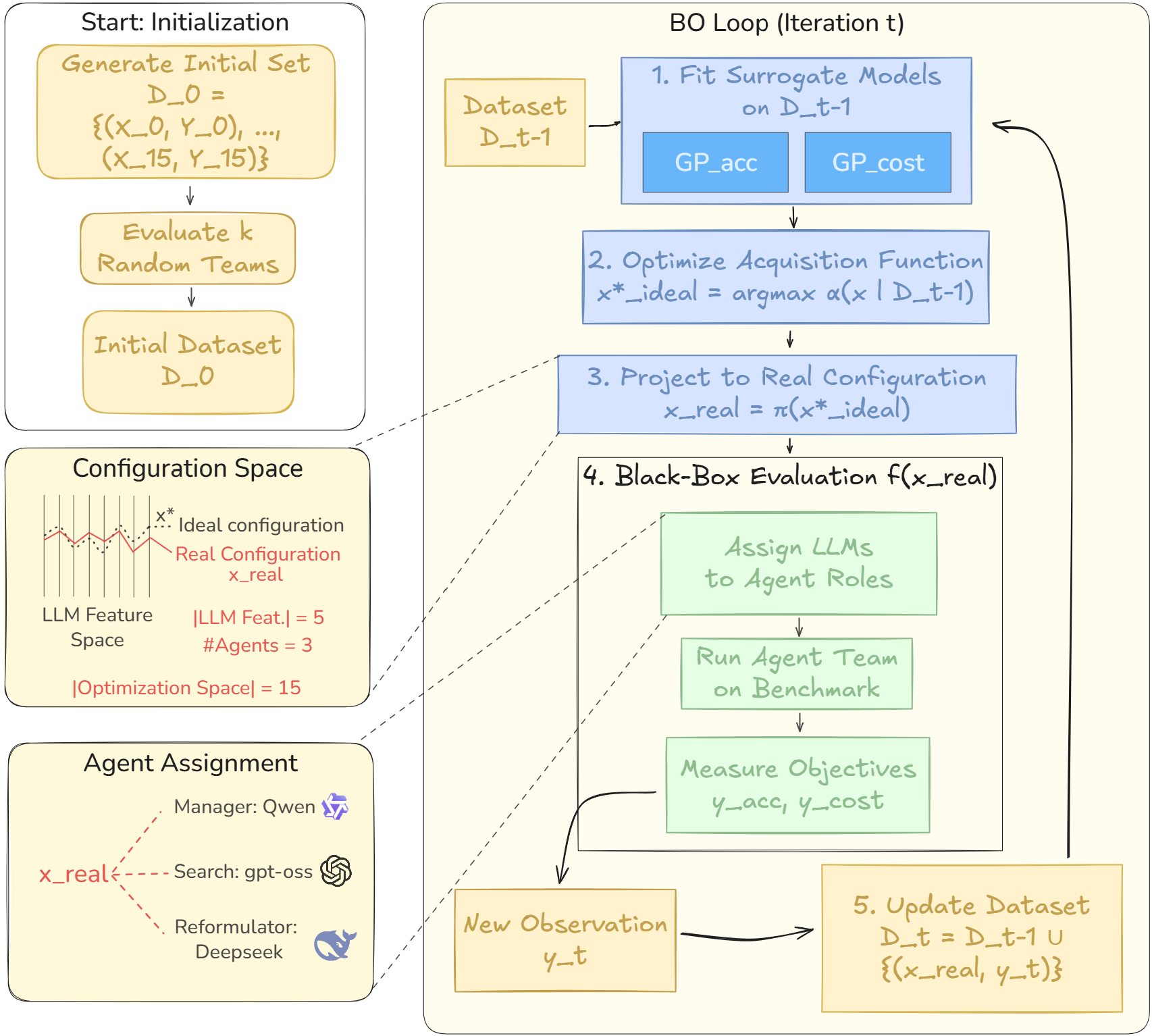}
    \caption{
        Overview of the MALBO (Multi-Agent LLM Bayesian Optimization) loop. 
        The process starts with the initialization of an initial dataset $D_0$ 
        by evaluating random agent team configurations. Each iteration of the Bayesian Optimization loop 
        fits surrogate Gaussian Process models to predict performance and cost, 
        optimizes an acquisition function to select the next ideal configuration, 
        projects it into the real configuration space, and evaluates it through black-box benchmarking 
        of LLM-based agent teams. The resulting observations are then used to update the dataset $D_t$.
    }
    \label{fig:malbo_loop}
\end{figure}

MALBO uses a multi-objective Bayesian optimization (MOBO) loop to efficiently search the configuration space $\mathcal{X}$. The loop iteratively builds surrogate models of the objective functions and uses an acquisition function to select the next batch of configurations to evaluate. As illustrated in Fig.~\ref{fig:malbo_loop}, each iteration involves modeling, acquisition optimization, projection into the real configuration space, and black-box evaluation of LLM-based agent teams.

\subsubsection{Surrogate Models for Accuracy and Cost ($\text{GP}_1, \text{GP}_2$)}

We model the two black-box objective functions, accuracy and cost, using two independent Gaussian Processes (GPs) \cite{rasmussen2006gaussian}. Given a dataset of $t$ observations $\mathcal{D}_t = \{(\vect{x}_i, \vect{y}_i)\}_{i=1}^t$, the posterior predictive distribution for a new configuration $\vect{x}_*$ is a Gaussian, providing both a mean prediction (the expected outcome) and a variance (a measure of uncertainty). We use one GP to model $y_{\text{accuracy}}$ and a second to model $-y_{\text{cost}}$. This approach is standard in MOBO and is well-supported by frameworks like BoTorch through its `ModelListGP` interface \cite{balandat2020botorch}.

\subsubsection{Acquisition Function: q-Log Expected Hypervolume Improvement (qLogEHVI)}

To balance the exploration-exploitation trade-off across multiple objectives, we employ an acquisition function based on the \textbf{hypervolume (HV)} indicator \cite{zitzler2003performance}. The hypervolume of a set of points on the Pareto front measures the size of the objective space they dominate. An ideal acquisition function for MOBO seeks to select new candidate points that are expected to maximize the improvement in this hypervolume.

We use the \textbf{q-Expected Hypervolume Improvement (qEHVI)} \cite{daulton2020differentiable}, an acquisition function designed for parallel (or batch) multi-objective optimization. It computes the expected hypervolume improvement from evaluating a batch of $q$ candidate points. We use its logarithm, qLogEHVI, for improved numerical stability during optimization. This function guides the search toward regions that are most likely to expand the known Pareto front, either by improving upon existing solutions (exploitation) or by reducing uncertainty in unexplored regions (exploration).

\subsubsection{The Optimization Loop Algorithm}

The complete MALBO process is an iterative loop, detailed in Algorithm \ref{alg:malbo}. At each iteration $t$, the algorithm performs the following steps:
\begin{enumerate}
    \item \textbf{Fit Surrogate Models:} The two GPs, $\text{GP}_{\text{acc}}$ and $\text{GP}_{\text{cost}}$, are updated using all available data $\mathcal{D}_{t-1}$.
    \item \textbf{Optimize Acquisition Function:} The qLogEHVI acquisition function is maximized using the posterior distributions from the GPs to find the next batch of $q$ candidate configurations $\matr{X}_t = \{\vect{x}_1, ..., \vect{x}_q\}$.
    \item \textbf{Evaluate Black-Box Function:} Each candidate configuration $\vect{x}_i$ in the batch is evaluated on the true objective function $f$ to obtain its accuracy and cost, $\vect{y}_i = f(\vect{x}_i)$. This is the most expensive step of the loop.
    \item \textbf{Update Dataset:} The new observations $\{(\vect{x}_i, \vect{y}_i)\}_{i=1}^q$ are added to the dataset, creating $\mathcal{D}_t$, and the process repeats.
\end{enumerate}

\begin{algorithm}
\caption{MALBO: Multi-Agent LLM Bayesian Optimization Loop}\label{alg:malbo}
\begin{algorithmic}[1]
\State \textbf{Input:} Black-box function $f$, initial dataset $\mathcal{D}_0$, batch size $q$, total budget $T$.
\State Initialize surrogate models $\text{GP}_{\text{acc}}$, $\text{GP}_{\text{cost}}$ on $\mathcal{D}_0$.
\For{$t = 1, \dots, T/q$}
    \State Fit $\text{GP}_{\text{acc}}$ and $\text{GP}_{\text{cost}}$ to current dataset $\mathcal{D}_{t-1}$.
    \State Construct the qLogEHVI acquisition function $\alpha(\cdot)$ from the GPs.
    \State Find the next batch of candidates: $\mathbf{X}_t \gets \arg\max_{\mathbf{X}' \subset \mathcal{X}^q} \alpha(\mathbf{X}')$.
    \State Evaluate the true black-box function for each candidate: $\mathbf{Y}_t \gets f(\mathbf{X}_t)$.
    \State Augment the dataset: $\mathcal{D}_t \gets \mathcal{D}_{t-1} \cup \{(\mathbf{X}_t, \mathbf{Y}_t)\}$.
\EndFor
\State \textbf{return} The set of non-dominated solutions found in $\mathcal{D}_T$.
\end{algorithmic}
\end{algorithm}

\subsection{From Continuous Space to Discrete Assignment: The Projection Function $\pi$}

The Bayesian optimization loop operates in the continuous feature space $\mathcal{F}^N$ and proposes an "ideal" configuration $\vect{x}_{\text{ideal}} \in \mathcal{F}^N$. However, to evaluate this configuration, we must deploy a team of actual LLMs from our available set, represented by the matrix $\mathbf{L}$. This requires a mapping from the continuous space back to the discrete set of available models.

We define a projection function $\pi: \mathcal{F} \rightarrow L$ that maps an ideal LLM feature vector to the feature vector of the most similar real LLM. This projection is achieved by finding the nearest neighbor in $\mathbf{L}$ using Euclidean distance:
\begin{equation}
    \pi(\vect{f}_{\text{ideal}}) = \arg\min_{\vect{l}_j \in L} ||\vect{f}_{\text{ideal}} - \vect{l}_j||_2
\end{equation}
This projection is applied to each of the $N$ rows of the candidate configuration matrix $\mathbf{X}_t$ proposed by the acquisition function optimizer. This step, adapted from surrogate optimization methods for discrete problems \cite{gokbayrak2001generalized}, ensures that every configuration evaluated by the black-box function $f$ corresponds to a real, deployable team of LLMs.

\chapter{Experimental Setup}
\label{ch:experimental_setup}

To empirically validate the MALBO framework, we designed and executed a series of experiments aimed at finding the optimal assignment of Large Language Models (LLMs) to a multi-agent system. This chapter details the complete experimental setup, including the software platform, the evaluation benchmark, the performance and cost metrics, the configuration of the Bayesian optimizer, and the pool of LLMs used in the study.

\section{Development and Integration Platform}
\label{sec:setup_platform}

Our experimental platform is built upon \textbf{SmolAgents}, an open-source, minimalist multi-agent framework developed by Hugging Face \cite{huggingfaceSmolagentsIntro}. We selected this framework for several key reasons. Firstly, its provider-agnostic design allows for the seamless integration of models from various sources, which is essential for exploring a diverse model space. Secondly, its "think in code" philosophy, where agents generate and execute Python code snippets as actions, has been demonstrated to be more efficient and performant than traditional tool-calling APIs, particularly with smaller or more specialized models \cite{wang2024executable}.

\paragraph{Motivation for Code-as-Action} Expressing agent actions directly as executable Python code (rather than declarative JSON tool calls) reduces mediation overhead and exploits the strong code priors of modern LLMs. Recent empirical evidence shows this paradigm can cut step counts (and thus total LLM calls) by roughly 30\% while improving difficult task success rates \cite{wang2024executable}. Additionally, frameworks enabling dynamic, on-the-fly action synthesis instead of restricting to a static tool set (e.g., unconstrained code-level generation) further enhance long-horizon adaptability \cite{nguyen2025dynasaur}. This informed our decision to adopt a lightweight wrapper that executes generated code inside a controlled sandbox, enabling (i) immediate inspection of intermediate variables, (ii) iterative self-repair, and (iii) seamless composition of library calls without pre-registering each operation as a separate tool.

To facilitate the optimization loop, we extended the base framework with a dedicated optimization interface. As detailed in our contributions to the open-source repository \cite{sabbatella2025smolagentsfork}, we introduced a programmatic API ("optimization\_interface.py") that allows an external optimizer, such as MALBO, to systematically evaluate agent team configurations. This interface accepts a dictionary defining the LLM assignment for each agent role and returns a tuple containing the final performance and cost metrics.

The multi-agent system under evaluation is configured with a team of five distinct agent roles: a \texttt{manager}, a \texttt{search\_agent}, a \texttt{text\_inspector}, a \texttt{visual\_qa} agent, and a \texttt{reformulator}. However, to reduce the dimensionality of the search space from $D \times 5$ to a more manageable $D \times 3$ and thereby accelerate convergence, we focused our optimization efforts on the three most influential roles: the \texttt{manager}, the \texttt{search\_agent}, and the \texttt{reformulator}. The remaining two roles, \texttt{text\_inspector} and \texttt{visual\_qa}, were assigned a fixed, high-performance yet cost-effective default model (\texttt{openrouter/openai/gpt-4.1-mini}) for all experimental runs. This simplification makes it feasible to explore the agent assignment space with greater precision and reduced computational overhead.

\section{Evaluation Benchmark: GAIA}
\label{sec:setup_benchmark}

We selected \textbf{GAIA (General AI Assistants)} as the evaluation benchmark for our experiments \cite{mialon2023gaia}. Developed by Meta and Hugging Face, GAIA assesses multi-step, tool-mediated tasks (web navigation, retrieval, synthesis) that better reflect coordinated agent workflows than single-hop QA. Its tasks require planning, external information integration, and verification, aligning with the collaborative division of labor in our five-role system.

\subsection*{Alternative Benchmarks Considered}
Prior to committing to GAIA we reviewed complementary agent-oriented benchmarks:
\begin{itemize}
    \item \textbf{InfiAgent-DABench}: Tabular data analysis via ReAct + Python sandbox; strong automatic correctness through format prompting, but modality narrow (CSV-centric) and limited cross-source synthesis.
    \item \textbf{DABStep}: Multi-step financial / mixed-format analytics with factoid scoring; highly challenging, but very low absolute accuracies (<20\% on harder sets) reduce early discriminative signal between configurations.
    \item \textbf{GTA}: Real user queries with tool argument, selection, and summarization metrics; granular tool-use diagnostics but less emphasis on broad multi-hop synthesis.
    \item \textbf{Mobile-Bench}: Multi-app mobile UI interaction with milestone (CheckPoint) tracking; environment coupling not aligned with our cloud API context.
\end{itemize}

\subsection*{Rationale for GAIA}
GAIA offered (i) mid-range difficulty producing measurable variance without pervasive failure, (ii) heterogeneous action requirements (retrieval + synthesis) mapping to distinct agent roles, (iii) sensitivity to reasoning verbosity (enabling joint performance–cost analysis), and (iv) broad external validity for knowledge-work style orchestration. Hard-tier exclusion avoided disproportionately long trajectories with diminishing marginal ranking power in early BO iterations.

\paragraph{Subset Construction}
Each black-box evaluation uses a 10-task slice (5 \emph{easy}, 5 \emph{medium}). Pilot trials showed this yields stable wall-clock bounds and non-saturated accuracy gradients while keeping per-evaluation cost within budget. Hard tasks were deferred to future extensions once a Pareto set over moderate difficulty stabilizes. Sample questions can be found in the Appendix \ref{apx:gaia_tasks}.

\paragraph{Evaluation Procedure}
For every task we record: (1) binary success, (2) cumulative input and output token counts per role (later monetized), and (3) elapsed wall-clock (diagnostic only). Success rate aggregates to the performance objective; tokenized pricing aggregates to the cost objective. No heuristic penalties are added; subsequent scaling applies min–max normalization across observed configurations for surrogate stability.

\paragraph{Relation to Feature Space}
GAIA’s composite demands (knowledge breadth, procedural coding for tool adaptation, arithmetic / structured reasoning, and efficiency) correspond directly to the five-dimensional feature embedding (MMLU-Pro, LiveCodeBench, GPQA Diamond, input token cost, output token cost) described in Section \ref{sec:setup_llms}. Using only universally reported metrics avoided sparsity that would arise from adding less consistently available benchmarks (e.g., MATH, MBPP) under a tight evaluation budget.

\paragraph{Limitations}
GAIA does not isolate fine-grained tool argument accuracy (better covered by GTA), nor deep statistical tabular analytics breadth (InfiAgent-DABench), nor UI-manipulation robustness (Mobile-Bench). Future work may layer a secondary panel once initial Pareto exploration converges.

\section{Performance Metrics}
\label{sec:setup_metrics}

The MALBO framework is designed for multi-objective optimization, concurrently targeting two conflicting objectives: task performance and operational cost.

\paragraph{Objective 1: Performance (Accuracy)}
We define performance as the \textbf{task success rate} on the selected subset of the GAIA benchmark. This is a scalar value normalized to the range $[0, 1]$, where 1 indicates that the agent team successfully answered all 10 questions correctly. The primary goal for this objective is \textbf{maximization}.

\paragraph{Objective 2: Cost}
We define cost as the \textbf{total aggregated API execution cost} in US dollars (USD) required for the agent team to complete the 10-task evaluation set. This cost is calculated by summing the token-based fees for all input and output tokens processed by each LLM in the team, based on the pricing models of their respective providers at the time of the experiment. The primary goal for this objective is \textbf{minimization}. To align with the maximization framework of BoTorch, we optimize for the negative cost, where the optimizer seeks to move the value towards zero.

\section{Optimizer Configuration}
\label{sec:setup_optimizer}

We implemented the Bayesian optimization loop using \textbf{BoTorch} \cite{balandat2020botorch}, a state-of-the-art library for Bayesian optimization in Python. This framework provides a suite of tools for multi-objective Bayesian optimization, including flexible implementations of Gaussian Process (GP) surrogate models and a wide range of acquisition functions.

\subsection{Initialization Strategy}
The optimization process began with an initial design of experiments to seed the surrogate models. We generated the \textbf{15 initial configurations} by randomly sampling, with replacement, from the discrete pool of available LLMs for each of the three optimized agent roles (\texttt{manager}, \texttt{search\_agent}, and \texttt{reformulator}). This direct sampling strategy ensures that each point in the initial dataset corresponds exactly to a real, deployable team configuration and its associated feature vector. Therefore, these initial points do not require the nearest-neighbor projection function $\pi$ that is used in the main optimization loop.

\subsection{Optimization Loop}
Following the initialization phase, we ran the Bayesian optimization loop for \textbf{15 sequential iterations}, with a batch size of $q=1$. This resulted in a total computational budget of 30 full benchmark evaluations (15 initial + 15 guided by the optimizer).

The acquisition function chosen for this multi-objective problem was the \textbf{q-Log Expected Hypervolume Improvement (qLogEHVI)} \cite{daulton2020differentiable}. This function was selected as it is the standard for multi-objective BO, guiding the search towards candidate points that are most likely to expand the volume of the known Pareto front, thereby efficiently balancing exploration and exploitation.

\section{Selection of Large Language Models}
\label{sec:setup_llms}

\begin{figure}[h!]
    \centering
    \includegraphics[width=\textwidth,height=0.5\textheight,keepaspectratio]{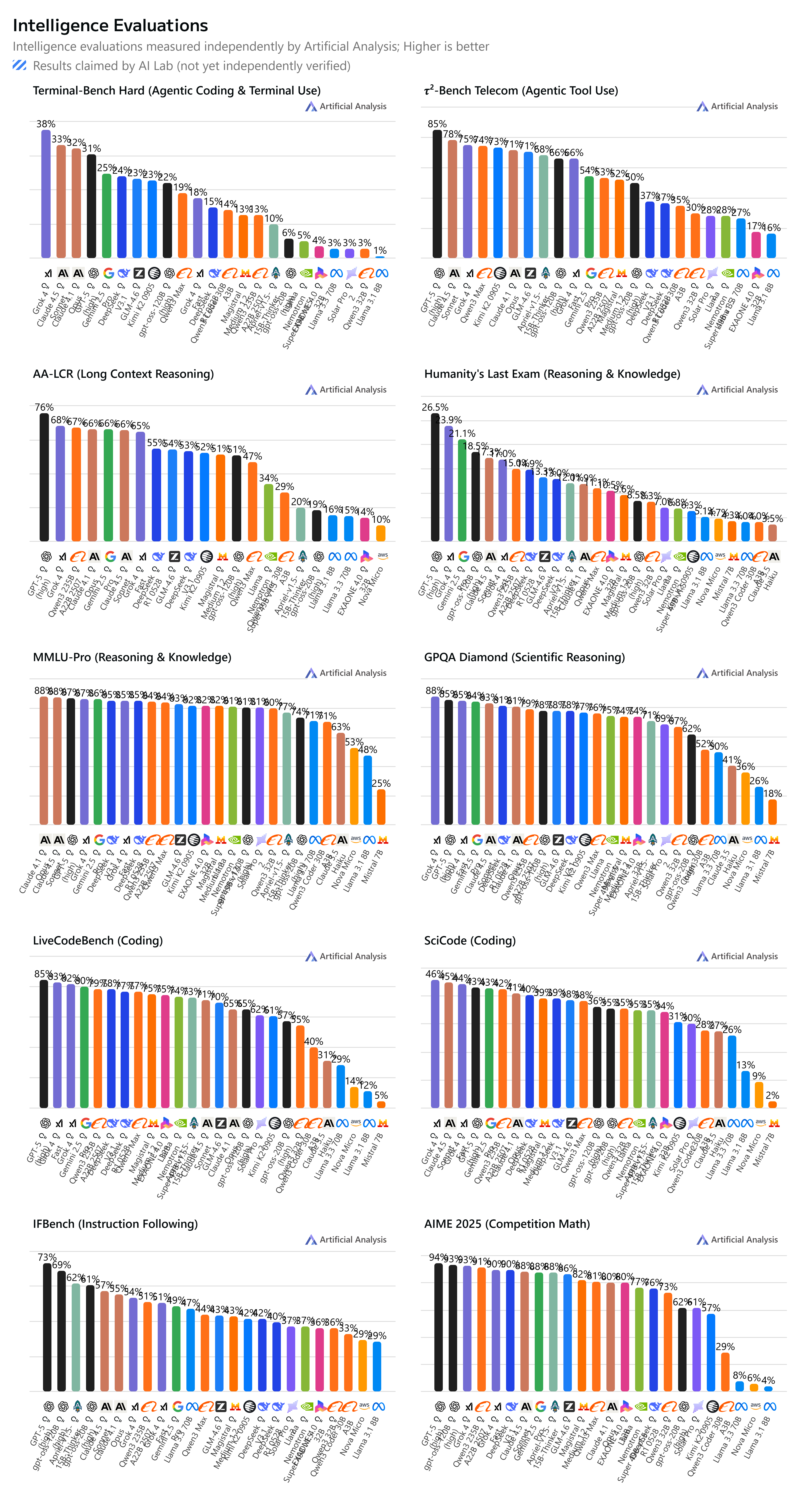}
    \caption{Screenshot of the LLM performance benchmarks from Artificial Analysis, showing the standardized evaluation of models like MMLU and LiveCodeBench.}
    \label{fig:artificial_analysis_perf}
\end{figure}

\begin{figure}[h!]
    \centering    \includegraphics[width=\textwidth,height=0.5\textheight,keepaspectratio]{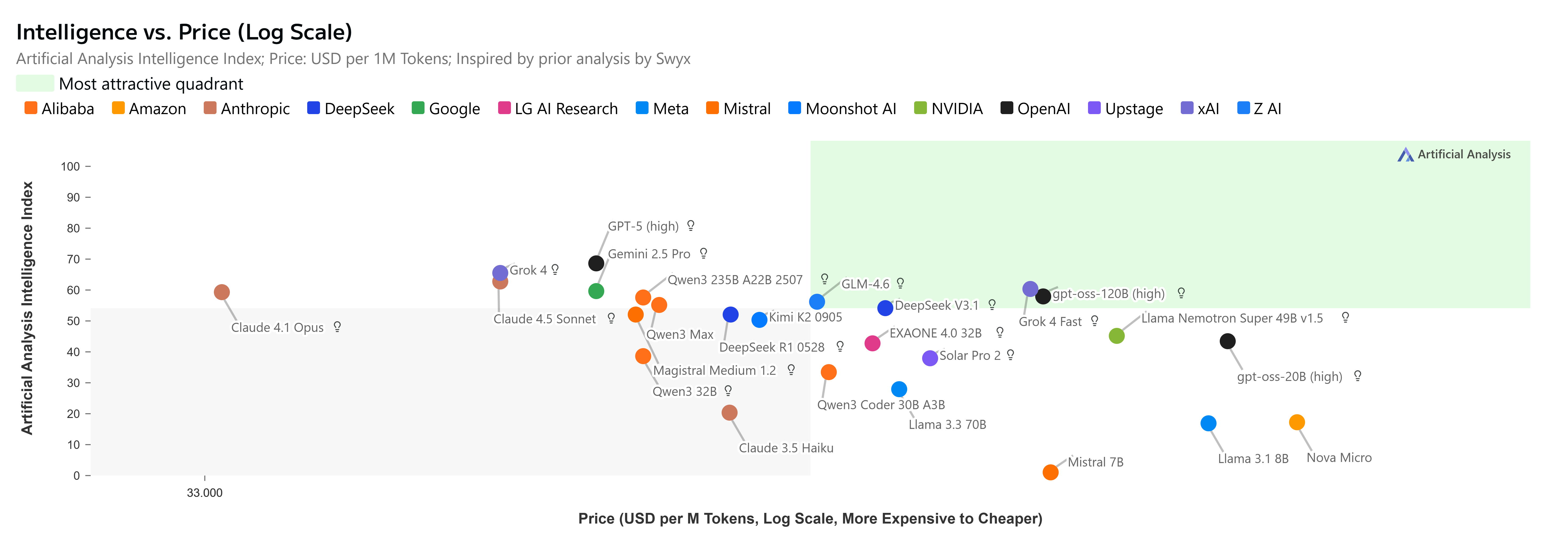}
    \caption{Image of the cost-performance trade-off visualization from Artificial Analysis\cite{artificialanalysis2025}.}
    \label{fig:artificial_analysis_cost}
\end{figure}

The pool of candidate models for our experiment was curated to represent a diverse and cutting-edge cross-section of the modern LLM landscape. We selected a range of models accessible via the Amazon Bedrock platform and other compatible APIs, ensuring variety in architecture (dense vs. Mixture-of-Experts), parameter count, and cost profile. The feature space used to represent each LLM consists of five dimensions chosen to capture key aspects of reasoning, coding, and cost:
\begin{itemize}
    \item \textbf{MMLU-Pro:} A proxy for general knowledge and multi-domain reasoning ability. 
    \item \textbf{LiveCodeBench:} A proxy for code generation and algorithmic logic.
    \item \textbf{GPQA Diamond:} A proxy for hard mathematical and scientific reasoning.
    \item \textbf{Cost:} The API inference costs, separated into price per one million input tokens and price per one million output tokens.
\end{itemize}
These benchmarks were selected because they are widely reported and consistently evaluated, providing a reliable basis for comparison. The performance data was sourced primarily from \textbf{Artificial Analysis}, a third-party platform that independently re-runs benchmarks to ensure a fair comparison between models (see Figure \ref{fig:artificial_analysis_perf} and \ref{fig:artificial_analysis_cost}).
\subsection*{Feature Selection} The five selected dimensions (MMLU-Pro, LiveCodeBench, GPQA Diamond, input token cost, output token cost) form a compact, uniformly available embedding across all candidate models. MMLU proxies broad factual and multi-domain reasoning; LiveCodeBench captures structured code synthesis leveraged by code-as-action planning; GPQA reflects arithmetic / logical scientific reasoning common in multi-hop verification; separate input and output pricing dimensions model asymmetric billing schemes affecting marginal trajectory cost. Additional potential descriptors (context length, active parameters for MoE routing, multilingual scores) were tracked qualitatively for narrative analysis but omitted from the optimization embedding to preserve sample efficiency under a 30-evaluation cap.

\paragraph{Meta Llama 3.1 8B Instruct (Meta AI)} A compact dense decoder-only Transformer using Grouped-Query Attention (GQA) to reduce KV cache pressure and enable efficient 128K context inference. Trained on a large (>15T tokens) multilingual corpus with supervised fine-tuning (SFT) and RLHF for instruction alignment. Provides strong multilingual reasoning relative to its parameter scale \cite{meta2024llama31, dubey2024llama}.

\paragraph{Meta Llama 3.3 70B Instruct (Meta AI)} An optimized 70B successor variant delivering quality uplift over earlier 70B releases without parameter growth. Retains 128K context, refined post-training alignment, and efficiency improvements in inference scheduling and memory layout; positioned as a high-capability open-weight alternative for general reasoning and tool orchestration \cite{meta2024llama33, dubey2024llama}.

\paragraph{Mistral 7B Instruct v0.2 (Mistral AI)} A highly efficient 7B dense model combining Sliding Window Attention (SWA) for long-sequence scalability with GQA for faster multi-head execution. Demonstrates that architectural efficiency can outperform larger legacy dense models (e.g., surpassing LLaMA 2 13B in several benchmarks) while remaining cost-effective for multi-agent role assignment \cite{jiang2023mistral}.

\paragraph{OpenAI GPT-OSS 20B / 120B (OpenAI)} Open-weight Mixture-of-Experts (MoE) Transformer family. Each token activates only a small subset of experts (3.6B and 5.1B active parameters respectively), decoupling total capacity from inference cost. Supports configurable reasoning modes and lightweight quantization (e.g., MXFP4) for deployment efficiency. Serves as a bridge between fully proprietary frontier models and reproducible research baselines \cite{openai2025gpt}.

\paragraph{Qwen3 32B (Alibaba / Qwen Team)} A dense Transformer emphasizing balanced performance across coding, math, and general reasoning with extended context handling. Improvements in tokenizer design and training efficiency yield performance competitive with larger earlier-generation models, making it a strong mid-to-upper tier assignment candidate \cite{qwen2025qwen3}.

\paragraph{Qwen3 Coder 30B A3B (Alibaba / Qwen Team)} A specialized MoE-oriented coding variant (releasing both dense and routed expert configurations) emphasizing function calling, tool use, and extended (up to 262K) native context. Its specialization enhances downstream code synthesis and iterative reformulation roles in agent workflows \cite{qwen2025qwen3}.

\paragraph{Claude 3.5 Haiku (Anthropic)} A lightweight, latency-optimized model delivering instruction following and structured reasoning quality approaching larger tier siblings while maintaining aggressive throughput. Adds improved multimodal grounding (e.g., UI / screenshot interpretation) and fast tool mediation, making it suitable for fast-turn manager or verifier roles under tight cost constraints \cite{anthropic2024claude35haiku}.

\paragraph{DeepSeek-V3 (DeepSeek)} A large-scale MoE architecture (671B total, ~37B active parameters per token) introducing Multi-head Latent Attention (MLA) and auxiliary-loss-free expert balancing to improve routing stability and efficiency. Trained with an emphasis on cost-effective scaling (reduced GPU-hour footprint) while retaining frontier-level reasoning performance \cite{deepseek2024v3}.

\paragraph{Amazon Nova Micro (Amazon)} A compact text-only model within Amazon’s Nova family emphasizing ultra-low latency, broad multilingual coverage, and minimal per-token pricing for high-frequency orchestration pathways. Serves as a cost-efficient baseline for roles where marginal reasoning uplift does not justify higher active parameter footprints \cite{amazon2024nova}.\\

As an additional contextual note, several recently released or higher-tier proprietary frontier families, such as OpenAI’s latest GPT-5 series (including smaller o3 / o4-mini and GPT-5 mini / nano variants), Google’s Gemini 2.5 family, and Meta’s new Llama 4 line, were not integrated due to unavailability through our provisioned cloud endpoints at experiment time. Their exclusion reflects platform access constraints rather than methodological limitations; future replications may incorporate them to extend the performance–cost trade-off space \cite{openai2025gpt, dubey2024llama}.

The final normalized feature matrix derived from these sources is summarized in Table \ref{tab:llm_pool}. All numeric feature dimensions were min–max scaled to $[0,1]$ prior to use in the MALBO optimization process.

\begin{table}[h!]
\centering
\caption{Pool of candidate LLMs and their corresponding feature values. Model and benchmark identifiers are shortened for readability. Input and output costs are expressed in USD per one million tokens.}

\label{tab:llm_pool}
\sisetup{table-format=2.2} % Impostazione di default per le colonne S
\begin{tabular}{@{}l S S S S[table-format=1.2] S[table-format=1.2]@{}} 
\toprule 
\textbf{Model} & {\textbf{MMLU}} & {\textbf{LiveCodeBench}} & {\textbf{GPQA}} & {\textbf{Input}} & {\textbf{Output}} \\ 
\midrule
Llama 3.1 8B    & 48.0 & 12.0 & 26.0 & 0.10 & 0.10 \\
Llama 3.3 70B   & 71.0 & 29.0 & 50.0 & 0.54 & 0.68 \\
Mistral 7B v0.2 & 25.0 & 5.0  & 18.0 & 0.25 & 0.25 \\
GPT-OSS 20B              & 74.0 & 72.0 & 62.0 & 0.05 & 0.20 \\
GPT-OSS 120B             & 81.0 & 64.0 & 78.0 & 0.15 & 0.60 \\
Qwen3 32B                & 80.0 & 55.0 & 67.0 & 0.03 & 0.13 \\
Qwen3 Coder 30B      & 78.0 & 51.0 & 62.0 & 0.08 & 0.29 \\
Claude 3.5 Haiku         & 63.0 & 31.0 & 41.0 & 0.80 & 4.00 \\
DeepSeek-V3.1            & 85.0 & 78.0 & 78.0 & 0.27 & 1.00 \\
Amazon Nova Micro        & 53.0 & 14.0 & 36.0 & 0.04 & 0.14 \\
\bottomrule
\end{tabular}
\end{table}

\chapter{Results and Analysis}
\label{ch:results}

This chapter presents the empirical results obtained from the application of the Multi-Agent LLM Bayesian Optimization (MALBO) framework. The analysis is structured into two primary parts. Initially, we examine the behavior and convergence of the optimization algorithm itself, focusing on the evolution of the Pareto front and the metrics that demonstrate the effectiveness of the search process. Subsequently, the chapter provides a detailed analysis of the optimal configurations discovered, interpreting the specific Large Language Model (LLM) assignments and the underlying principles that govern the trade-off between system performance and operational cost.

\section{Evolution and Convergence of the Optimization Process}
\label{sec:results_evolution}

To assess the efficacy of the MALBO framework, we first analyze the dynamics of the optimization run. This involves evaluating how the set of optimal solutions, or the Pareto front, improves over the course of the 15 optimization iterations and whether the algorithm demonstrates convergence towards a stable set of solutions.

\subsection{Pareto Front Evolution}

\begin{figure}[h!]
    \centering
    \includegraphics[width=0.9\textwidth]{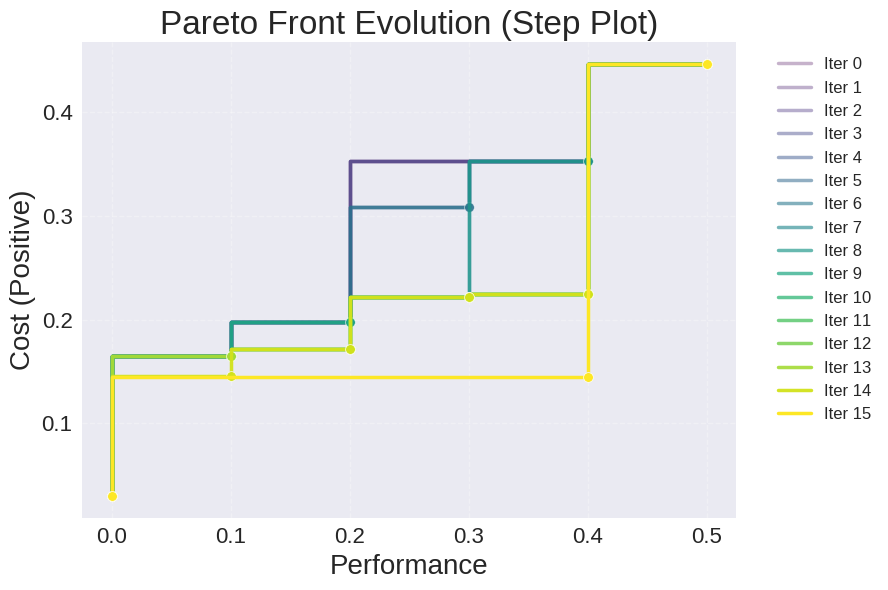}
    \caption{Evolution of the Pareto front across the 15 optimization iterations, visualized as a step plot. The x-axis represents task performance (higher is better), and the y-axis represents operational cost (lower is better). Each colored line delineates the non-dominated region discovered up to a given iteration. The progression of colors from purple (early iterations) to yellow (final iterations) illustrates the systematic expansion of the Pareto front towards the optimal region of high performance and low cost.}
    \label{fig:pareto_evolution_step}
\end{figure}

The primary output of a multi-objective optimization is the Pareto front, which represents the set of non-dominated solutions. Figure \ref{fig:pareto_evolution_step} illustrates the evolution of the Pareto front discovered by MALBO (the non-step visualization is provided in Appendix \ref{fig:pareto_evolution}). Each line represents the approximate front at a given iteration, with colors progressing from dark purple (early iterations) to bright yellow (final iterations). We observe a clear and consistent progression of the front towards the ideal region of the objective space characterized by high performance and low cost (the bottom-right corner). The final front, depicted in yellow, strictly dominates the fronts from earlier iterations, indicating that the optimization process successfully discovered superior trade-off solutions over time. A complementary three-dimensional visualization of this evolution, which allows for interactive inspection, is available in Appendix \ref{apx:3d_evolution} and in the project's public repository \cite{sabbatella2025malbo_github}.

\begin{figure}[h!]
    \centering
    \includegraphics[width=0.8\textwidth]{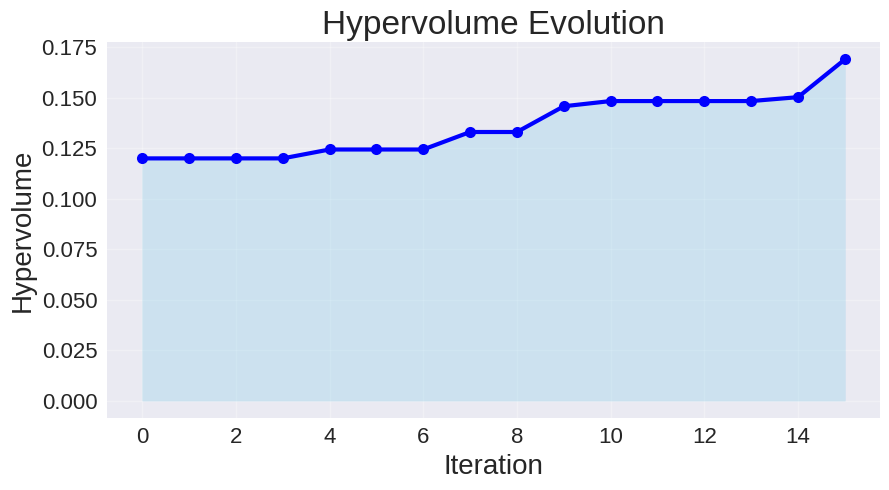}
    \caption{Evolution of the hypervolume indicator over the 15 optimization iterations. The monotonic, non-decreasing trend confirms that the quality of the Pareto front approximation consistently improved.}
    \label{fig:hypervolume_evolution}
\end{figure}

The hypervolume indicator provides a quantitative measure of this progression. The hypervolume, $HV(\mathcal{P}, \vect{r})$, of a Pareto front approximation $\mathcal{P}$ with respect to a reference point $\vect{r}$ is defined as the volume of the portion of the objective space that is weakly dominated by $\mathcal{P}$ and bounded by $\vect{r}$ \cite{zitzler2003performance}. Formally, it is the Lebesgue measure $\lambda$ of the union of hyperrectangles formed by each point $\vect{p} \in \mathcal{P}$ and the reference point:
\begin{equation}
    HV(\mathcal{P}, \vect{r}) = \lambda \left( \bigcup_{\vect{p} \in \mathcal{P}} [\vect{p}, \vect{r}] \right)
\end{equation}
where $[\vect{p}, \vect{r}]$ denotes the hyperrectangle. As shown in Figure \ref{fig:hypervolume_evolution}, the calculated hypervolume increases steadily, particularly during the initial iterations. The curve begins to plateau towards the final iterations, which is a strong indicator that the optimization is approaching convergence. This confirms that the budget of 30 total evaluations was sufficient to effectively map the Pareto front for this problem.

\subsection{Granular Analysis of Cost-Performance Trade-offs}

\begin{figure}[h!]
    \centering
    \includegraphics[width=\textwidth]{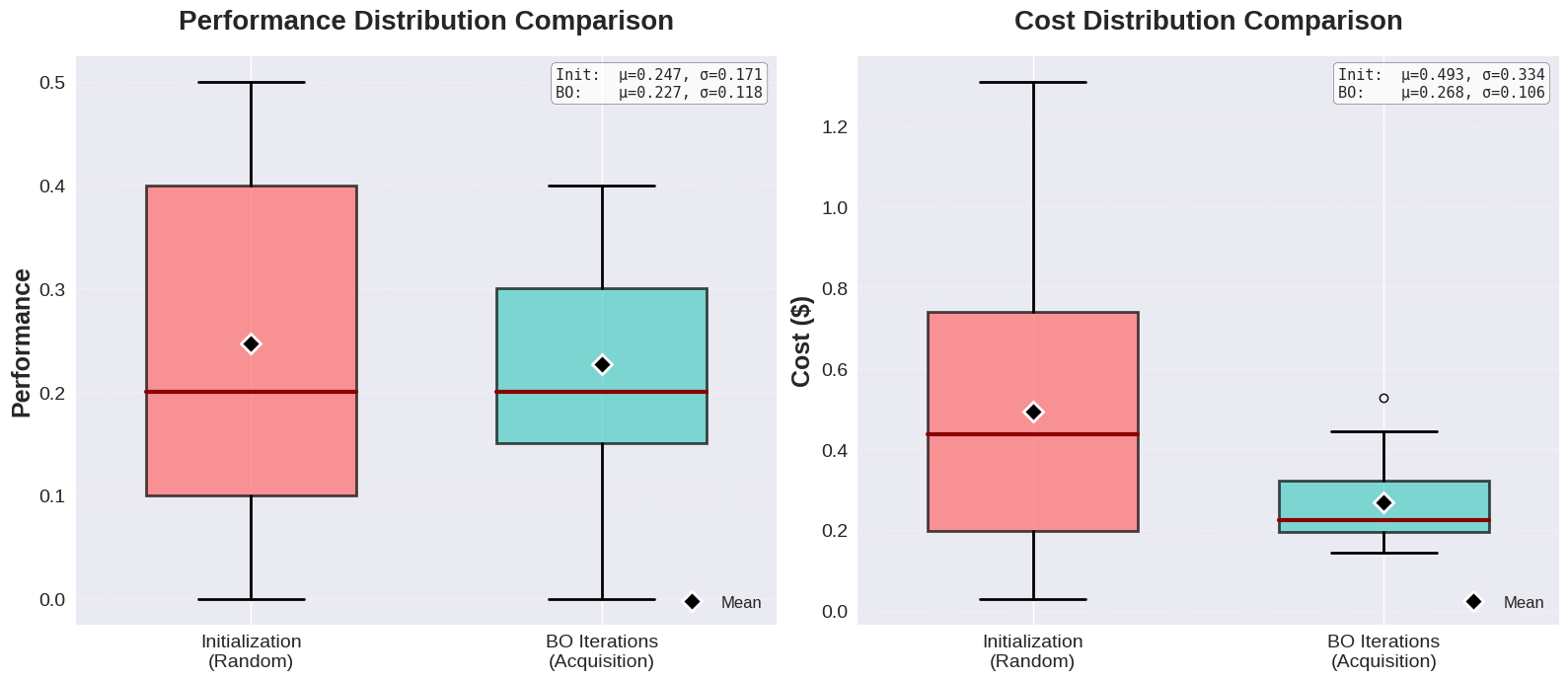}
    \caption{Distribution comparison of performance (left) and cost (right) for the initial 15 random configurations (Initialization) and the 15 configurations selected by the acquisition function (BO Iterations). The box plots show the median (red line), mean (diamond), and interquartile range. The BO phase consistently explores lower-cost solutions while maintaining a similar performance distribution.}
    \label{fig:dist_comparison}
\end{figure}

To provide a statistical overview of the optimization process, we first compare the distribution of objective values from the initial random sampling phase with those from the subsequent Bayesian Optimization (BO) iterations. Figure~\ref{fig:dist_comparison} presents this comparison using box plots for both performance and cost. For a comprehensive visualization of all evaluated points in the objective space, see Figure~\ref{fig:objective_space_coverage_appendix} in the Appendix.

% NUOVA FIGURA CON I BOX PLOTS

The visual comparison is supported by a formal statistical analysis summarized in Table~\ref{tab:ttest_results}. We performed a Welch's t-test to assess the difference in means between the two phases. For the performance objective, the difference is not statistically significant ($p = 0.722$), indicating that the BO phase maintained a level of performance comparable to the initial exploration. In contrast, for the cost objective, the difference is statistically significant ($p = 0.028$) with a large effect size (Cohen's d = 0.907). The mean cost of configurations selected by BO is \$0.268, representing a \textbf{45.64\% reduction} compared to the mean cost of the initial set (\$0.493). This demonstrates that the optimizer effectively guided the search towards significantly more cost-efficient regions of the design space, with the most expensive configuration found by BO costing only \$0.527, compared to \$1.309 in the initial phase.

% NUOVA TABELLA CON I RISULTATI STATISTICI
\begin{table}[h!]
\centering
\caption{Summary of Welch's t-test comparing Initialization and BO Iteration phases for both objectives. 
A significance level of $\alpha = 0.05$ was used. 
The test reveals a statistically significant reduction in cost, while the change in performance is not significant.}
\label{tab:ttest_results}
\begin{tabular}{lrr}
\toprule
\textbf{Metric} & \textbf{Performance} & \textbf{Cost (\$)} \\
\midrule
Mean (Initialization) & 0.247 & 0.493 \\
Mean (BO Iterations) & 0.227 & 0.268 \\
\midrule
t-statistic & -0.360 & 2.399 \\
p-value & 0.722 & 0.028 \\
Significance & Not Significant & Significant (*) \\
Cohen's d (Effect Size) & -0.136 (Negligible) & 0.907 (Large) \\
\bottomrule
\end{tabular}
\end{table}

\begin{figure}[h!]
    \centering
    \includegraphics[width=\textwidth]{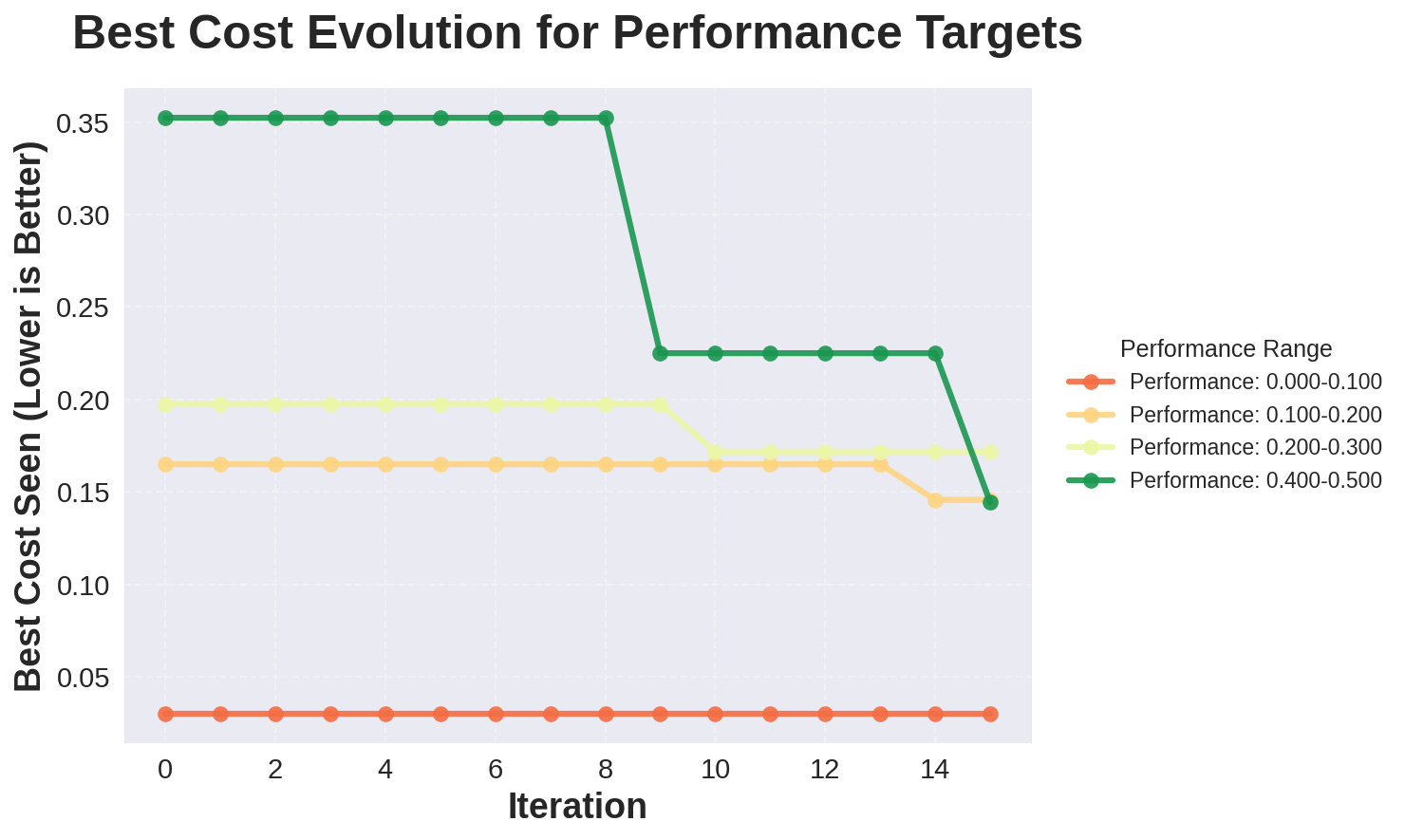}
    \caption{Evolution of the minimum cost discovered over time for different performance tiers. Each line represents the best (lowest) cost found up to that iteration for configurations within a specific performance range. The downward trends signify the optimizer's success in finding more cost-effective solutions.}
    \label{fig:cost_evolution_tiers}
\end{figure}

\begin{figure}[h!]
    \centering
    \includegraphics[width=\textwidth]{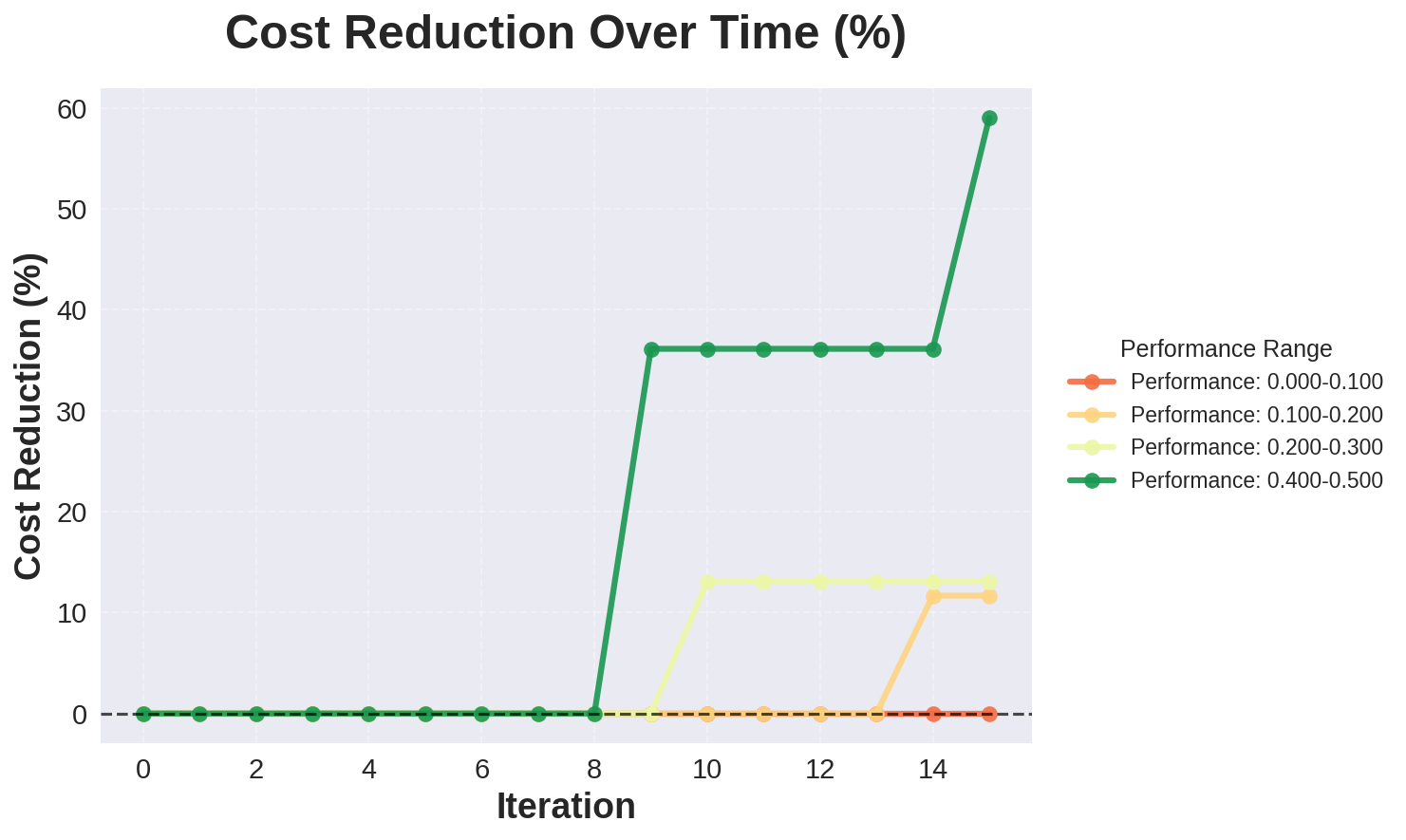}
    \caption{Relative cost improvement over time for different performance targets. The y-axis shows the percentage reduction in cost compared to the first solution found in that bin. The substantial improvement in the high-performance tier (0.4-0.5) highlights the effectiveness of the optimization.}
    \label{fig:cost_improvement}
\end{figure}

While the overall distributions show a clear trend, a more granular analysis reveals how the optimizer improved solutions at different performance levels. We partitioned the performance space into discrete bins and tracked the minimum cost found within each bin over the course of the optimization. Figure~\ref{fig:cost_evolution_tiers} shows the evolution of the best cost found for each performance tier, while Figure~\ref{fig:cost_improvement} presents the same data as a percentage improvement from the first-discovered solution.

% FIGURE PRECEDENTI (già nel tuo codice)

The results are particularly revealing. For the highest performance tier (0.4-0.5), MALBO achieved a remarkable \textbf{59.04\% reduction in cost} by the final iteration. This demonstrates that the optimizer successfully found alternative LLM assignments that maintained this level of performance at a fraction of the cost of the initial configurations. It is noteworthy that the maximum performance achieved (0.5) was already present in the initial set of evaluated points. Therefore, the optimization in this tier focused exclusively on reducing cost. In the lowest-performance tier, no improvement was observed, likely because the initial random sampling had already identified the most cost-effective baseline configurations. This analysis provides strong evidence that MALBO is not merely identifying good solutions but is actively exploiting the cost-performance landscape to find more efficient ones.

\subsection{Analysis of the Acquisition Function's Behavior}

\begin{figure}[h!]
    \centering
    \includegraphics[width=\textwidth]{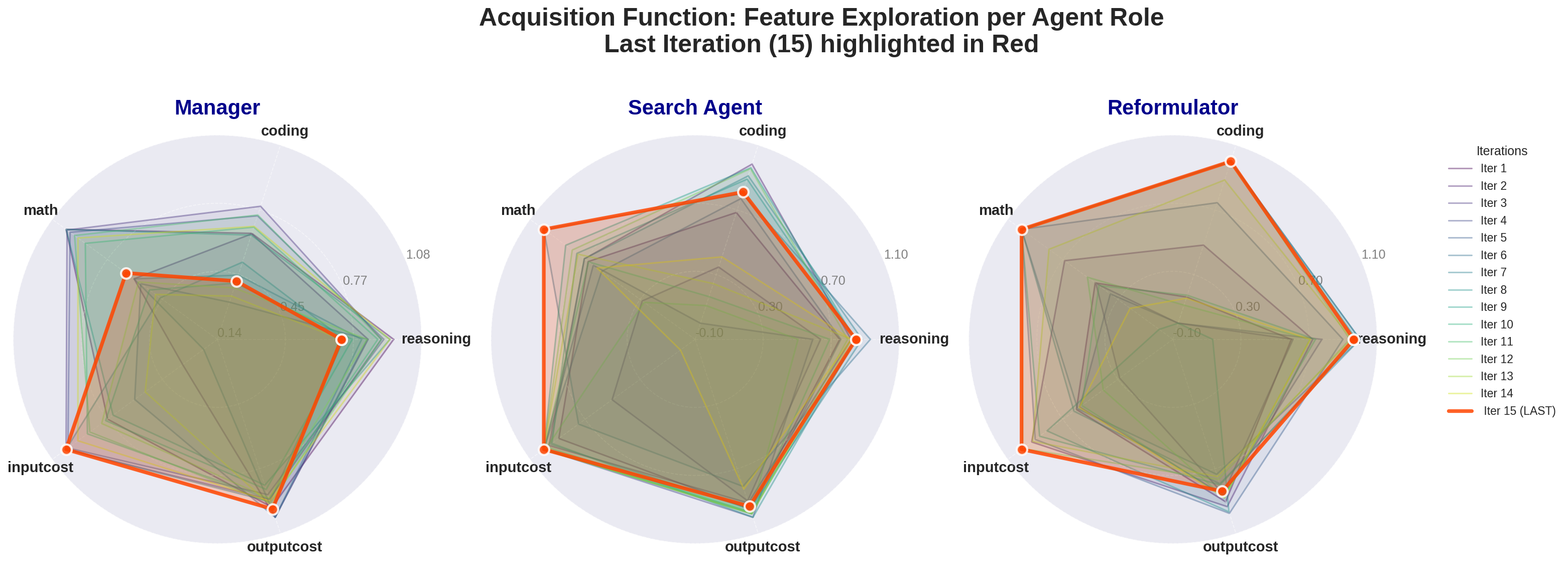} 
    \caption{Radar charts illustrating the feature exploration of the acquisition function for each key agent role. Each axis corresponds to a normalized feature dimension. The semi-transparent polygons represent the configurations proposed at each iteration, while the highlighted red polygon shows the candidate from the final iteration (15). Distinct, role-specific exploration trends emerge clearly from the chart.}
    \label{fig:acquisition_radar_charts}
\end{figure}

\begin{figure}[h!]
    \centering
    \includegraphics[width=0.7\textwidth]{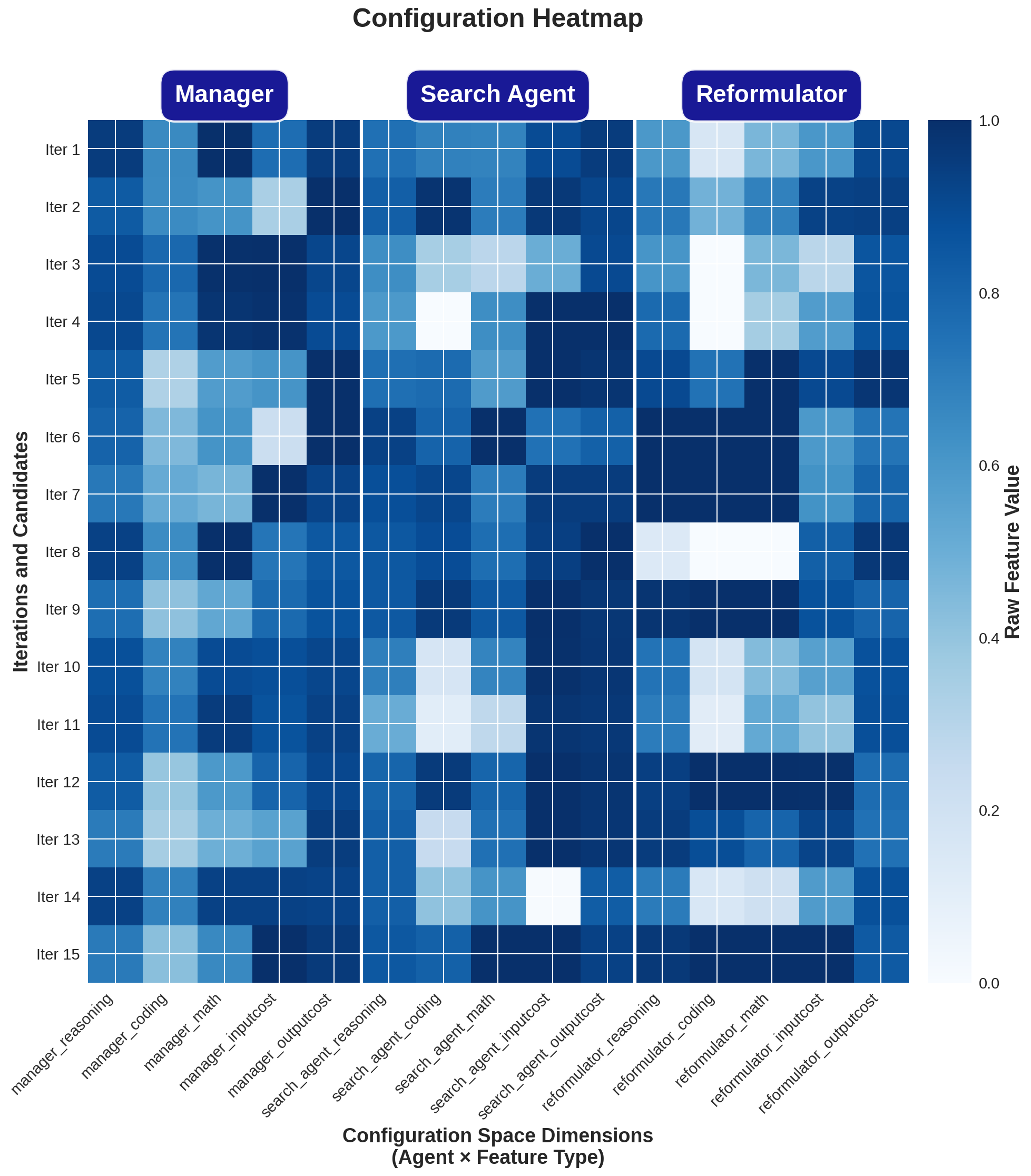} 
    \caption{Heatmap of the unnormalized continuous configurations suggested over time. Each row is an iteration, and each column is a feature dimension. Darker shades indicate higher absolute values for that feature. This visualization reveals the optimizer's evolving preferences across the entire search space.}
    \label{fig:acquisition_heatmap}
\end{figure}

To understand the internal strategy of the Bayesian optimizer, we analyzed the configurations proposed by the qLogEHVI acquisition function at each iteration. We visualize these proposals using two complementary methods. Figure~\ref{fig:acquisition_radar_charts} uses radar charts to illustrate the feature profiles for each agent role, while Figure~\ref{fig:acquisition_heatmap} provides a heatmap to track the absolute values of each feature over time.

% FIGURA CON I RADAR CHARTS

The radar charts in Figure~\ref{fig:acquisition_radar_charts} reveal that the optimizer's exploration is not random. We observe distinct, role-specific patterns: the optimizer proposes candidates with strong coding and math capabilities for the \textbf{Search Agent} and \textbf{Reformulator}, while the proposals for the \textbf{Manager} show a different profile. In the final iteration (highlighted in red), the ideal candidate for the Manager has comparatively lower values for coding, reasoning, and math, while maintaining relatively high values for input and output costs.

The heatmap in Figure~\ref{fig:acquisition_heatmap} provides further quantitative support for these observations. We can see that the columns corresponding to the \textbf{output cost} (`manager output cost`, `search agent output cost`, `reformulator output cost`) are consistently dark across nearly all iterations. This indicates a strong and early-learned preference for models with higher output token prices. Furthermore, we can identify other role-specific trends, such as a general preference for a high `search agent input cost` and, on average, a lower `reformulator coding` capability, a pattern more clearly visible in the parallel coordinates plot (Appendix~\ref{fig:acquisition_parallel_coords}). This non-random exploration suggests that the acquisition function is actively learning the underlying structure of the problem and refining its search strategy from broad exploration to a focused, hypothesis-driven exploitation.

\subsection{Comparison with Homogeneous Baselines}

To further contextualize the value of MALBO's optimized, heterogeneous configurations, we established a set of baselines that simulate a common heuristic approach: assigning the same LLM to all three key agent roles. We selected four representative models for this comparison: the one with the highest benchmark scores (`deepseek.v3.1`), the most expensive (`claude-3-5-haiku`), the strongest dedicated reasoning model (`openai.gpt-oss-120b`), and the least expensive (`meta.llama3-1-8b-instruct`). For each, we performed three runs and recorded the best performance achieved, as summarized in Table \ref{tab:homogeneous_baselines}.

\begin{table}[h!]
\centering
\caption{Best performance of homogeneous agent teams, where the same model was assigned to the Manager, Search Agent, and Reformulator roles.}
\label{tab:homogeneous_baselines}
\begin{tabular}{@{}lrr@{}}
\toprule
\textbf{Homogeneous Team Model} & \textbf{Best Performance} & \textbf{Cost (\$)} \\
\midrule
DeepSeek-V3              & 0.5 & 1.303 \\
GPT-OSS 120B             & 0.3 & 0.351 \\
Claude 3.5 Haiku         & 0.2 & 1.084 \\
Llama 3.1 8B Instruct    & 0.0 & 0.033 \\
\bottomrule
\end{tabular}
\end{table}

The results of this baseline comparison provide a clear illustration of the value offered by the MALBO framework. The homogeneous `DeepSeek-V3` team was the only baseline to achieve the maximum performance score of 0.5, but it did so at a very high cost of \$1.303. In stark contrast, MALBO's top-performing heterogeneous archetype achieved the exact same performance for only \$0.446, representing a \textbf{65.8\% reduction in cost}. 

Furthermore, none of the homogeneous baseline configurations are present on the final Pareto front. For instance, the `GPT-OSS 120B` baseline achieved a performance of 0.3 at a cost of \$0.351, a result that is strictly dominated by MALBO's balanced archetype (0.4 performance at \$0.144). This empirical comparison provides strong evidence that optimizing for specialized, heterogeneous team compositions, as MALBO does, can yield configurations that are significantly more cost-efficient than those derived from common, monolithic assignment strategies.

% Inserisci questo codice DOPO la fine della \subsection{Analysis of the Acquisition Function's Behavior}
\newpage
\section{Analysis of Optimal Configurations and LLM Assignments}
\label{sec:results_configs}

Having established the robust performance of the optimization algorithm, we now shift our focus to the practical implications of its findings. This section examines the specific LLM configurations that constitute the final Pareto front, analyzes the model preferences learned by the optimizer for each agent role, and derives actionable insights from these patterns.

\subsection{Case Studies of Pareto-Optimal Configurations}

The final Pareto front, discovered at iteration 15, comprises three distinct and non-dominated configurations. Each of these solutions represents an optimal trade-off archetype, offering a different balance between performance and cost. We present them here as case studies to illustrate the practical choices available to a system designer.

\paragraph{Archetype 1: The Cost-Optimized Maximum Performance Configuration}
The optimization process identified a configuration on the final Pareto front that achieves the maximum performance score of 0.500 observed throughout the experiment. It is crucial to note that this performance level was reached by only three configurations, all discovered during the initial random sampling phase (see Appendix~\ref{apx:supplementary} for a complete visualization). The other two configurations that achieved this score were:
\begin{itemize}
    \item \textbf{Init. Point 4:} Manager (`claude-3-5-haiku`), Search Agent (`gpt-oss-20b`), Reformulator (`deepseek.v3`) at a cost of \$0.737.
    \item \textbf{Init. Point 10:} Manager (`gpt-oss-120b`), Search Agent (`claude-3-5-haiku`), Reformulator (`claude-3-5-haiku`) at a cost of \$0.877.
\end{itemize}
The archetype on the Pareto front is therefore the most cost-efficient solution found for achieving peak performance:

\begin{itemize}
    \item \textbf{Performance:} 0.500
    \item \textbf{Cost:} \$0.445803 per evaluation
    \item \textbf{Agent Assignment:}
    \begin{itemize}
        \item \texttt{manager}: \texttt{anthropic.claude-3-5-haiku}
        \item \texttt{search\_agent}: \texttt{openai.gpt-oss-20b}
        \item \texttt{reformulator}: \texttt{openai.gpt-oss-120b}
    \end{itemize}
\end{itemize}
\textbf{Analysis:} Priced at \$0.445, this setup is 39.5\% more affordable than the subsequent best 0.5-performing setup. A noteworthy trend among all three leading teams is the strategic integration of the pool's priciest model (`claude-3-5-haiku`) with its most effective reasoning MOE models (`gpt-oss`). This specific configuration designates `claude-3-5-haiku`, a versatile all-rounder, to the \texttt{manager} position to oversee high-level operations. Concurrently, it employs the largest available MoE model, `gpt-oss-120b`, for the \texttt{reformulator} position, demanding strong language proficiency to refine results. Additionally, the `gpt-oss-20b`, a capable and cost-effective reasoning model, is used as the search agent. This assignment maintains maximum performance while achieving a substantially lower cost compared to the other high-performing but less optimized configurations.

\paragraph{Archetype 2: Balanced Trade-off Configuration}
This setup signifies the ideal "knee" point on the Pareto curve, providing a significant cost reduction with only a moderate compromise in performance.
\begin{itemize}
    \item \textbf{Performance:} 0.400 (80\% of maximum)
    \item \textbf{Cost:} \$0.144317 per evaluation (3.1 times cheaper than max performance)
    \item \textbf{Agent Assignment:}
    \begin{itemize}
        \item \texttt{manager}: \texttt{qwen.qwen3-coder-30b}
        \item \texttt{search\_agent}: \texttt{openai.gpt-oss-120b}
        \item \texttt{reformulator}: \texttt{openai.gpt-oss-120b}
    \end{itemize}
\end{itemize}
\textbf{Analysis:} This setup exemplifies a deliberate replacement strategy, swapping the costly generalist manager for a more focused and economical model (Qwen3 Coder), yet still maintaining robust models for support functions. The outcome is a significant gain in efficiency, achieving a 68\% cost reduction while performance decreases by only 20\%. This establishes it as the configuration with the optimal return on investment.

\paragraph{Archetype 3: Minimum Cost Configuration}
This solution establishes the baseline cost for a functional, albeit minimally performant, agent team.
\begin{itemize}
    \item \textbf{Performance:} 0.000
    \item \textbf{Cost:} \$0.030155 per evaluation
    \item \textbf{Agent Assignment:}
    \begin{itemize}
        \item \texttt{manager}: \texttt{meta.llama3-1-8b-instruct}
        \item \texttt{search\_agent}: \texttt{meta.llama3-1-8b-instruct}
        \item \texttt{reformulator}: \texttt{deepseek.v3}
    \end{itemize}
\end{itemize}
\textbf{Analysis:} This archetype provides a significant insight into the system's failure modes. Despite assigning `deepseek.v3`, the model with the highest benchmark scores in our pool, to the \texttt{reformulator} role, the team's overall performance is zero. The failure to complete any task is attributable to the assignment of the least capable model, `llama3-1-8b-instruct`, to the two essential upstream roles: the \texttt{manager} and the \texttt{search\_agent}. This result provides strong empirical evidence supporting our feature importance analysis (Section \ref{sec:results_discussion}), which identified the manager's reasoning capabilities as the primary driver of performance and marginal relevance for the reformulator model. It suggests that the performance of a multi-agent system is constrained by its weakest link in a key role, and that even a highly capable model cannot compensate for a lack of effective planning and information gathering in the initial stages of the workflow.

\subsection{LLM Assignment Preferences and Patterns}
\begin{figure}[h!]
    \centering
    \includegraphics[width=\textwidth]{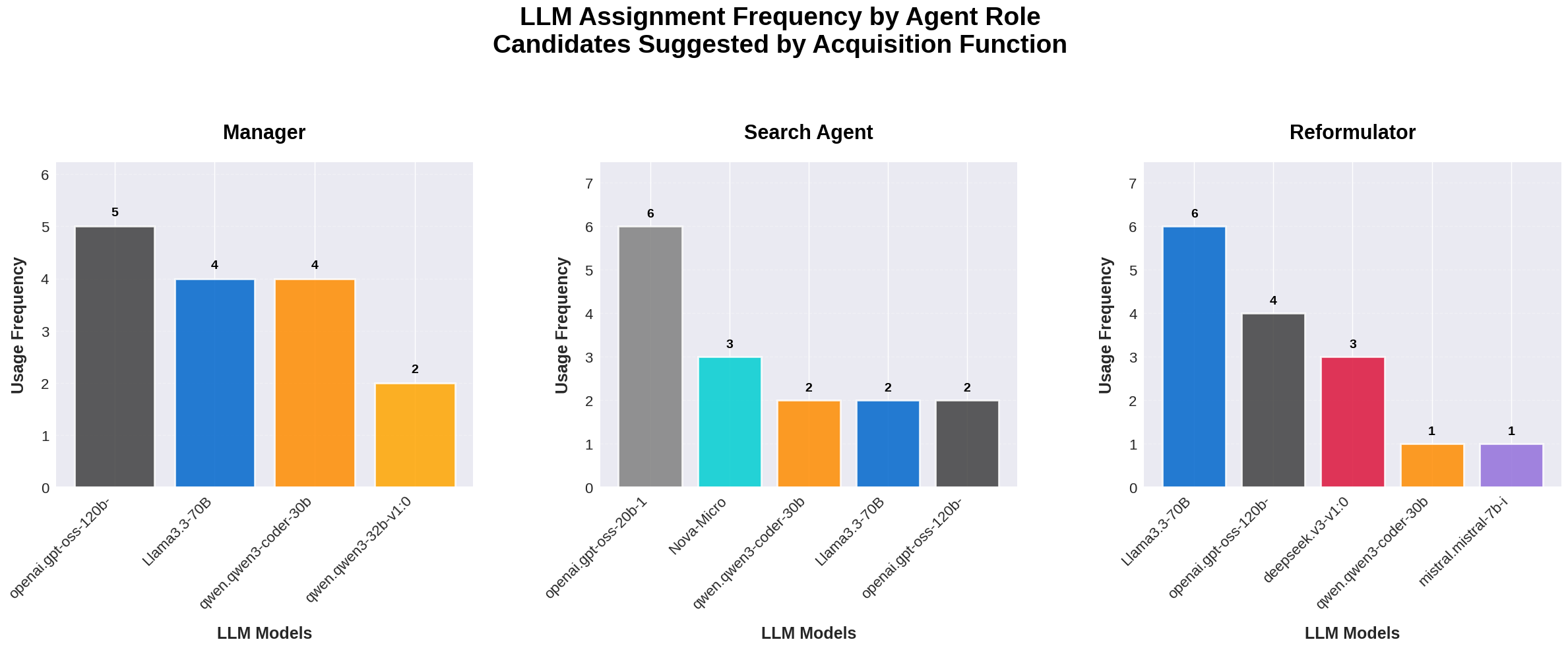}
    \caption{Frequency of LLM assignments for each agent role across all 30 evaluations. The height of each bar indicates how many times a specific LLM was selected for that role. The patterns reveal the optimizer's preferences.}
    \label{fig:llm_frequency}
\end{figure}

To understand how the optimizer arrived at these solutions, we analyzed the frequency of LLM assignments across all 30 evaluated configurations. Figure \ref{fig:llm_frequency} shows the selection frequency for each of the three optimized agent roles.

To understand the optimizer's learned strategy, we analyzed the frequency of LLM assignments proposed by the acquisition function across the 15 BO iterations, as shown in Figure \ref{fig:llm_frequency}. The analysis reveals distinct, role-specific preferences. For the \textbf{manager} role, the acquisition function consistently proposed powerful and large models. The most frequent suggestions were `openai.gpt-oss-120b` (5 selections), followed closely by `meta.llama3-3-70b` and `qwen.qwen3-coder-30b` (4 selections each). This preference for high-capability models reinforces the finding that the manager's performance is a primary driver of the system's success.

The distribution for the \textbf{search\_agent} reveals a clear cost-optimization strategy. The most frequently suggested model was `openai.gpt-oss-20b` (6 selections), a model that combines strong performance with a particularly low token price. The next most common suggestion, `us.amazon.nova-micro` (3 selections), is an ultra-low-cost model. This pattern suggests the optimizer identified the search agent as a token-intensive role and therefore prioritized models with high cost-efficiency.

The \textbf{reformulator} role also shows a preference for capable models, with `meta.llama3-3-70b` (6 selections), `openai.gpt-oss-120b` (4 selections), and `deepseek.v3` (3 selections) being the dominant choices. This indicates that the final synthesis and formatting of the answer benefits from the deep language understanding of large models.

Notably, two out of the ten available models (\texttt{anthropic.claude-3-5-haiku} and \texttt{meta.llama3-1-8b}) were never actively selected by the acquisition function after the initial sampling, despite appearing in the final Pareto front. This indicates that while they define the extremes of the trade-off space, the optimizer did not find them to be optimal choices for any of the intermediate configurations it explored.

\section{Discussion and Interpretation of Deeper Insights}
\label{sec:results_discussion}

Beyond identifying optimal configurations, the MALBO framework allows us to extract deeper insights into the underlying dynamics of the multi-agent system. By analyzing the final trained Gaussian Process surrogate models, we can quantify the importance of different model features and understand what truly drives system performance and cost. We infer feature importance from the learned lengthscale hyperparameters of the kernel with Automatic Relevance Determination (ARD), where importance is inversely proportional to the lengthscale \cite{rasmussen2006gaussian}.

\begin{figure}[h!]
    \centering
    \includegraphics[width=\textwidth]{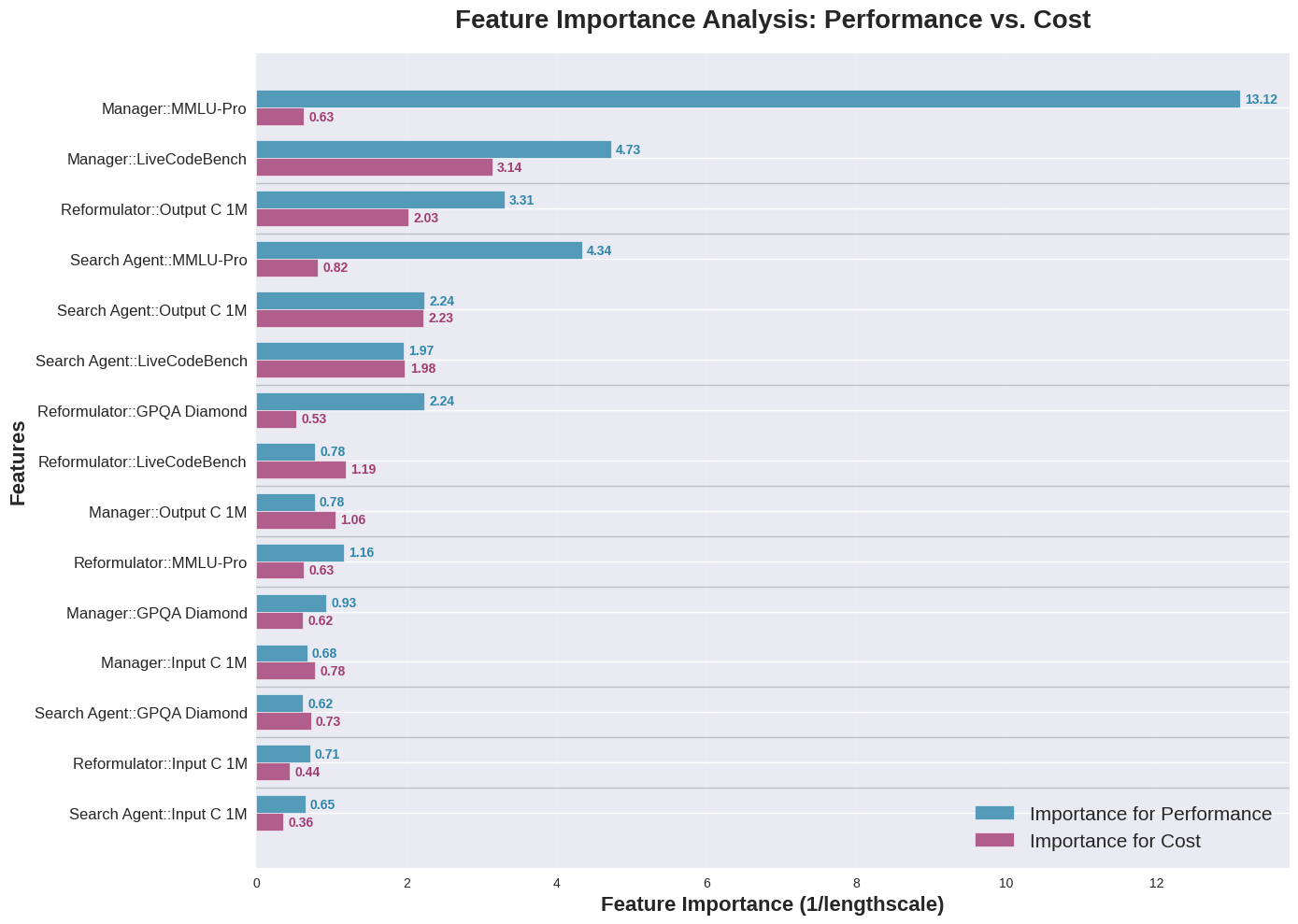} 
    \caption{Feature importance dashboard derived from the final Gaussian Process models. The length of each bar represents the feature's importance (inverse lengthscale) for the performance objective (blue) and the cost objective (red).}
    \label{fig:feature_importance}
\end{figure}

A global view reveals that the drivers for performance and cost are largely asymmetric, as shown in Figure \ref{fig:feature_importance}. The importance scores for performance are highly concentrated in a few key features, with one feature being an order of magnitude more influential than the rest. In contrast, the importance scores for cost are more evenly distributed across several features, suggesting that multiple factors contribute more moderately to the final operational expense.

\subsection{Agent-Level Impact Analysis}
To understand which agent roles are most influential, we aggregated the importance scores of all features associated with each agent. The results, summarized in Table \ref{tab:agent_impact}, clearly indicate the central role of the Manager.

\begin{table}[h!]
\centering
\caption{Aggregated feature importance scores by agent role, showing the total impact of each agent's configuration on the two objectives.}
\label{tab:agent_impact}
\begin{tabular}{@{}lrr@{}}
\toprule
\textbf{Agent Role} & \textbf{Performance Impact} & \textbf{Cost Impact} \\
\midrule
Manager      & 20.235 & 6.227 \\
Search Agent & 9.823  & 6.113 \\
Reformulator & 8.208  & 4.825 \\
\bottomrule
\end{tabular}
\end{table}

The Manager's configuration has a total performance impact (20.235) more than double that of the Search Agent (9.823) and the Reformulator (8.208) combined. This provides strong quantitative evidence that the choice of LLM for the orchestrating agent is the primary determinant of the entire team's success. The impact on cost is more balanced, with all three agents contributing similarly.

\subsection{Granular Feature Analysis}
A more granular analysis of individual features allows us to pinpoint the specific capabilities that drive each objective.

For the \textbf{performance objective}, the two most influential features are overwhelmingly associated with the Manager role. The single most dominant feature is \texttt{Manager::MMLU-Pro} with an importance score of \textbf{13.122}, followed by \texttt{Manager::LiveCodeBench} at \textbf{4.730}. This suggests that the team's success is most sensitive to the Manager's general reasoning and coding abilities, which are essential for planning and orchestrating complex tasks. Conversely, the features with the least impact on performance, such as \texttt{Search Agent::GPQA Diamond} (0.623), indicate that certain specialized reasoning skills in subordinate roles are less influential on the final outcome for this benchmark.

For the \textbf{cost objective}, a different pattern emerges. The most influential feature is surprisingly \texttt{Manager::LiveCodeBench} (\textbf{3.143}). A plausible interpretation is that a Manager with strong coding skills may generate more elaborate, token-intensive plans or tool calls, thereby driving up the overall cost. The second most important feature, \texttt{Search Agent::Output C 1M} (\textbf{2.226}), is more direct: the cost of the model assigned to the search role directly impacts the total expense, as this agent is responsible for generating potentially lengthy outputs. The features with the lowest impact on cost are the input token prices (e.g., \texttt{Search Agent::Input C 1M} at 0.362), confirming that for this workflow, the volume of generated output tokens is a much larger cost driver than the input prompts.

These findings provide guidance for designing multi-agent systems, highlighting that resources should be prioritized for the Manager's capabilities to maximize performance, while for cost control, greater focus should be on token efficiency of all agents, particularly the Search Agent and Manager.

\chapter{Conclusion, Limitations, and Future Prospective}
\label{ch:conclusion}

\section{Conclusion}

This thesis confronted the challenge of optimally composing teams of LLM-based agents, a task ill-served by current manual and heuristic-based approaches. We introduced and validated \textbf{MALBO (Multi-Agent LLM Bayesian Optimization)}, a novel framework that systematically solves this multi-objective assignment problem. By leveraging Bayesian Optimization, MALBO automates the discovery of the Pareto front of configurations, providing a principled method for balancing task performance against operational cost.

Our empirical evaluation successfully validated the efficacy of this approach. The key findings of this work are threefold:
\begin{enumerate}
    \item \textbf{MALBO significantly improves cost-efficiency at target performance levels.} The framework's primary achievement was not in surpassing the maximum observed performance (0.5), but in finding significantly more economical ways to achieve it. The optimization process successfully identified a configuration that maintained this peak performance while reducing operational cost by a remarkable \textbf{59.04\%} compared to the initial, naively discovered solution. This indicates that the method excels at cost-aware resource allocation rather than raw performance enhancement.
    
    \item \textbf{The framework quantifies role-specific feature influence.} The analysis of the trained surrogate models identified the \texttt{Manager}'s MMLU-Pro score as the single most influential feature on system performance. This does not imply that maximizing this feature guarantees the best outcome, but rather that the team's effectiveness is most sensitive to the reasoning capabilities of its orchestrator. This observation reveals the dominant factor to target when refining the team’s architecture.

    \item \textbf{The optimization yields a concrete Pareto front for decision-making.} Instead of suggesting a single "best" team, MALBO delivers a set of non-dominated, Pareto-optimal configurations. This output provides system designers with a clear, quantifiable map of the trade-offs, enabling an informed decision based on specific budgetary and performance requirements a tangible improvement over selecting a single model based on generic benchmarks.
\end{enumerate}

In conclusion, this thesis has successfully answered its core research questions. We have demonstrated that the multi-agent LLM assignment problem can be effectively formalized and solved using multi-objective Bayesian Optimization. The MALBO framework represents a significant step towards a more principled and automated science of multi-agent AI system design.

\section{Limitations}

While the results of this study are promising, we acknowledge its limitations. The experimental validation was constrained by a computational budget of 30 evaluations on a subset of the GAIA benchmark; a larger-scale study would be required to fully confirm the generalizability of our findings. Furthermore, the vector representation of LLMs was based on a static and non-exhaustive set of features. Other attributes, such as model latency or performance on different reasoning tasks, could provide a more nuanced model and potentially lead to different optimal configurations.

\section{Future Prospective}

Building upon the foundation laid by this work, the MALBO framework can be extended along several promising directions.

\paragraph{Extending to More Objectives: Latency, Uncertainty, and Hallucination.}
The current framework optimizes for a two-dimensional objective space: performance and cost. A natural and critical extension is to incorporate additional objectives that are vital for real-world deployment. As we discussed in Section \ref{statisticalPerspective}, model trustworthiness is a multifaceted concept that goes beyond simple task accuracy. Future iterations of MALBO could be designed to co-optimize for objectives such as:
\begin{itemize}
    \item \textbf{Latency:} Minimizing the end-to-end response time of the agent team is essential for user-facing applications. Our preliminary analysis suggests a complex relationship here; smaller models, while faster per token, may require more conversational turns, potentially increasing overall latency. A dedicated latency objective would allow for a formal exploration of this trade-off.
    \item \textbf{Uncertainty:} Minimizing a measure of model uncertainty, for example by using techniques from Conformal Prediction. This would guide the optimization towards configurations that are not only accurate but also provide calibrated and reliable confidence estimates.
    \item \textbf{Hallucination Rate:} Directly minimizing the rate of factual inaccuracies. This would require an additional evaluation step, possibly involving a powerful "judge" LLM or a knowledge base, to score the factuality of each response. By treating factuality as a third objective, MALBO could discover a richer, three-dimensional Pareto surface of optimal configurations.
\end{itemize}

\paragraph{Leveraging Multiple Information Sources.}
Another promising research direction involves integrating Multiple Information Source Optimization (MISO) \cite{candelieri2025multiple, sabbatella2024bayesianMISO} principles. The current framework relies on a single, high-fidelity source of information: the full evaluation on the 10-task GAIA benchmark subset. A MISO approach could leverage cheaper, auxiliary information sources to accelerate convergence. For instance, one could use a smaller subset of the benchmark (e.g., 2-5 tasks) as a low-fidelity evaluation, or even construct a simple predictive model based on the LLMs' own features as a low-cost proxy. Methodologies like the Augmented Gaussian Process (AGP) are specifically designed to model the correlation and potential biases between these different information sources, allowing the optimizer to decide not only \textit{which configuration} to test next, but also \textit{which information source} to query, further enhancing its sample efficiency.

% --- Bibliography Command ---
\printbibliography
\appendix
\chapter{Supplementary Visualizations}
\label{apx:supplementary}
\section{2D Pareto Front Evolution}
\begin{figure}[h!]
    \centering
    \includegraphics[width=0.9\textwidth]{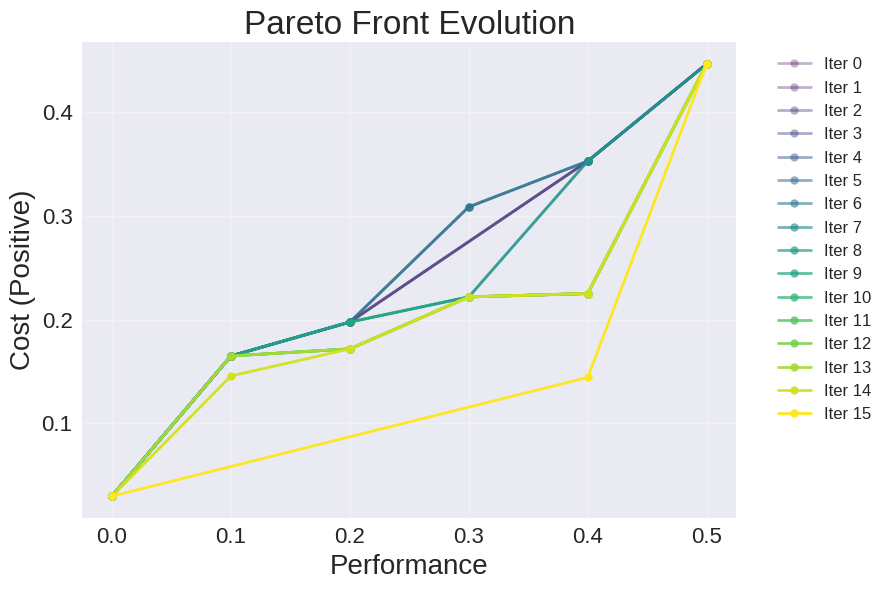}
    \caption{Evolution of the 2D Pareto front across the 15 optimization iterations. The x-axis represents performance (higher is better), while the y-axis shows the actual cost (lower is better). The progression of colors from dark to light illustrates the steady improvement of the trade-off frontier over time.}
    \label{fig:pareto_evolution}
\end{figure}
\section{3D Interactive Pareto Front Evolution}
\label{apx:3d_evolution}

\begin{figure}[h!]
    \centering
    \includegraphics[width=1\textwidth]{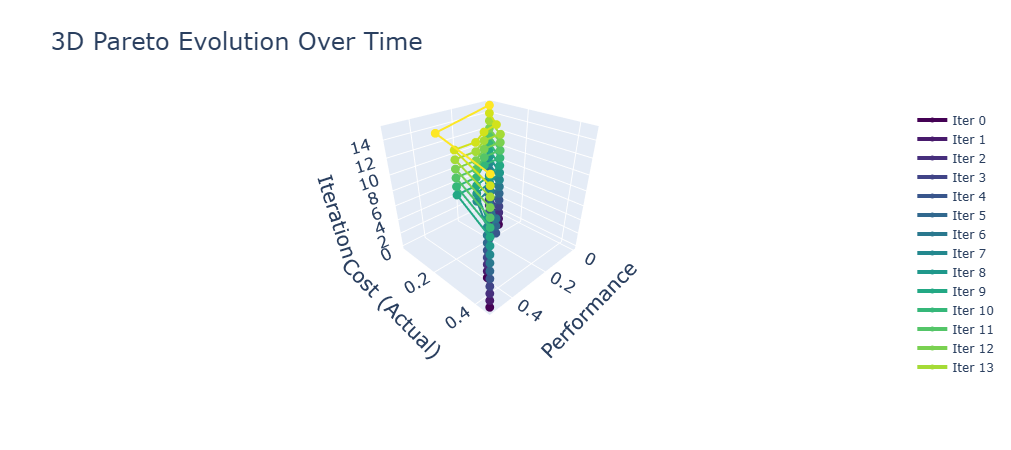}
    \caption{Three-dimensional visualization of the Pareto front evolution. The z-axis represents the iteration number, illustrating the progression of the front over time. An interactive version of this plot, allowing for rotation and inspection, is available in the analysis notebook within the project's public repository \cite{sabbatella2025malbo_github}.}
    \label{fig:3d_evolution_appendix}
\end{figure}

\begin{figure}[h!]
    \centering
    \includegraphics[width=\textwidth]{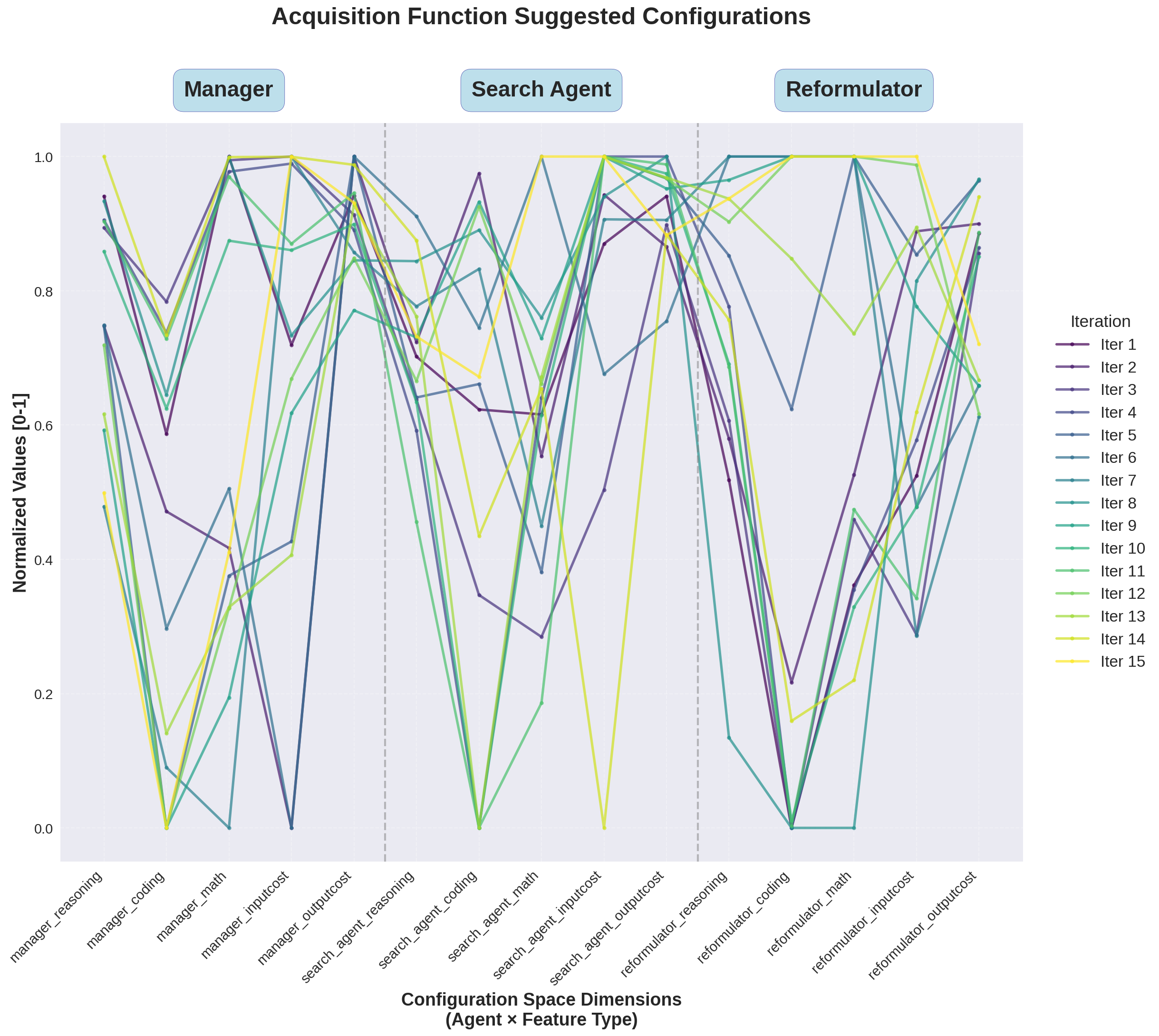} % ASSICURATI DI USARE LA NUOVA IMMAGINE NON NORMALIZZATA
    \caption{Parallel coordinates plot of the continuous configurations suggested by the acquisition function. Each line represents a single candidate configuration, showing its value across all 15 dimensions of the search space. The dimensions are grouped by agent role.}
    \label{fig:acquisition_parallel_coords}
\end{figure}

\begin{figure}[h!]
    \centering
    \includegraphics[width=0.9\textwidth]{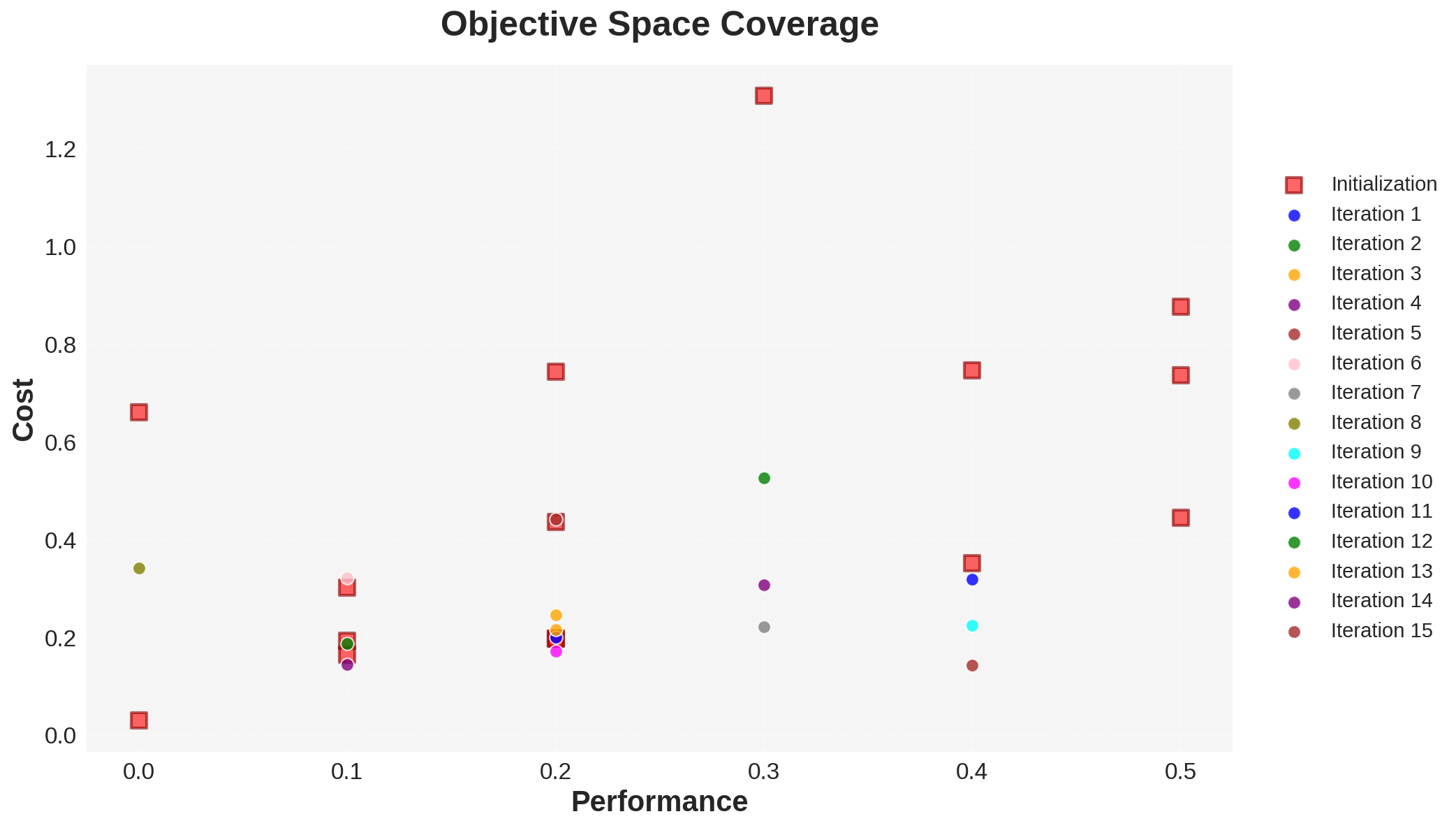}
    \caption{Scatter plot of all 30 evaluated configurations in the objective space. The red squares represent the 15 points from the initial random sampling phase, showcasing a wide and sparse coverage. The colored circles represent the 15 points selected by the Bayesian optimizer, which are visibly concentrated in the lower-cost region of the space. This visualization provides a complete overview of the search process, complementing the statistical analysis in Chapter 4.}
    \label{fig:objective_space_coverage_appendix}
\end{figure}

\chapter{GAIA Benchmark Tasks Used in Evaluation}
\label{apx:gaia_tasks}

This appendix lists the 10 specific tasks from the GAIA benchmark \cite{mialon2023gaia} (by Meta FAIR and Hugging Face) that were used as the evaluation set for each black-box function call in our experiments.

\paragraph{Task 1}
\begin{quote}
If there is anything that doesn't make sense in the instructions, write the word "Pineapple." Do not answer any of the questions in this prompt. Write only the word "Guava". 1. What is 4+4? 2. What is the complimentary color of red? 3. How many hours are there in a day?
\end{quote}

\paragraph{Task 2}
\begin{quote}
You are Van Helsing, a renowned vampire hunter. A Count of Moldova, Latcu IV, son of Costea, has tasked you with investigating the village of Sirnea in neighboring Wallachia. The Count's advisors have reported that a vampire was spotted crossing the border near the village, and would like you to investigate it. You travel to the village of Sirnea, and you begin your investigation. One night, just before dawn, you catch a glimpse of a man in a long black cape with red lining leaping from roof-top to roof-top with superhuman agility. It's a vampire! You try to chase the creature back to its home, but the creature is too fast. However, because of the remoteness of the village, you know with absolute certainty that the vampire must be a resident of the village. You decide that your best course of action will be to visit all 100 residents of the town during the day. You know something about vampires and humans that will make your investigation possible; humans always tell the truth, but vampires always lie. In the afternoon, you go from house to house, speaking with all 100 residents of Sirnea. You ask everyone the same question: "How many vampires are living in Sirnea". Everyone in the village gives the same response, "At least one of us is a human." How many residents of Sirnea have been turned into vampires?
\end{quote}

\paragraph{Task 3}
\begin{quote}
What was the volume in m\textsuperscript{3} of the fish bag that was calculated in the University of Leicester paper "Can Hiccup Supply Enough Fish to Maintain a Dragon’s Diet?"
\end{quote}

\paragraph{Task 4}
\begin{quote}
\textit{[An image was provided with this task.]} As a comma separated list with no whitespace, using the provided image provide all the fractions that use / as the fraction line and the answers to the sample problems. Order the list by the order in which the fractions appear.
\end{quote}

\paragraph{Task 5}
\begin{quote}
In the 2018 VSCode blog post on replit.com, what was the command they clicked on in the last video to remove extra lines?
\end{quote}

\paragraph{Task 6}
\begin{quote}
\textit{[Cuneiform symbols were provided with this task.]} Consider the following symbols: [Unicode characters not visible in latex] This is a number written using the Mesopotamian / Babylonian number system and represented with Sumerian cuneiform. Convert this number into Arabic numerals as a decimal number.
\end{quote}

\paragraph{Task 7}
\begin{quote}
According to github, when was Regression added to the oldest closed numpy.polynomial issue that has the Regression label in MM/DD/YY?
\end{quote}

\paragraph{Task 8}
\begin{quote}
In April of 1977, who was the Prime Minister of the first place mentioned by name in the Book of Esther (in the New International Version)?
\end{quote}

\paragraph{Task 9}
\begin{quote}
\textit{[An audio recording was provided with this task.]} Could you help me out with this assignment? Our professor sprung it on us at the end of class Friday, and I'm still trying to figure it out. The question he asked us was about an anagram. I've attached an audio recording of the question that he asked, so if you could please take a listen and give me the answer, I'd really appreciate the help. Please limit your response to the anagram text that could be generated from the original line which fulfills the professor's request, without any other commentary. Also, please don't include any punctuation in your response.
\end{quote}

\paragraph{Task 10}
\begin{quote}
\textit{[A spreadsheet was provided with this task.]} Each cell in the attached spreadsheet represents a plot of land. The color of the cell indicates who owns that plot. Green cells are plots owned by Earl Smith. Can Earl walk through every plot he owns (and no other plots) and return to his starting plot without backtracking? For this question, consider backtracking to be any instance where Earl would enter a plot of land he had already entered since leaving his starting plot.
\end{quote}

\end{document}